# CEP-IP: An Explainable Framework for Cell Subpopulation Identification in Single-cell Transcriptomics

**Kah Keng Wong**

Department of Immunology, School of Medical Sciences, Universiti Sains Malaysia, 16150 Kubang Kerian, Kelantan, Malaysia

**Background and Objective**

Single-cell RNA sequencing (scRNA-seq) frameworks lack explainable approaches for identifying cell subpopulations harboring strong pairwise monotonic gene-module relationships between a gene of interest (GOI) and its co-expressed genes. In this study, CEP-IP is introduced as a novel explainable machine learning framework to address this gap.

**Methods**

Prostate cancer (PCa) scRNA-seq dataset was used as the initial dataset, whereby *TRPM4* served as the GOI and its co-expressed ribosomal genes (Ribo) were identified via Spearman-Kendall dual-filter (*i.e.*, dual-filtered gene, DFG). Next, generalized additive modeling quantified the strength of *TRPM4*-Ribo relationship, represented by deviance explained (DE). *TRPM4*-Ribo's DE was then assigned to individual cells via cell explanatory power (CEP) classification, identifying cells harboring the *TRPM4*-Ribo module [*i.e.*, top-ranked explanatory power (TREP) cells]. *TRPM4*-Ribo transcriptional space was then stratified into pre-IP and post-IP regions using inflection point (IP) analysis, producing four distinct cell subpopulations per patient for pathway analysis. Validation was performed in the Allen middle temporal gyrus (MTG) and Neftel glioblastoma multiforme (GBM) transcriptomically heterogeneous datasets.

**Results**

*TRPM4*-Ribo modeling outperformed alternative gene set modules (FDR<0.05). In each PCa patient, CEP-IP yielded four cell subpopulations, where pre-IP TREP cells showed enrichment of immune-related processes, and post-IP TREP cells were enriched for ribosomal, translation, and cell adhesion pathways. In the MTG validation dataset (*CARM1P1*-DFGs module), post-IP TREP cells showed enrichment of neuron projection ontologies. In the GBM dataset, *FOXM1* was the sole GOI yielding mesenchymal-state DFGs, with *FOXM1*-DFGs post-IP TREP cells enriched for cell division and microtubule pathways; 3D trajectory analysis demonstrated continuous trajectories of TREP cells that were obscured in 2D embeddings.

**Conclusions**

CEP-IP identifies biologically distinct cell subpopulations in three independent scRNA-seq datasets. The framework may generalize to other pairwise GOI-DFGs module in single-cell transcriptomics beyond the datasets investigated in this study.

***Keywords:*** *CEP-IP; cell explanatory power; inflection point; generalized additive model; explainable AI.*

Correspondence: kahkeng@usm.my, kahkeng3@gmail.com

## 1. Introduction

Prostate cancer (PCa) is one of the most commonly diagnosed cancers, a predominant cancer type in elderly men with increasing incidence, and the second highest cancer-related mortality in males [1, 2]. While the five-year overall survival for PCa cases has improved to approximately 90% [3], aggressive PCa characterized by metastasis confers five-year survival rate of 30% [1, 4, 5], highlighting the ongoing needs to improve the outcomes of invasive PCa.

Transient receptor potential melastatin 4 (TRPM4), a member of the TRP superfamily, is a $Ca^{2+}$-activated, non-selective cation channel that is impermeable to $Ca^{2+}$ but transports monovalent cations including $Na^+$ and $K^+$ [6, 7]. TRPM4 has multiple physiological roles, particularly in regulating the physiological processes of cardiac tissues [8-11]. TRPM4 is overexpressed in multiple cancer types such as breast cancer [12, 13], lymphomas [14, 15], and PCa [16-18]. TRPM4 is an established oncoprotein in PCa as the ion channel is required for PCa tumor growth and its aggressive phenotypes including extravasation, invasion, and metastasis [19-22]. In particular, higher TRPM4 protein expression is associated with metastatic progression of PCa patients [23] and increased risk of biochemical recurrence in PCa [17]. There is a lack of single-cell RNA sequencing (scRNA-seq) analysis of *TRPM4* expression and its potential functions at single-cell level in PCa cases.

Genes whose expression moves in the same direction as the gene of interest (GOI) represent a consistent and co-regulated gene module in individual cells [24, 25]. Such monotonic co-expression is quantifiable via rank-based correlation measures such as Spearman's $r_s$ and Kendall's $\tau$. These non-parametric metrics capture non-linear but directionally consistent relationships between these genes, potentially reflecting a shared cellular program or state. When a GOI's monotonic partner genes are identified, their pairwise relationships can reveal which individual cells most strongly exhibiting their co-expression signal. However, existing machine learning (ML) techniques lack approaches to map this signal back to the individual cells, which may harbor distinct biology.

Generalized additive model (GAM) is an extension of generalized linear models where it models non-linear relationships between predictors and response variable, through the use of flexible splines [26-28]. The splines' flexibility allows modeling of numerous types of predictor-responder relationship, and GAM is widely used in various fields such as environmental sciences [29-31], engineering [32, 33], public health and biomedicine [34-36]. GAMs have also been utilized for scRNA-seq analysis, particularly for cell trajectory analysis [37, 38], attributable to GAM's strengths in modeling non-linear relationships of continuous variables without assuming predetermined functional forms. However, conventional GAM applications lack methods for assigning overall model performance into the level of individual data points. Moreover, certain methods for cell subpopulation identification rely on black-box algorithms that may not be as explainable as GAMs. This explainability gap represents a challenge in explainable artificial intelligence (XAI) applications in medicine, where mapping for the specific cancer cells that drive model predictions is key for potential therapies.

In this study, the potential functions of TRPM4 at single-cell level were investigated in scRNA-seq dataset of invasive PCa patients (GSE185344; 10x Genomics Chromium) [39]. By leveraging the flexibility and explainability of GAMs, TRPM4 and its potential response variables were modeled, with specific focus on optimizing the GAM models, mapping for individual cells with strong *TRPM4*-Ribo relationship to identify their distinct biological features, as well as the explainability of GAMs. Through the transparent and explainable GAM modeling, this study presents the cell explanatory power with inflection point (CEP-IP) framework, an explainable ML technique that assigns GAM performance into individual cell contributions. The framework introduces cell explanatory power (CEP) that determines which cells are best predicted by the model, combined with inflection point (IP) analysis that stratifies transcriptional space into biologically distinct quadrants. To assess generalizability beyond PCa and *TRPM4*-Ribo module, the framework was subsequently validated in two independent brain datasets, representing distinct tissue contexts, cell types, and sequencing platforms.

## 2. Methods

### 2.1 scRNA-seq dataset processing workflow

The scRNA-seq dataset of invasive and intraductal cribriform PCa cases (n=7) with matched non-cancerous samples (NonCa; n=7) of benign-enriched prostate cells (BP), annotated by the original study [39] as HYW_4701 (assigned as Pt.1 in this study), HYW_4847 (Pt.2), HYW_4880 (Pt.3), HYW_4881 (Pt.4), HYW_5386 (Pt.5), HYW_5742 (Pt.6), and HYW_5755 (Pt.7), were obtained from Gene Expression Omnibus (GSE185344).

Standard Seurat processing workflow was performed [40], and multiple quality control (QC) steps were performed to exclude low quality cells before downstream modeling analysis. Initially, cells with <500 unique genes expression were removed. Ribosomal gene filtering was then conducted by mapping for ribosomal genes with RP or MRP prefixes, before excluding top 10% of cells with the highest ribosomal content, minimizing potential bias in downstream analysis introduced by cells with high ribosomal expression. Next, mitochondrial gene filtering was performed by identifying genes with MT- prefix for the exclusion of top 10% of cells with highest mitochondrial content, filtering out stressed or dying cells. Cell cycle effects were then regressed out to remove cell cycle-driven variation that can compromise the main biological signals. Doublets were then removed using `scDblFinder` algorithm that identified potential doublets for exclusion. The number of cells before and after each of these QC steps are detailed in **Supplementary Table 1**. Batch effects correction was then conducted using the `SCTransform` function as the final QC step. Subsequently, dimensionality reduction was performed using principal component analysis (PCA) and uniform manifold approximation and projection (UMAP). This enabled clustering analysis by constructing k-nearest neighbor graph, and for the identification of the top 50 markers representing each cluster (**Supplementary Table 1**) using the `FindAllMarkers` function. The top markers for each individual cluster were then utilized to determine the cell type according to the enrichment of the Coexpression, Coexpression Atlas, and ToppCell Atlas gene sets available on the ToppGene database [41]. UMAP plots for both PCa and BP cells, with annotated cluster numbers or according to *TRPM4* expression levels, were generated.

### 2.2 Features selection for GAM modeling and CEP-IP framework

Genome-wide correlation analysis was conducted using Spearman ($r_s$) or Kendall ($\boldsymbol{\tau}$) correlation with the base R function `cor()`, of BP cluster 3 (from NonCa cases) versus all PCa clusters 0-22 (from PCa cases), and of PCa clusters versus each other clusters. In addition, $r_s$ and $\boldsymbol{\tau}$ correlation analysis of *TRPM4* with all genes was conducted in clusters 6, 9, 11, 14, and 19 combined in the PCa cases, and in cluster 3 (BP cells) from the NonCa cases, in pre-integration Seurat objects to avoid potential batch correction artifacts that may influence correlation patterns. Computation for $\boldsymbol{\tau}$ was accelerated utilizing 23 CPU cores (Intel Core i9-14900KF) by using the `doParallel` package. Genes that passed the correlation dual-filter, $r_s$ >0.6 and $\boldsymbol{\tau}$ >0.5, termed as dual-filtered gene (DFG) were shortlisted for gene set enrichment analysis.

Heatmap to compare the expression of the seven dual-filtered ribosomal (*RPL10*, *RPL27*, *RPL28*, *RPS2*, *RPS8*, *RPS12*, and *RPS26*) and seven AR-related (*KLK4*, *KLK2*, *KLK3*, *PDLIM5*, *ABHD2*, *ALDH1A3*, and *SORD*) genes, as well as each of the seven genes' averaged expression representatives termed as Ribo and AR in this study, in PCa and NonCa clusters. In each cluster, cells were ordered according to ascending *TRPM4* expression level. Expression was represented as z-score using the mako palette of the `viridis` package. To assess whether averaging the expression of the seven ribosomal or AR genes was suitable for downstream modeling, the internal reliability of these two gene sets was assessed for PCa and NonCa cases. For each sample and gene set combination, the consistency among individual genes within the averaged gene set was tested. The metrics used for the evaluation were Cronbach's alpha ($\alpha$), McDonald's omega ($\omega$), and the Kaiser-Meyer-Olkin (KMO) measure of sampling adequacy, using the `psych` package. A consolidated score was calculated by averaging the values of these three metrics.

### 2.3 Optimization and explainability of *TRPM4*-ribosomal (*TRPM4*-Ribo) relationship modeling with GAM

Modeling with GAM was conducted using the `mgcv` package [28, 42, 43]. The GAM model was initialized to take the form $f(x_i)=\beta_0 + \beta_1 x_i + b_1\varphi_1(x_i) + \ldots + b_n\varphi_n(x_i)$ where each cell's *TRPM4* value $x_i$ was implemented to predict the expression value $f(x_i)$ of seven gene sets separately (*i.e.*, Ribo, AR, GSK-3β, mTOR, NF-κB, PI3K/AKT, and Wnt, with the Ribo and AR gene sets utilized for optimization and explainability). Optimization steps were conducted for the modeling to minimize potential overfitting or underfitting of the fitted GAM curve.

To achieve this, selection of the k value (maximum number of smooth basis functions) was optimized by minimizing penalized residual sum of squares (PRSS) via two phases *i.e.*, initial exploration phase (PRSS iterations 1-8), and convergence verification phase (PRSS iterations 9 up to maximum 100 with early stopping). In the exploration phase, cells were extracted from specific clusters (*e.g.*, the five PCa clusters 6, 9, 11, 14, and 19) and samples (*e.g.*, PCa samples) using Seurat objects, and average gene set expression versus *TRPM4* expression were calculated before $\log_2$ transformation of the expression values.

Then, the exploration phase tested different k values (k=3-10) in the first eight iterations, followed by PRSS convergence verification phase that alternated between using the best k value and sampling from neighboring k values (±1 of best k) for up to 92 iterations. The alternation occurred every third iteration (*i.e.*, iterations 9, 12, 15, etc. sampled randomly from the ±1 range, while other iterations used the best k value). Early stopping was initiated if no improvement in PRSS value was observed for 20 consecutive iterations, indicating PRSS value convergence. The key lines of code illustrating the exploration of k values in both phases of PRSS optimization are as follows:

```
# Exploration phase: In the first eight iterations, explore k values
ranging from 3 to 10
max_k <- min(10, unique_trpm4 - 1)
k <- 3 + ((i - 1) %% (max_k - 2))

# PRSS convergence verification phase: Up to 92 iterations (iterations
no. 9 to 100), tested random ±1 best k value every third iteration,
early stopping after 20 consecutive iterations without PRSS
improvement
if (best_k == max_k) {
    k_range <- c(max_k - 1, max_k)
```

```
} else if (best_k == 3) {
    k_range <- c(3, 4)
} else {
    k_range <- c(best_k - 1, best_k, best_k + 1)
}

# Try best_k±1 range every third iteration
if (i %% 3 == 0) {
   k <- sample(k_range, 1)
} else {
    k <- best_k
}
```

In each PRSS iteration, GAM was fitted using thin-plate regression spline (TPRS). For each PRSS iteration, the fitted spline was penalized by the smoothing parameter λ derived from restricted maximum likelihood (REML). Hence, PRSS was the outer loop (running both the exploration and verification phases described earlier, where the lowest PRSS value yielded the best k value) and REML was the inner loop (where the lowest REML score yielded the best λ value) in the optimization process to yield optimal GAM fitting. For each PRSS iteration (outer loop), the PRSS iteration number, PRSS value, k value, REML iteration count, and final REML score, were captured. The summarized code is as follows:

```
# PRSS and REML as the outer and inner loop, respectively
for (i in 1:num_iterations) {
    # Fit GAM model with PRSS-optimized k value
    gam_model <- gam(Expression ~ TRPM4 + s(TRPM4, bs = "tp", k = k),
      data = sample_data, method = "REML", select = TRUE, gamma = 1.5)

    # Calculate PRSS for outer loop optimization
    prss_components <- calculate_prss(gam_model, sample_data)
    current_prss <- prss_components$PRSS
}
```

For each REML iteration (inner loop), the REML score, λ value, iteration number, REML convergence, gradient and Hessian values were extracted. Gradient extraction was achieved by accessing `model$outer.info$grad`, where `mgcv` stored the first three iterations. Hessian extraction was challenging as `mgcv` stored only the final convergence Hessian and not for earlier REML iterations during REML optimization. To address this, inverse Hessian ($H^{-1}$) value was backcalculated using the Newton-Raphson method by utilizing the rest of the available components including the initial λ value ($\lambda_{old}$), its consecutive λ value ($\lambda_{new}$), and gradient value. Using the Newton-Raphson method formula $\log(\lambda_{new}) = \log(\lambda_{old}) - H^{-1}g$, the inverse Hessian was backcalculated as $H^{-1} = (\log(\lambda_{old}) - \log(\lambda_{new}))/g$

For REML, monotonic decrease in REML score across iterations was assessed to verify that REML score reduced with each successive iteration. In addition, REML convergence was determined by either of these criteria:

i) Maximum number of iterations reached, following the default set by `model$control$maxit`

ii) Gradient-based convergence: when gradient norm (GN) fell below threshold ($<1 \times 10^{-5}$).

iii) Score-based convergence: when relative REML score change dropped below threshold ($<1 \times 10^{-6}$).

iv) Fallback convergence: when `mgcv` reported convergence for reasons other than the aforementioned.

The PRSS and REML optimization processes were visualized in line plots showing how their values changed across iterations, and their convergence. For the best model (*i.e.*, fitted with k and λ derived from the best PRSS iteration), multiple model information was extracted including each cell's coefficient values for the linear and smooth terms (*i.e.*, spline basis functions $\varphi_1$, $\varphi_2$, etc.), the knot locations, model and null deviances, smooth terms' effective degrees of freedom and their *p*-values adjusted for multiple testing by the Benjamini-Hochberg method (false discovery rate, FDR), testing the significance of non-linear relationships, individual cell-level *TRPM4* expression, and gene set expression values (actual or predicted values by the model).

Multi-gene set analysis was performed in batch for all samples. The fitted full model of the relationship between *TRPM4* expression (x-axis, $\log_2$ scale) and various gene sets (Ribo, AR, GSK-3β, mTOR, NF-κB, PI3K/AKT, and Wnt; y-axis, $\log_2$ scale) was visualized in scatter plots with fitted GAM curves. To adjust for optimal smoothness of the fitted curves, the γ parameter in REML that controls penalization of smooth terms was tested in 0.5 increments (γ range tested: 0.5-3.0), and the resulting GAM curves were visually inspected for signs of overfitting, underfitting, or optimal model fit.

### 2.4 Visualization of TPRS basis functions formation

To visualize the smooth basis functions (TPRS) formation, *TRPM4*-Ribo modeling in Pt.4 was used as the representative and GAM fitting was performed with optimized parameters (k=10, λ=0.528). Initially, basic statistical analysis (*e.g.*, data distribution) was demonstrated before showing representative knots for the formation of radial basis functions into actual spline shapes. Individual smooth basis functions ($\varphi_1$-$\varphi_8$) in unpenalized forms were then extracted from the model's prediction matrix. Coefficients were applied, yielding penalized smooth basis functions that were then cumulatively combined to demonstrate how the complete smooth terms ($\varphi_1$-$\varphi_8$ combined) differed from those with incomplete smooth basis functions (*e.g.*, $\varphi_1$ and $\varphi_8$ only).

The final GAM curve was formed by adding complete smooth terms with the linear terms, and three data points with low (5th percentile), medium (50th percentile), and high (95th percentile) *TRPM4* values were selected as representatives for the calculation of the full model's predicted Ribo expression values. The variance contribution of each term to the full model was extracted to compare the magnitude of their contributions to the model.

**2.5 CEP-IP framework: The CEP classification**

The CEP-IP framework was developed and introduced here to identify biologically distinct cell subpopulations. This was achieved by combining the CEP classification for identifying well-predicted cells, and IP analysis for spatial stratification of transcriptional space. Deviance explained (DE) was the performance metrics used in this study to measure how much of variability in a specific gene set's expression was captured by *TRPM4*. As the spline in GAM was fitted according to the distribution of cells in *TRPM4*-gene set transcriptional space, specific cells with strong *TRPM4*-gene set relationship would be well-predicted by the fitted model, and this population of cells may be biologically relevant. Hence, attempts to decompose DE into cell-level were conducted, where Gaussian GAM's DE aggregate sums were decomposed into cell-level consisting of null deviance contribution (NDC), model deviance contribution (MDC), and explanatory power (EP) as follows:

***DE (Gaussian GAM):***

Null deviance (ND) = $\Sigma(y_i - \bar{y})^2$ (*sum of squared null residuals from all cells)*

Model deviance (MD) = $\Sigma(y_i - f(x_i))^2$ *(sum of squared model residuals from all cells)*

DE = 1 - (MD / ND) *(proportion of variability explained by the model)*

***EP:***

NDCi = $(y_i - \bar{y})^2$ *(i-th cell ND contribution)*

MDCi = $(y_i - f(x_i))^2$ *(i-th cell MD contribution)*

$EP_i = 1 - (MDC_i / NDC_i)$ *(proportion of i-th cell-level variability explained)*

When all individual cell's EP had been computed, the cells were then ranked according to EP values from the highest to the lowest. Based on this ranking, the top DE% were selected as top-ranked EP (TREP) cells. For instance, if total of 100 cells and DE was 45%, then cells ranked first to 45th by EP were assigned as TREP cells, while cells ranked 46th to 100th were non-TREP cells. When DE resulted in fractional cell numbers (*e.g.*, DE=45.6% with 100 cells), the exact number was determined by rounding $n \times$ DE to the nearest integer.

**2.6 CEP-IP framework: Monte Carlo cross-validation (MCCV) of CEP classification**

To validate the accuracy of CEP classification in decomposing DE into cell-level assignment of TREP and non-TREP cells, MCCV was conducted where PCa cells of each patient were divided into training:test sets in 70:30 ratio for 20 randomized iterations. In the training set, the optimal *TRPM4*-Ribo GAM model determined from prior analysis (with optimized k, $\lambda$, and $\gamma$ derived from the minimized PRSS, REML, and GAM curves visual inspection, respectively) was fitted for each case. The cells were then CEP-classified into TREP and non-TREP as described previously *i.e.*, according to EP and DE values.

In the test set, the trained GAM model was applied to predict the Ribo values of the cells required to generate their EP values, and training set's DE was then applied to binarize test set cells into TREP and non-TREP. Root mean squared error (RMSE) was calculated for both TREP and non-TREP groups, where each cell's actual versus predicted Ribo value residuals were computed, and RMSE was the square root of the mean of TREP (or non-TREP) cells' squared residuals [`√(mean(test_residuals[TREP_cells]^2))`]. Difference in the RMSE values for both TREP and non-TREP groups were then computed. Significance of the RMSE differences between both groups across 20 iterations was computed using one-sample t-test (against a null hypothesis of zero RMSE difference) or Wilcoxon signed-rank test if RMSE differences had normal or non-normal distribution, respectively, using Shapiro-Wilk test.

For comparison with negative controls, the cells were randomly assigned as TREP and non-TREP cells (maintaining the same proportion as the training set's DE) instead of CEP classification and underwent identical analysis. For comparison with an additional control group, leverage-based classification was adopted to assign cells into TREP and non-TREP, whereby statistical leverage (hat values) was first computed to generate influence scores [combination of leverage and standardized residuals, `influence_score = (leverage × standardized_residuals^2)^(1/2)`] used to rank the cells, replacing EP-based ranking before selecting the top DE% as leverage-based TREP cells. Downstream RMSE calculations and comparisons between both cell groups for random or leverage-based classification were computed as described above for CEP-based classification.

## 2.7 CEP-IP framework: Spatial stratification with IP analysis, distribution pattern, and Gene Ontology (GO) analysis of cell subpopulations

IP in a GAM plot represented the *TRPM4* expression value, determined visually, where distribution pattern of CEP-classified TREP cells (colored in purple; non-TREP cells colored in gray) shifted near the midpoint of *TRPM4* expression range, where the number of TREP cells were immediately more above the GAM curve. IP binarized the GAM plot into the pre-IP (*TRPM4*<IP) and post-IP (*TRPM4*≥IP) regions on the x-axis scale (*TRPM4* expression). Pre-IP was characterized by a pattern of decreasing TREP cells frequency toward the IP, while the post-IP region exhibited increasing frequency of TREP cells away from the IP. Differences in the proportion of TREP versus non-TREP cells above and below the GAM curve was assessed using chi-square test or Fisher's exact test (when any expected count was <5 in the contingency table), adjusted by FDR and represented in mosaic plots. Ribo values of pre-IP cells were compared with post-IP cells in raincloud plots, and overlap between both regions was compared using overlap coefficient (OVL).

Differential gene expression (DEG) analysis was performed in TREP versus non-TREP cells within pre-IP or post-IP regions separately, using Seurat's `FindMarkers` function with Wilcoxon rank-sum tests. DEG detection required at least 10% gene expression frequency, $\log_2$ fold-change ($\log_2$FC) threshold of 0.1, and minimum three cells per comparison group. DEGs with $p<0.01$ and $\log_2$FC>0.2 were shortlisted for further analysis, except in Pt.5 adopting DEGs with $p<0.05$ and $\log_2$FC>0.1 due to insufficient DEGs for downstream analysis with prior thresholds. For cases with >500 shortlisted DEGs, the top 500 DEGs ranked according to the most significant *p*-values were selected to avoid false positives beyond 500 DEGs. GO enrichment analysis was performed using the ToppGene platform with the full gene set used as the background gene set. Each patient's pre-IP and post-IP regions consisted of GOs upregulated in TREP or non-TREP cells. The top 50 GOs with FDR<0.05 were shortlisted. GOs with similar annotations were compiled into a consensus functional group, and GO with the most significant FDR within a functional group was selected for comparison with other functional groups' most significant GO.

## 2.8 CEP-IP framework: Automated IP detection and IP reliability score (IPRS)

After establishing the CEP-IP framework utilizing the *TRPM4*-Ribo dataset, the framework was applied to validation datasets that required automated IP detection. To achieve this, TREP cell residuals were computed against the fitted main GAM curve $[R_i = y_i - f(x_i)]$ in each sample. A secondary GAM, R(x), was fitted to these residuals as a smooth function of GOI expression using TPRS with REML optimization, applied to TREP cells only. R(x) models the expected signed residual of TREP cells at any given GOI expression level. For visualization, the mean signed residual within each GOI expression interval was represented by equal-width bins. The IP was defined as the first negative-to-positive zero-crossing of the smoothed R(x) curve, evaluated at 2,000 equally-spaced GOI expression values (*i.e.*, grid points on the x-axis) of the observed expression range. Due to REML penalisation of the R(x) curve, this crossing may occur later than apparent sign changes in the binned residual visualization. The precise crossing location was determined between adjacent grid points by solving for R(x)=0 via linear interpolation: $x_{IP} = x_0 - \mathrm{R}(x_0) \times (x_1 - x_0) / (\mathrm{R}(x_1) - \mathrm{R}(x_0))$

IP quality was quantified using the IP reliability score (IPRS), a composite value (0-1) averaged from five components: C1 (95% CI width relative to GOI expression range at IP); C2 (pre- vs. post-IP residual shift significance by FDR); C3 [R(x) steepness at IP via central difference]; C4 [mean R(x) confidence band width normalized to amplitude]; C5 [R(x) total variation and post-IP sign changes penalization]. IPRS is divided into five tiers: Very strong (IPRS ≥0.80), strong (0.60-0.79), moderate (0.40-0.59), weak (0.20-0.39), and failed (<0.20). Moderate and weak IPRS samples were flagged for manual visual review to confirm post-IP TREP enrichment above the main GAM curve (or below if inverse relationship), while failed samples were excluded from downstream analysis. Samples with any C1-C5 component scoring zero were also flagged for manual visual review, as this typically indicates insufficient TREP cells. For instance, C2=0 occurs when insufficient TREP cells in pre- vs. post-IP preventing significance test entirely.

### 2.9 CEP-IP framework: Monocle3 trajectory analysis of mapped TREP and non-TREP cells

Seurat-processed PCa and NonCa objects were converted to Monocle3 format for trajectory analysis, followed by dimensionality reduction with PCA and UMAP embedding. Trajectory roots were defined as cells with the highest GOI expression, and cells were clustered with trajectory graphs. Pseudotime ordering was performed to capture cellular transitions along the learned trajectories. Quantitative assessment of TREP cells separation between pre-IP and post-IP regions, in terms of UMAP1 coordinate distributions, was performed using t-test (normal distribution) or Mann-Whitney U test (non-normally distributed data) with Cliff's delta ($\delta$) effect size estimation. Ridgeline plots were generated to visualize UMAP1 coordinate distributions across all four cell groups (pre-IP TREP, pre-IP non-TREP, post-IP TREP, post-IP non-TREP), and *p*-values of the UMAP1 comparisons between both regions (pre-IP versus post-IP) were computed and FDR-corrected for all patients.

### 2.10 Validation of CEP-IP framework: Allen Human Middle Temporal Gyrus (MTG) dataset

Subsequently, the CEP-IP framework was examined in two independent brain datasets. The first was the Allen Human MTG SMART-seq dataset [44] (15,928 nuclei, 8 donors, aged 24-66 years) obtained from Allen Institute for Brain Science database (https://brain-map.org/) and processed using the same Seurat pipeline. Two MTG samples, H200.1030 (termed in this study as MTG.1) and H200.1023 (MTG.2) were shortlisted for further analysis, while the rest of the six samples were excluded due to each had <50 cells per sample post-processing. The Spearman (sensitivity screen) and Kendall (specificity filter due to concordance stringency) dual-filter correlation analysis ($r_s$ >0.5 and $\tau$ >0.4; thresholds reduced by 0.1 as no gene survived the stricter $r_s$ >0.6 and $\tau$ >0.5 thresholds in the MTG dataset; The lower thresholds retained the dual-filter requirement, while reflecting that scRNA-seq datasets inherently vary in expression dynamics; If no genes pass the reduced thresholds, further lowering of thresholds is not recommended) was applied to identify genes monotonically correlated with *CARM1P1* (the GOI) in the highest-expressing cluster, yielding DFGs consisting of *CLMN*, *EPHA3*, *EPHA6*, *LOC101928964*, and *ROBO2* as the response variables for GAM modeling. The housekeeping gene (HKG) gene set (*ACTB*, *GAPDH*, *PPIA*, *RPL13A*, and *TBP*) was devised as the comparator.

In contrast with the PCa dataset where all DFGs were ribosomal and averaged into a single Ribo composite, the MTG dataset yielded distinct gene categories. Thus, individual genes within each set were averaged into composite scores *i.e.*, composite DFG (cDFG) and composite HKG (cHKG), following the same averaging rationale applied to Ribo and AR in the PCa dataset. GAM was fitted using the same PRSS-REML optimization framework (*e.g.*, PRSS iterations, $\gamma$=1.5, TPRS) for both gene sets. CEP classification was applied to assign model-level DE into cell-level EP, classifying the top DE% of cells as TREP cells, followed by automated IP detection. DEG analysis between TREP and non-TREP cells within pre-IP and post-IP regions was conducted for downstream GO enrichment analysis, and Monocle3 trajectory analysis was performed with pseudotime rooted at the highest *CARM1P1*-expressing cell.

### 2.11 Validation of CEP-IP framework: Glioblastoma multiforme (GBM) dataset

The CEP-IP framework was subsequently examined in the Neftel GBM SMART-seq2 dataset [45] obtained from Single Cell Portal (https://singlecell.broadinstitute.org/). The dataset consisted of adult *IDH*-wildtype malignant GBM cells processed using the same standard Seurat workflow. Unlike the MTG dataset, GBM cells were stratified by four canonical cell states *i.e.*, neural-progenitor-like (NPC), oligodendrocyte-progenitor-like (OPC), astrocyte-like (AC), and mesenchymal-like (MES), established by Neftel *et al*. [45]. Dual-filter correlation analysis ($r_s$ >0.6 and $\tau$ >0.5) was applied within each canonical cell state pool independently to identify genes monotonically correlated with *FOXM1* (the GOI), with emphasis on the MES canonical cell state. This yielded a frequency-weighted DFG composite (wcDFG: *BIRC5*, *MKI67*, *CENPF*, *TOP2A*, *PBK*, *TROAP*, and *NUSAP1*; weights proportional to the number of cell states each gene was identified in) vs. cHKG (*ACTB*, *PGK1*, *PPIA*, *RPL13A*, and *SDHA*) as comparator.

Next, within-positive monotonicity of the wcDFG was validated by restricting Spearman and Kendall correlations to *FOXM1*$^+$ cells only ($\log_2$ >0), confirming genuine co-expression relationships independent of zero-concordance inflation. The study's standard GAM fitting was performed in *FOXM1*$^+$ cells. Subsequently, CEP classification, automated IP detection with IPRS scoring, DEG analysis, GO enrichment, and Monocle3 trajectory analysis were applied following the same procedures. Monocle3 trajectory analysis was also performed using 3D UMAP embeddings, with TREP cell separation quantified on each UMAP axis (UMAP1-3), and 3D centroid test, PERMANOVA with permutations on Euclidean distance matrices (`adonis2`), and PERMDISP homogeneity of dispersion testing (`betadisper`).

### 2.12 Statistical analysis

The Shapiro-Wilk test was conducted to test normality of the distribution of continuous variables. For comparison of continuous variables between two groups, t-test or Mann-Whitney U test was conducted for normal or non-normally distributed data, respectively. For continuous variables' comparison between more than two groups, ANOVA with Holm-Šídák's post hoc test or Kruskal-Wallis test with Dunn's post hoc test was conducted for normal or non-normally distributed data points, respectively. The Benjamini-Hochberg method was used to correct the *p*-values for multiple testing, yielding FDR, where FDR<0.05 was considered as significant. Interpretation of Cliff's $\delta$ effect size was according to established cut-offs [46]: negligible ($\delta$ <0.15), small ($\delta$ ≥0.15), medium ($\delta$ ≥0.33), and large ($\delta$ ≥0.47). All analysis was conducted using RStudio, except the boxplot comparison of DE values between different gene sets were conducted using GraphPad Prism v10 (CA, USA).

## 3. Results

### 3.1 Identification and characterization of five distinct PCa cell clusters

In the investigated scRNA-seq dataset (GSE185344) [39], two different groups of cases derived from PCa patients were assessed *i.e.*, benign adjacent (NonCa) and PCa groups, each consisting of seven cases. In the NonCa group, *TRPM4* was highly expressed in BP cells (cluster 3; n=890 cells), where its levels were significantly higher in BP than other cell clusters (FDR<0.01) such as immune cells (*e.g.*, NK, NKT, helper T cells, B cells, macrophages), endothelial cells, and fibroblasts (**Figure 1A** and **Supplementary Table 2**).
In the PCa group, UMAP clustering showed that *TRPM4* levels were most elevated, based on median levels, in five distinct PCa clusters *i.e.*, clusters 6 (n=1,869 cells), 9 (n=1,618 cells), 11 (n=1,222 cells), 14 (n=824 cells), and 19 (n=322 cells) (**Figure 1B** and **Supplementary Table 2**). *TRPM4* levels were also elevated, although to a lesser extent, in clusters 16 [potential internal BP (IBP) cells *i.e.*, benign cells found within cancerous tissues; n=733 cells] and 22 (epithelial cells; n=38 cells).

As *TRPM4* was most highly expressed in cluster 14, *TRPM4* levels in cluster 14 were compared with each of the other cell clusters, showing that *TRPM4* levels were significantly higher in cluster 14 (FDR<0.01) except when compared with cluster 11 or 19. As the PCa group consisted of seven distinct PCa patients, PCa cell heterogeneity was captured by the UMAP clustering represented as five PCa clusters (clusters 6, 9, 11, 14, and 19), with each cluster positioned adjacent or close to each other in the UMAP plot and with similar *TRPM4* expression levels. Cluster 16 was potentially IBP cells based on its similar *TRPM4* median levels (*TRPM4* value: 0.96) with those of BP cells (NonCa group cluster 3; *TRPM4* value: 1.02) but lower compared with the aforementioned five PCa clusters (*TRPM4* value: all >1.5; **Figure 1B**). To assess this further:

i) Qualitatively: The top 50 markers representing each cluster (**Supplementary Table 1**) were examined and each of the five PCa clusters demonstrated markers known to be overexpressed in PCa cells *i.e.*, PCa group cluster 6 (*OR51E2*), 9 (*PMEPA1*, *GLIPR1*), 11 (*SCHLAP1*, *CTAG2*), 14 (*NNMT*, *TSPAN8*), and 19 (*PRAC1*). These markers were absent from the top 50 markers significantly associated with the BP cluster (NonCa group cluster 3) or the IBP cluster (PCa group cluster 16). For both BP and IBP clusters, conventional BP markers were present including *MME* and *SCGB1D2*, and these markers were absent from the top 50 markers representing each PCa cluster (clusters 6, 9, 11, 14, and 19).

ii) Quantitatively: $r_s$ and $\tau$ correlation matrix was constructed to compare the transcriptome profile similarities between BP versus the five PCa and IBP clusters, and epithelial cell cluster (PCa group cluster 22, as controls) (**Figure 1C**). BP and IBP clusters shared high transcriptome similarities ($r_s$=0.99, $\tau$=0.91), while BP cluster showed consistently lower $r_s$ (<0.95) and $\tau$ (<0.80) with the rest of the PCa group clusters.

### 3.2 Spearman-Kendall dual-filter reveals *TRPM4* correlation with ribosomal gene expression in PCa cells

Next, $r_s$ and $\tau$ correlation values of *TRPM4* with all genes in IBP cells (y-axis) and PCa cells from the five PCa clusters (hereby termed as PCa cells for simplicity) were examined, and genes passing the Spearman-Kendall dual-filter ($r_s$ >0.6 and $\tau$ >0.5) were shortlisted (in **Figure 1D**). No gene passed the Spearman-Kendall dual-filter in IBP cells, but in PCa, seven genes (*RPL10*, *RPL27*, *RPL28*, *RPS2*, *RPS8*, *RPS12*, and *RPS26*) passed the dual-filter (**Figure 1D** and **Supplementary Table 3**). As these seven were ribosomal genes, ribosomal gene sets were significantly enriched (FDR <0.01).

In order to gain preliminary insights on the potential functions of TRPM4 on those seven ribosomal genes in PCa cells, correlation values ($r_s$ and $\tau$) of all genes with each of the seven ribosomal genes individually were computed, and genes that passed the Spearman-Kendall dual-filter were shortlisted. For instance, 342 genes showed $r_s$ >0.6 and $\tau$ >0.5 (the dual-filter's cut-offs) with *RPL10*, and these genes were considered as *RPL10*-monotonic genes in PCa cells. Subsequently, the DFGs across each of the seven ribosomal genes were compared for consensus DFGs, yielding a total of 56 and 25 ribosomal and non-ribosomal genes, respectively. *TRPM4* was one of the 25 non-ribosomal genes that passed the dual-filter for each of the seven ribosomal genes. GO enrichment of the 25 non-ribosomal genes (excluding ribosomal genes to uncover functions regulating or supporting ribosomal processes) was conducted and GOs containing *TRPM4* in their gene lists were shortlisted, demonstrating that regulation of protein or cellular localization, and transporter complex GOs were significantly enriched (FDR<0.01) (**Figure 1E** and **Supplementary Table 4**).

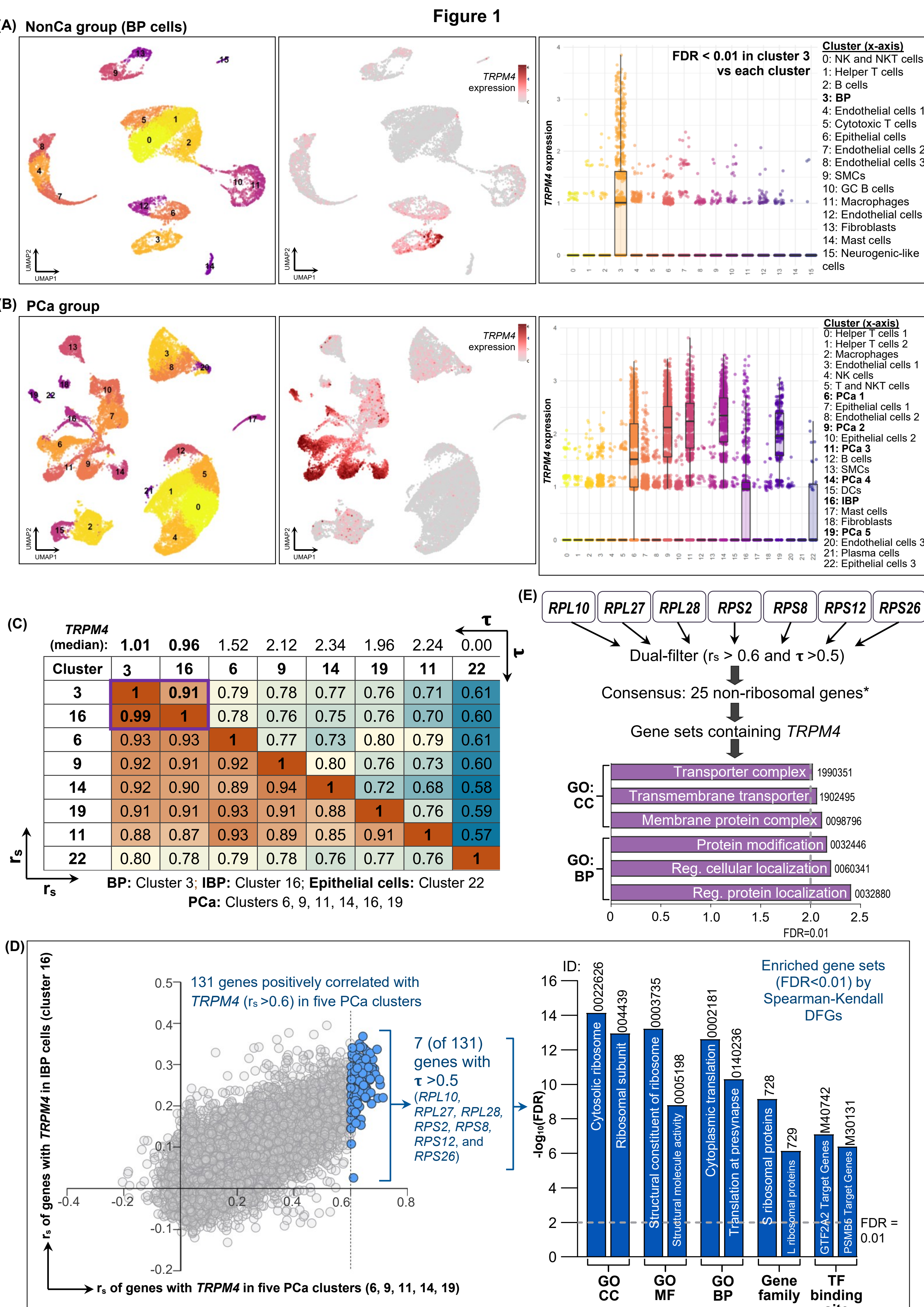

**Figure 1**. *TRPM4* expression profile and enriched gene sets in scRNA-seq dataset (GSE185344) of PCa and NonCa (n=7 each group). (A) UMAP plot of sixteen cell clusters (0-15) in NonCa group (n=12,205 cells; left panel), *TRPM4* expression levels in the cell clusters (middle panel), and comparison of *TRPM4* expression levels in BP cluster 3 versus each other cluster (right panel); (B) UMAP plot of 23 cell clusters (0-22) in PCa group (n=30,932 cells; left panel), *TRPM4* expression levels in the cell clusters (middle panel), and boxplot with jitter plot comparison of *TRPM4* levels in each cluster (right panel); (C) $r_s$ and $\tau$ correlation matrix of BP cluster 3, IBP cluster 16 (from PCa group), and five PCa clusters 6, 9, 11, 14 and 19, and epithelial cells cluster 22. Median levels of *TRPM4* in each cluster are shown on top of the matrix; (D) Scatter plot of $r_s$ values of all genes in relation to *TRPM4* expression in IBP cells (cluster 16 of PCa group) versus five PCa clusters (clusters 6, 9, 11, 14 and 19). Spearman-Kendall dual filters were applied to shortlist for *TRPM4*-monotonic genes and the corresponding enriched gene sets (FDR<0.01) shown in bar plot; (E) For each of the seven *TRPM4*-monotonic genes (all ribosomal genes), dual filters were applied to identify consensus genes in the five PCa clusters *i.e.*, the group of genes with $r_s$ >0.6 and $\tau$ >0.5 in relation to each of the seven ribosomal genes in PCa cells. A total of 25 non-ribosomal consensus genes were identified and the enriched gene sets (FDR<0.01) containing *TRPM4* are shown in bar plot. *There were 56 ribosomal consensus genes. GO BP: GO Biological Process; GO CC: GO Cellular Component; GO MF: GO Molecular Function; TF: Transcription factor.

### 3.3 Validation of gene set representatives and distribution family for GAM analysis

Next, the seven ribosomal genes' expression values were averaged to yield a representative Ribo expression, for each patient separately, for downstream analysis. This excluded the need to model *TRPM4* with each of the seven ribosomal genes separately. Multiple analyses were conducted to test the reliability of averaging the seven ribosomal genes into Ribo as the representative. Cronbach's α (0.969-0.992) and McDonald's ω (0.971-0.991) were >0.95 for all patients, indicating internal consistency of each ribosomal gene expression values across the patients. All patients' KMO values were >0.90 (0.934-0.961), supporting sampling adequacy for downstream factor analysis utilizing Ribo. The CS score (average of Cronbach's α, McDonald's ω, and KMO scores) for averaging into Ribo was >0.95 and >0.85 in PCa and BP cells, respectively (**Supplementary Table 5**).

AR signaling is a key pathway that controls prostate-specific gene expression in both BP and PCa cells [47], and TRPM4 is expressed in androgen-sensitive PCa cells potentially involved in their AR signaling [18, 48]. Hence, seven common AR pathway genes were included (*i.e.*, *ABHD2*, *ALDH1A3, KLK2*, *KLK3*, *KLK4, PDLIM5*, and *SORD*) and averaging them to represent AR as a control group to compare with Ribo in downstream modeling. The CS for averaging into AR was >0.85 (except Pt.5 CS of 0.814) and >0.74 (except Pt.5 CS of 0.663) in PCa and BP cells, respectively (**Supplementary Table 5**). Heatmap of each of the seven ribosomal and AR pathway genes, and their averaged representatives Ribo and AR, in PCa and BP cells are presented in **Figure 2**.

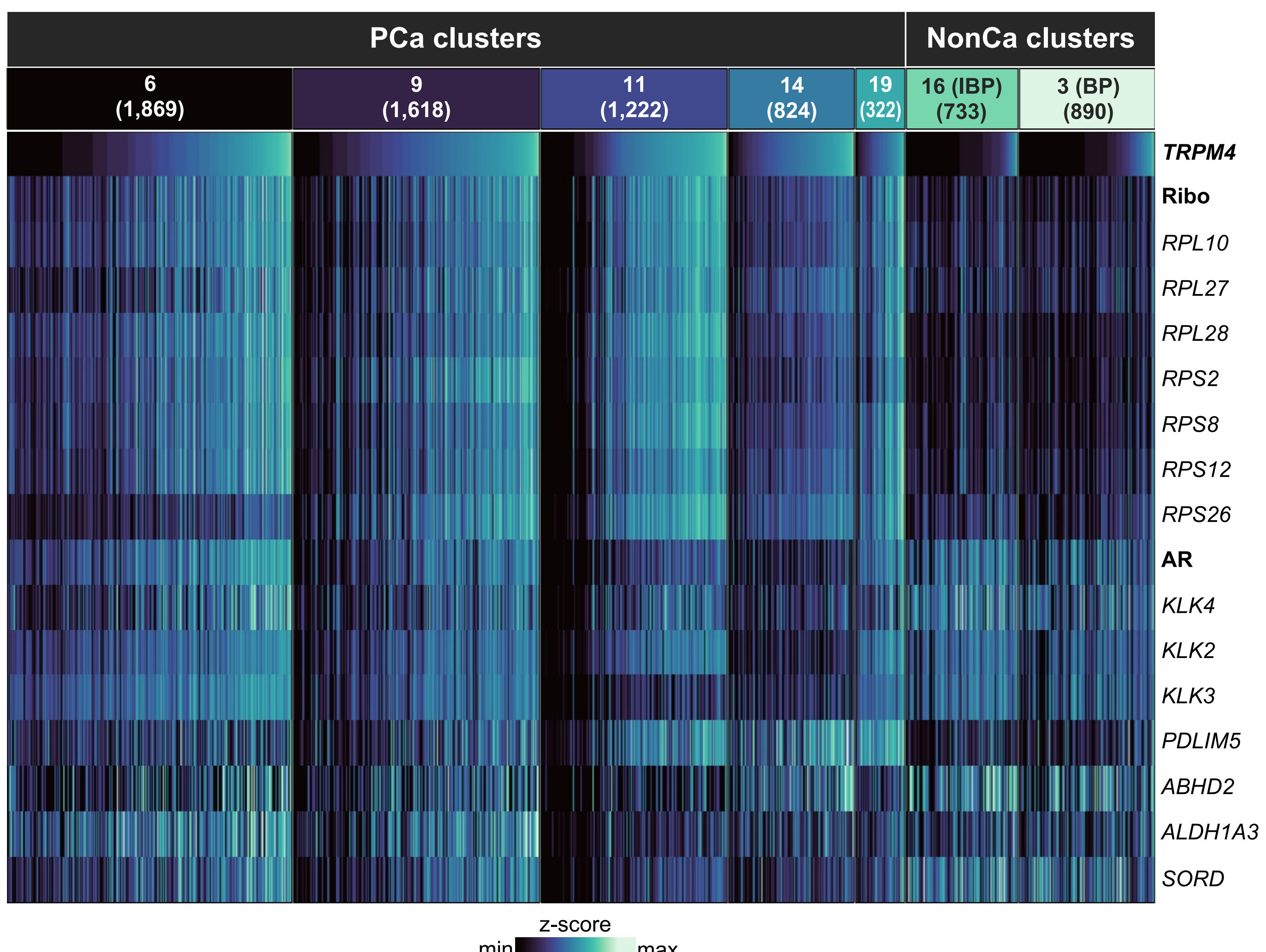

**Figure 2**. Heatmap visualization of mean ribosomal genes' expression (Ribo), mean AR signaling expression, individual ribosomal and AR signaling genes in PCa and NonCa clusters. The cells were ordered according to ascending TRPM4 expression in each cluster. Number of cells within each cluster are shown in brackets.

As the gene expression data had been pre-processed by removing cells with zero gene expression, subsequently log-transformed, while Ribo and AR were aggregates (*i.e.*, average expression of seven genes each), the data distribution may have been transformed to a more continuous and less skewed pattern, approximating a Gaussian distribution. Hence, the `gam.check()` standard diagnostic plots were generated to qualitatively assess the suitability of model fitting with Gaussian GAM.

For *TRPM4*-Ribo and *TRPM4*-AR modeling in PCa cells (the main models in this study), Q-Q plots showed that residuals aligned closely with the diagonal line (*i.e.*, normal distribution) although with deviations at both ends of the data, while the histogram of deviance residual plots demonstrated bell-shaped distribution centered around zero (**Supplementary Figure 1**). The deviance residuals versus fitted values plot, and response versus fitted values plot qualitatively suggested homoscedasticity and sufficient model fitting, respectively.

To validate these observations using independent metrics, the Akaike information criterion (AIC), Bayesian information criterion (BIC), and DE were computed after GAM fitting with Gaussian, negative binomial, gamma, inverse Gaussian, and quasi-Poisson family of distributions. Gaussian GAM yielded the lowest AIC and BIC, and the highest DE, across all modeling (except for Pt.4 *TRPM4*-AR modeling in BP cells where all families tied with the same DE) compared with the other distribution families (**Supplementary Table 6**). Thus, Gaussian GAM was utilized for modeling in this study.

### 3.4 Convergence and explainability of REML and PRSS in *TRPM4*-Ribo modeling

As explainability is vital for the implementation and optimization of any ML model, particularly in medical AI applications where model decisions should be explainable for clinical translation, the overall workflow and mechanism of GAM in this study are presented in **Figure 3**. The main objective was to determine how much *TRPM4* expression can explain the variability in Ribo expression based on GAM, with minimal overfitting or underfitting. The initial GAM model was specified to contain linear terms (intercept and a linear term) and smooth terms (with multiple smooth basis functions), with each term containing its own coefficient (magnitude of their effects) determined and optimized by REML and PRSS (**Figure 3**).

For each PRSS iteration, optimization of the smoothing parameter λ was performed according to REML, where REML iterated until convergence according to the Newton-Raphson method. Across 98 GAM models [7 samples × 7 gene sets × 2 cell types (PCa and BP)], gradient-based convergence (60/98 instances) was the most common form of REML convergence, followed by score-based convergence (37/98 instances), fallback convergence (1/98 instances occurring in Pt.7 *TRPM4*-PI3K/AKT modeling in BP cells), and none with maximum number of iterations reached (**Supplementary Table 7**).

The REML iteration with the lowest REML score was selected by the algorithm as the best model, and this selected REML iteration provided the optimized λ. Hence, each PRSS conducted its own λ optimization by minimizing REML (the inner loop *i.e.*, REML nested within PRSS), and the resulting best λ was then used to determine PRSS value (the outer loop). The earliest PRSS iteration with the lowest PRSS value was selected as the final model (extended descriptions and example results of REML and PRSS iterations are presented in the next

Results subsection). The REML-optimized λ was used to penalize the splines (*i.e.*, TPRS basis functions) in PRSS, where the product of λ and integral of the squared second derivative $f''(x)$ (the curvature) represents the penalized spline. More curvature leads to higher integral value, and the curvature is penalized by λ to avoid overfitting (*i.e.*, fitted GAM curve that is jagged or "chases" every data point), resulting in a smoother GAM curve as exemplified in **Figure 3** (section III). In terms of GAM performance metrics, the DE in percentage was adopted to quantify how much of Ribo expression variability was captured by *TRPM4* expression (**Figure 3**, section IV). To identify individual cells with strong *TRPM4*-Ribo relationship, CEP classification of the cells into TREP and non-TREP cells were conducted based on individual cell's EP value and ranking. These cells were subsequently binarized into TREP or non-TREP cells based on the overall modeling DE%, before the biology of TREP and non-TREP cells were investigated (**Figure 3**, section V).

Optimal GAM curvature penalization, without being overly restrictive (underfitting) or permissive (overfitting) in the penalization, was finetuned by minimizing REML (inner loop) and PRSS (outer loop) until convergence. To illustrate this more clearly, *TRPM4*-Ribo modeling in Pt.4 was demonstrated as an example in **Figure 4A**. For each PRSS iteration, REML iterations until REML convergence occurred to obtain the best REML score (*i.e.*, lowest REML score, as the loss function of the algorithm was to minimize REML), and precise calculations of REML iteration 1 (λ=0.264) to iteration 2 (λ=0.333) according to Newton-Raphson method operating on gradient and inverse Hessian values are illustrated in **Figure 4A**. Upon REML convergence, that specific REML iteration yielded the best REML score, along with its optimized λ value that was subsequently used to penalize GAM curvature (PRSS iteration 8 in this example).

In PRSS (the outer loop), it consisted of two phases: (i) Exploration phase testing k=3-10 in increment (k denotes the maximum number of spline basis functions, $\varphi$, for GAM modeling); (ii) Convergence verification phase that tested 20 additional iterations with different k values, where a third of the iterations tested best k value obtained from the exploration phase and the rest of the iterations tested best k's vicinity. PRSS convergence was considered when there was no further improvement in its value and all modeled relationships' PRSS converged within this phase. As the loss function was to minimize PRSS value, the earliest iteration that yielded the lowest PRSS value was selected as the final, best model. This PRSS iteration contained its REML-optimized λ used to penalize the spline basis functions. The consolidated REML and PRSS results (*e.g.*, k, λ, REML and PRSS values) for all iterations and multiple gene sets

**Figure 3**

**Questions**

- How much does *TRPM4* explain variability in Ribo expression?
- Which cells hold strong *TRPM4*-Ribo relationship?

## (I) Initial Model Specification

Linear terms | Smooth terms

$$f(x_i) = \beta_0 + \beta_1 x_i + b_1\varphi_1(x_i) + \cdots + b_n\varphi_n(x_i)$$

$f(x_i)$ = Spline model to predict responder values (*i.e.*, Ribo); $x_i$ = *TRPM4* value; $\beta_n$ = Linear term coefficient; $\varphi_n(x_i)$: Smooth term basis function; $b_n$ = Smooth term coefficient, each $b_n$ is optimized to avoid an overfitting/underfitting model

## (II) PRSS (to optimize k) and REML (to optimize λ): Loss function to minimize PRSS and REML

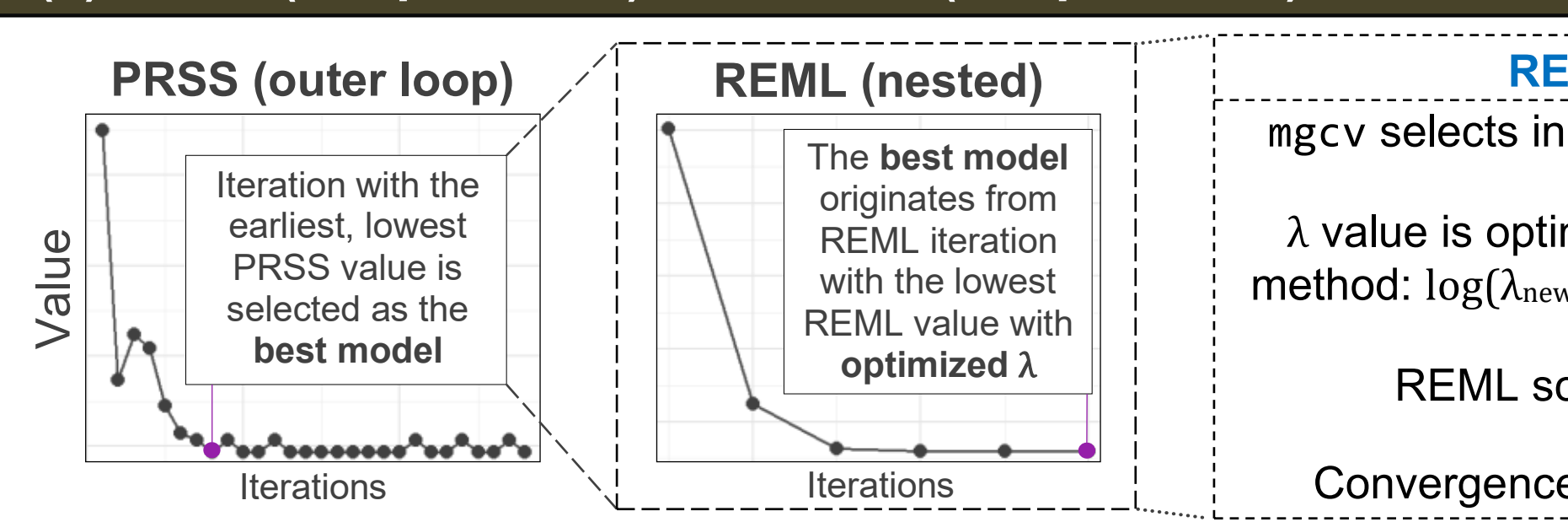

**REML iterations to improve λ**

mgcv selects initial λ value (adjusted by model structure)

↓

λ value is optimized iteratively by the Newton-Raphson method: $\log(\lambda_{new}) = \log(\lambda_{old}) - H^{-1}g$ (H: Hessian; g: gradient)

↓

REML score is minimized at each iteration

↓

Convergence (*e.g.*, $<10^{-6}$ REML value improvement)

## (III) Fitting with Penalized Splines: Applies optimized λ in PRSS to fit the model

### PRSS

RSS (how well the model fits the data)

REML-optimized λ

$$\text{PRSS} = \sum_{i=1}^{n}(y_i - f(x_i))^2 + \lambda\int(f''(x))^2dx$$

$f''(x_i)$ represents curvature; λ × integral of squared curvature is the TPRS penalty

- Higher curvature → Higher integral
- Higher γ (in REML) → Higher λ → Reduced $b_n$ (smooth terms' coefficients) as the algorithm minimizes PRSS → Less curvature
- Minimized PRSS → Fitting of data with penalized splines, resulting in optimized $f(x)$

***Example f(x) and its derivatives***

*f(x)*: Function | *f'(x)*: First derivative (slope)

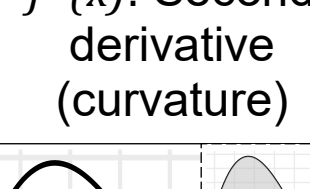

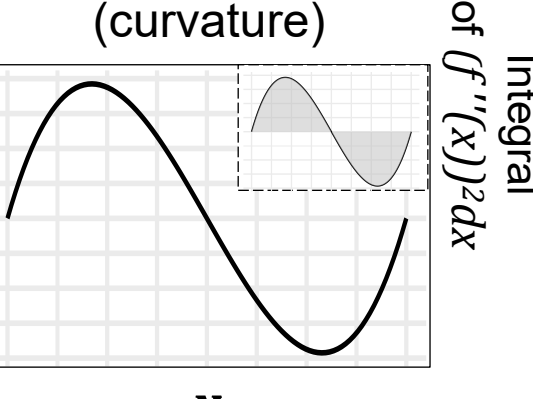

*f(x), f'(x), f''(x)*

*x* | *x* | *x*

***Example modeling according to λ***

Inferior λ | Suboptimal λ | Optimal λ

*f(x)*

*x* | *x* | *x*

Optimal λ → Penalized splines → Smoother model

## (IV) Performance Metrics: Deviance explained (DE)

### Null deviance (ND)

$$\text{ND} = \sum_{i=1}^{n}((y_i - \bar{y})^2)$$

Observed value ($y_i$); Mean (all observations) ($\bar{y}$)

- Null model without predictors (intercept only *i.e.*, null residuals)

### Model deviance (MD)

$$\text{MD} = \sum_{i=1}^{n}((y_i - f(x_i))^2)$$

Observed value ($y_i$); Predicted value (model) ($f(x_i)$)

- How well the $f(x)$ model predicts the observed value (*i.e.*, model residuals)

### DE

$$\text{DE} = 1 - \frac{\text{MD}}{\text{ND}}$$

- How much of the variability (%) in the data is captured by $f(x)$ model
- Higher DE = better performance

## (V) CEP-IP: To map for top-ranked EP (TREP) and non-TREP cells (and their biology)

### ND contribution (NDC) by *i*-th cell

$$\text{NDC}_i = (y_i - \bar{y})^2$$

ND is the sum of NDC from all cells

- How much *i*-th cell contributes to the total ND (individual null residuals)

→

### MD contribution (MDC) by *i*-th cell

$$\text{MDC}_i = (y_i - f(x_i))^2$$

MD is the sum of MDC from all cells

- How much *i*-th cell contributes to the total MD (individual model residuals)

→

### Explanatory power (EP) of *i*-th cell

$$\text{EP}_i = 1 - \frac{\text{MDC}_i}{\text{NDC}_i}$$

- How well the model performs for *i*-th cell; Higher EP = Model fits much better than null model for *i*-th cell

↓

$\text{EP}_1$, $\text{EP}_2$, $\text{EP}_3$, ⋮, $\text{EP}_n$

- Rank cells from the highest EP ($\text{EP}_1$) to the lowest ($\text{EP}_n$)
- Select the top DE% as TREP cells; the remaining as non-TREP cells

*Example:* If total 100 cells and DE is 45%, then $\text{EP}_1$ to $\text{EP}_{45}$ are TREP cells, $\text{EP}_{46}$ to $\text{EP}_{100}$ are non-TREP cells

**TREP mapping**

←

### Validation and biology

- Validate CEP vs other methods
- Investigate TREP vs non-TREP for:
  - Spatial patterns in GAM plots
  - GO enrichment and trajectory

**Figure 3**. Workflow of Gaussian GAM in this study. The study was initiated by specifying the function with linear and smooth terms. Next, REML optimization was conducted to obtain the optimal λ value. Each PRSS iteration minimized REML score until convergence, and PRSS was treated as a hyperparameter space for optimization with different k values. The best model (with the optimal k and λ) was determined by the earliest, lowest PRSS value. In the PRSS step, each spline term was penalized by the REML-optimized λ (*i.e.*, TPRS penalty), leading to a smoother GAM fit. The optimized model with penalized splines was then selected to calculate DE (*i.e.*, how much *TRPM4* expression can capture the variability in Ribo expression). CEP classification was performed where EP values of each individual cell were ranked from the highest to the lowest, and the top DE% were selected as TREP cells. Validation of the cell classification methodology, and biology of the classified cells were subsequently investigated. By preserving transparency in both model optimization and cell-level predictions, this workflow enables biological interpretation of ML outputs and serves as an example of XAI modeling.

investigated (*e.g.*, Ribo, AR) in PCa or BP cells separately, and each of their best model's REML iteration parameters (gradient and Hessian) are presented in **Supplementary Table 7**.

To test if the best model chosen by the converged PRSS and REML was valid, an independent GAM fitting without nesting REML within PRSS (*i.e.*, without the outer and inner loops structure found in the original PRSS and REML optimization workflow) was performed using fixed k values (k=3-10) and λ values derived from the original optimization (Pt.4: λ=0.264, 0.333, 0.419, and 0.528) with additional λ values beyond the original best model's λ (*e.g.*, λ=0.6, 0.75, 1, 1.25, 1.5, 1.75, and 2). The aim was to observe if this independent GAM fitting with fixed k values and wider range of λ values resulted in the same best k and best λ values with the original REML and PRSS optimization process. As shown for *TRPM4*-Ribo modeling in Pt.4, the independent GAM fitting with fixed k and λ values showed that the lowest PRSS and REML scores yielded k=10 and λ=0.528, respectively, aligning with the best k and best λ values derived from the original GAM fitting (**Figure 4B**).

In the rest of the patients, the independent *TRPM4*-Ribo GAM fitting also resulted in best k and best λ that tallied with the original REML and PRSS optimization, except Pt.6 where the independent GAM fitting yielded different best λ (original best λ=5099.360 vs. independent best λ=9000). However, the REML scores for these two distinct GAM fittings differed by <0.001 (rounded up to 0.0002; **Supplementary Table 7**), where both of their REML scores were 1,229.607 (**Figure 4C**), suggesting that a huge range of λ values may be within a flat region of REML optimization surface where the objective function was insensitive to λ changes. Monotonic decrease in REML value of <0.001 for each successive iteration was observed, thus REML minimization still occurred but with diminishing returns in further λ optimization.

## Figure 4

(A)

**PRSS: Outer Loop**

Exploration phase (k=3-10 increment)

k=3

k=4

k=5 6 7 8 9

Best k, 10

PRSS convergence verification phase
• 2/3 iterations tested best k
• 1/3 iterations tested best k's vicinity
• Different random starting point (even for the same k value)
• 20 iterations, stop if no PRSS improvement

PRSS value

620 610 600 590 580

0 10 20

PRSS iteration

**REML: Nested in PRSS, k = 10**

λ=0.264

λ= 0.333

λ=0.419

λ=0.528

REML Convergence GN < 1 × $10^{-5}$ or REML improvement < 1 × $10^{-6}$

REML value

870 869 868 867

1 2 3 4

REML iteration

**Example REML iterations**

**REML iteration 1 ($\lambda_{old}$) to 2 ($\lambda_{new}$)**

Initial λ value ($\lambda_{old}$) = 0.264

Inverse Hessian ($H^{-1}$): -2,150.52

Gradient (g): 1.074 × $10^{-4}$

$\log(\lambda_{new}) = \log(\lambda_{old}) - H^{-1}g$

$\therefore \log(\lambda_{new}) = \log(0.264) - (-2{,}150.52 \cdot 1.074 \times 10^{-4}) = -1.101$

$\therefore \lambda_{new} = e^{-1.1011} = 0.333$

This repeats until REML convergence *i.e.*, iteration 4 with best model's **λ = 0.528**. This λ value was used to penalize curvature (PRSS iteration 8)

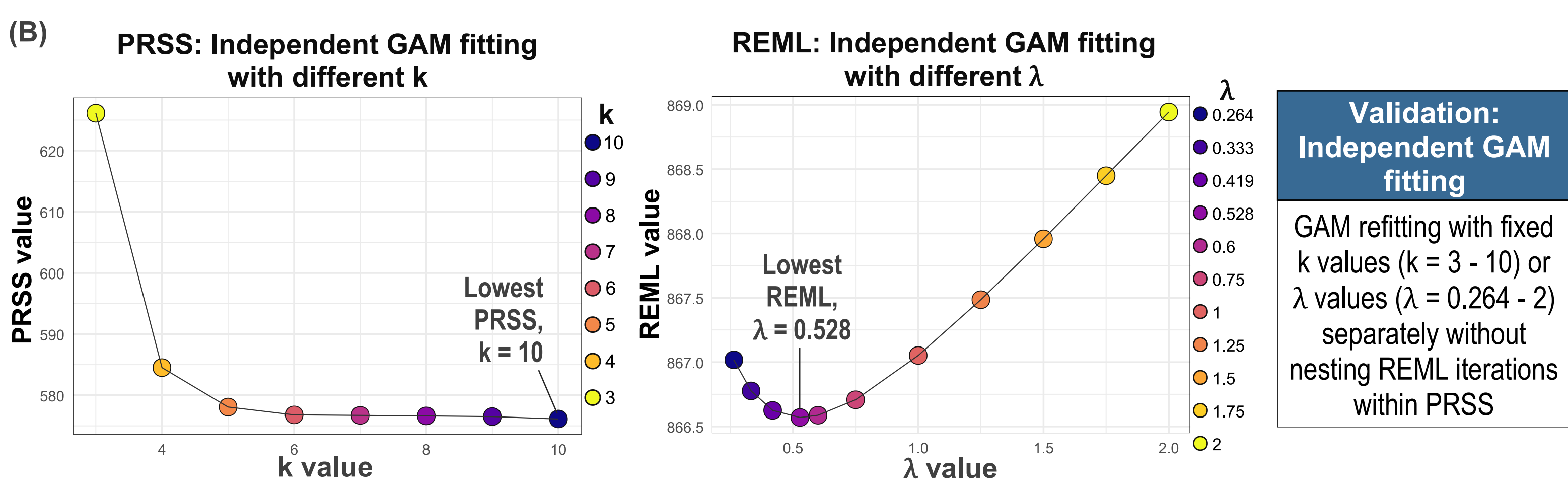

(C)

| PCa (Ribo gene set) | PRSS: Best k | REML (nested in PRSS): Best λ | Independent GAM fitting | |
|---|---|---|---|---|
| | | | Best k | Best λ |
| **Pt.1** | 6 | 1.570 | 6 | 1.570 |
| **Pt.2** | 3 | 0.079 | 3 | 0.079 |
| **Pt.3** | 6 | 0.139 | 6 | 0.139 |
| **Pt.4** | 10 | 0.528 | 10 | 0.528 |
| **Pt.5** | 4 | 0.287 | 4 | 0.287 |
| **Pt.6** | 4 | 5099.360 | 4 | 9000 |
| **Pt.7** | 3 | 1.287 | 3 | 1.287 |

**Pt.6 REML: Independent GAM fitting with different λ**

REML value

1229.607 1229.607 1229.607 1229.607

4000 6000 8000

λ value

REML value difference is <0.001 for λ value 5099.360 vs. 9000

λ = 5099.360

λ = 9000

λ: 2549.67887, 2928.811924, 3364.32144, 3864.590503, 4439.248754, 5099.357741, 5500, 6000, 7000, 8000, 9000

**Figure 4**. PRSS and REML iterations of *TRPM4*-Ribo modeling represented by PCa cells in Pt.4 as the representative, and validation of PRSS and REML best models. (A) PRSS consisted of 28 iterations, divided into exploration phase (testing k=3-10 increment) and PRS convergence verification phase (additional 20 iterations and stopped if no further reduction in PRSS). Each PRSS iteration minimized REML via the Newton-Rhapson method until REML convergence as defined by GN <1 × $10^{-5}$ or REML improvement <1 × $10^{-6}$. The inverse Hessian value was backcalculated using $\lambda_{old}$, $\lambda_{new}$, and gradient value (**Supplementary Table 7**); (B) GAM refitting with fixed k values (3-10) optimized by PRSS (left panel), or with different λ

values optimized by REML (right panel), using Pt.4-specific λ range (0.264-2) as the representative example; (C) Best model's PRSS, REML, k, and λ values for each patient, and Pt.6 was highlighted showing minimal difference (<0.001) in REML value when fitted using λ value (5099.360) optimized by REML nested in PRSS method versus a fixed λ value (9000).

### 3.5 Explainable spline penalization and formation of the full *TRPM4*-Ribo model

For further explainability, a schematic representation of GAM splines formation in *TRPM4*-Ribo modeling (Pt.4) with `mgcv` is demonstrated in **Figure 5A**. Initially, basic statistical assessment was performed on the data (*e.g.*, mean, range) to guide the construction of representative knots (dotted lines) (**Figure 5A**). The splines (TPRS basis functions) were then initialized surrounding the knots, and in this case there were eight splines ($\varphi_1$ to $\varphi_8$) determined by the best k value (k=10 basis functions total, with $\varphi_1$ to $\varphi_8$ representing the eight main splines shown here, while $\varphi_9$ to $\varphi_{10}$ had been penalized out of the final model).

The splines were subsequently processed based on the data distribution and data density at knots, forming unweighted and unpenalized splines in their raw form with obvious curvatures, particularly $\varphi_3$ to $\varphi_7$. The coefficient of each spline was estimated by REML, with the λ value controlling the penalty strength and applied in PRSS to weight and penalize each spline, reducing their curvature and minimizing overfitting. Coefficient values below |1.0| resulted in reduced curvature or magnitude of the individual spline, while negative coefficient value caused the individual spline's direction to invert.

Next, the weighted and penalized splines were combined to collectively form a smooth curve (representing the smooth terms), and the combination of linear terms with smooth terms formed the final full GAM model, where *TRPM4* values as the predictor of Ribo values for each cell such as Cell #1 [linear terms + smooth terms = 5.15 + (-1.92) = 3.23, representing the predicted Ribo value given the actual *TRPM4* value of Cell #1] (**Figure 5A**). The variance contribution of each linear and smooth terms was also computed, and $\varphi_1$ showed the largest contribution (49.7%) (**Figure 5B**), aligning with the shape of the penalized and weighted collective smooth curve that reflects the characteristic sigmoidal shape of penalized and weighted $\varphi_1$ individual spline.

### 3.6 Final model optimization through γ parameter tuning and visual assessment

Higher REML's γ value typically leads to higher λ value, resulting in more penalization on spline's curvature. Hence, spline smoothing to yield the optimal GAM curves, avoiding overfitting or underfitting, can be achieved by adjusting for the ideal γ value. To this end, *TRPM4*-Ribo or *TRPM4*-AR modeling was conducted for PCa, BP, and IBP cells in all patients

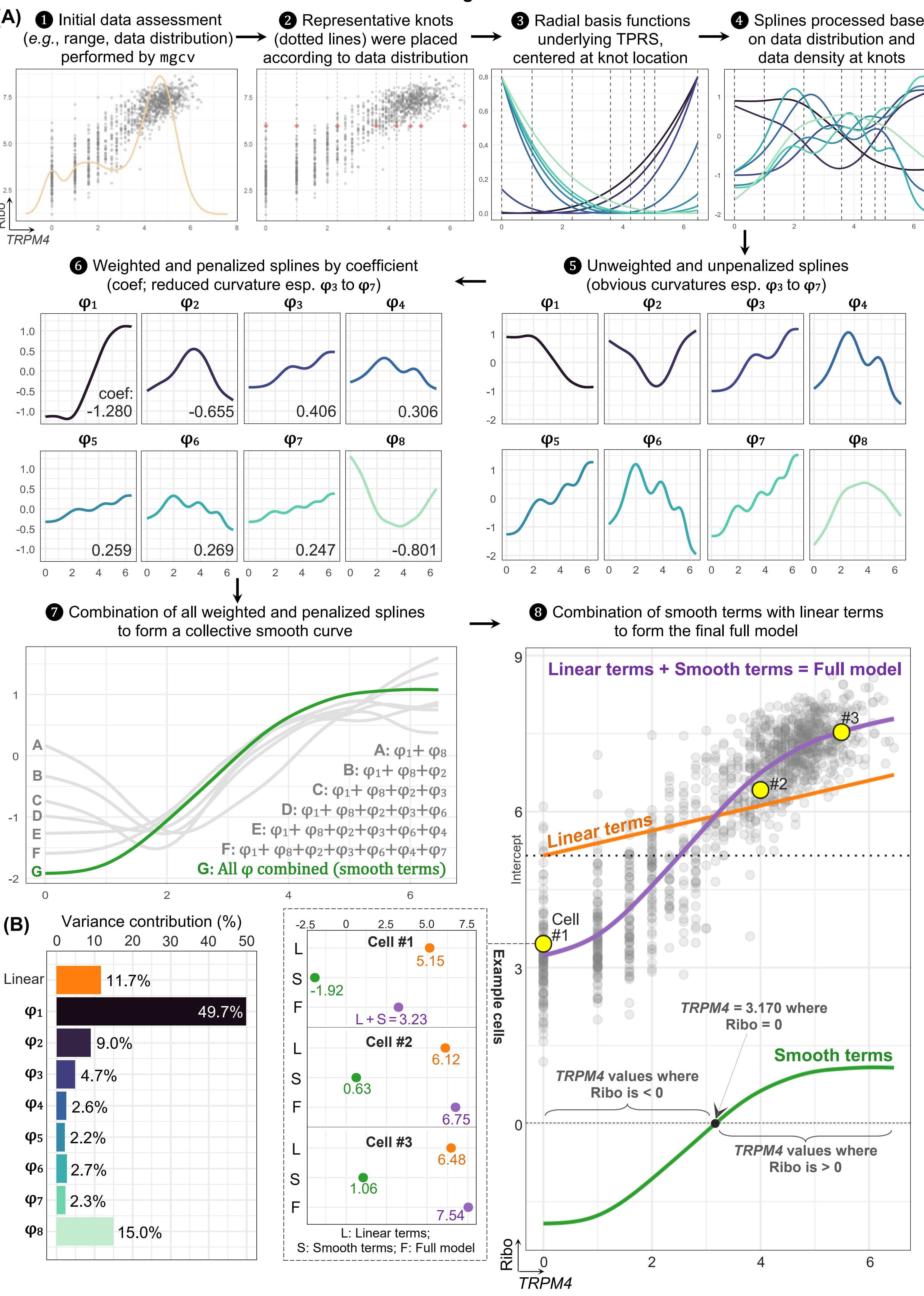

Figure 5
(A)
1 Initial data assessment (e.g., range, data distribution) performed by mgcv
2 Representative knots (dotted lines) were placed according to data distribution
3 Radial basis functions underlying TPRS, centered at knot location
4 Splines processed based on data distribution and data density at knots
Ribo
TRPM4
5 Unweighted and unpenalized splines (obvious curvatures esp. φ3 to φ7)
6 Weighted and penalized splines by coefficient (coef; reduced curvature esp. φ3 to φ7)
φ1
φ2
φ3
φ4
φ5
φ6
φ7
φ8
coef:
-1.280
-0.655
0.406
0.306
0.259
0.269
0.247
-0.801
7 Combination of all weighted and penalized splines to form a collective smooth curve
A: φ1+ φ8
B: φ1+ φ8+φ2
C: φ1+ φ8+φ2+φ3
D: φ1+ φ8+φ2+φ3+φ6
E: φ1+ φ8+φ2+φ3+φ6+φ4
F: φ1+ φ8+φ2+φ3+φ6+φ4+φ7
G: All φ combined (smooth terms)
8 Combination of smooth terms with linear terms to form the final full model
Linear terms + Smooth terms = Full model
Linear terms
Intercept
#3
#2
Cell #1
Example cells
TRPM4 = 3.170 where Ribo = 0
Smooth terms
TRPM4 values where Ribo is < 0
TRPM4 values where Ribo is > 0
Ribo
TRPM4
(B)
Variance contribution (%)
Linear 11.7%
φ1 49.7%
φ2 9.0%
φ3 4.7%
φ4 2.6%
φ5 2.2%
φ6 2.7%
φ7 2.3%
φ8 15.0%
Cell #1
L 5.15
S -1.92
F L + S = 3.23
Cell #2
L 6.12
S 0.63
F 6.75
Cell #3
L 6.48
S 1.06
F 7.54
L: Linear terms;
S: Smooth terms; F: Full model

**Figure 5**. Visual representation of GAM's TPRS formation represented by Pt.4 *TRPM4*-Ribo modeling. (A) Schematic representation of the formation of the TPRS basis functions and final full model. Number of splines was determined by the best model's k value, and each was penalized by REML-optimized $\lambda$ as reflected by each spline's coefficient. Combination of linear and smooth terms formed the final full model, and three representative cells (Cell #1, #2, and #3) were selected as examples illustrating that combination of linear and smooth terms' Ribo values, yielding the full model's predicted Ribo value. Panel ❸ shows the theoretical radial basis functions that form the mathematical basis of TPRS, and panel ❹ shows the resulting actual unpenalized TPRS basis functions after `mgcv`'s transformations; (B) Variance contribution of linear and each spline ($\varphi_1$ to $\varphi_8$) terms to the full model.

(Pt.1-Pt.7) using `mgcv` default $\gamma=1$, and to observe overfitting (fluctuating) or underfitting (overly smooth) characteristics in the resulting GAM curves.

In PCa cells, all the patients showed relatively stable *TRPM4*-Ribo and *TRPM4*-AR GAM curves without obvious signs of overfitting or underfitting. However, in BP cells, Pt.6 (*TRPM4*-Ribo) and Pt.7 (*TRPM4*-AR) GAM curves showed overfitting characteristics with fluctuations (**Supplementary Figure 2**). In IBP cells, *TRPM4*-AR also exhibited signs of overfitting in Pt.1 and Pt.3, but four of the patients, Pt.2, Pt.4, Pt.5, and Pt.7, did not have sufficient IBP cells for the modeling. In view of this, PCa and BP cells were subsequently prioritized for downstream optimization by attempting multiple different $\gamma$ values in 0.5 increment (*i.e.*, $\gamma=0.5$, 1, 1.5, 2, 2.5, and 3) in order to improve the characteristics of the GAM curves via visual inspection. Assessment with the range of $\gamma$ values showed that $\gamma=1.5$ yielded optimal smoothing for *TRPM4*-Ribo (Pt.6) and *TRPM4*-AR (Pt.7) GAM curves of BP cells, mitigating the overfitting characteristics observed previously with $\gamma=1$ (**Figure 6A**). Hence, $\gamma=1.5$ was adopted for all subsequent GAM modeling for comparable results between different models. The optimization results and statistical performance metrics (*e.g.*, DE, model equation) utilizing different $\gamma$ values are presented in **Supplementary Table 8**.

### 3.7 *TRPM4*-Ribo outperforms alternative cancer pathways in PCa

Modeling of *TRPM4* with Ribo and AR, as well as with other pathways reported to involve TRPM4 in cancers including GSK-3β [49], mTOR [50], NF-κB [13], PI3K/AKT [50, 51], and Wnt [19] (gene sets averaged from seven typical genes implicated in each pathway; **Supplementary Table 9**), were conducted with $\gamma=1.5$. The resulting GAM curves of *TRPM4*-Ribo and *TRPM4*-AR showed higher absolute gene set expression values (y-axis) than the rest of the gene sets investigated in PCa or BP cells (**Figure 6B**).

Additionally, *TRPM4*-Ribo and *TRPM4*-AR GAM curves demonstrated similar shape in PCa cells across all patients (except Pt.5), such as sigmoidal shape in Pt.4. For explainability and transparency, the calculation of the model's predicted Ribo expression (*i.e.*, prediction made by *TRPM4*-Ribo modeling) for a single PCa cell, with known Ribo ($y_1$=7.529) and *TRPM4* ($x_1$=5.492) expression values, was illustrated in **Figure 6C**. The predicted Ribo expression value was the sum of two linear terms and eight smooth terms (weighted and penalized by their coefficient). DE (83.46%) of the *TRPM4*-Ribo modeling in Pt.4 was calculated as the performance metrics of how much variability in Ribo expression was captured by *TRPM4*. Comparison of *TRPM4*-Ribo DE with other modeled gene sets (AR, GSK-3β, mTOR, NF-κB, PI3K/AKT, and Wnt) in BP cells (**Figure 6D**) or IBP cells (**Supplementary Figure 3**) showed that none demonstrated significant difference, but in PCa cells *TRPM4*-Ribo DE was significantly higher than other modeled gene sets (GSK-3β, mTOR, PI3K/AKT, Wnt, and NF-κB; all FDR <0.05) except with *TRPM4*-AR DE (FDR=0.128; **Figure 6D**). The complete set of performance metrics, observed and model-predicted values, and smooth basis functions' coefficients of *TRPM4* modeling with all investigated gene sets in PCa cells ($\gamma$=1.5) are presented in **Supplementary Table 9**.

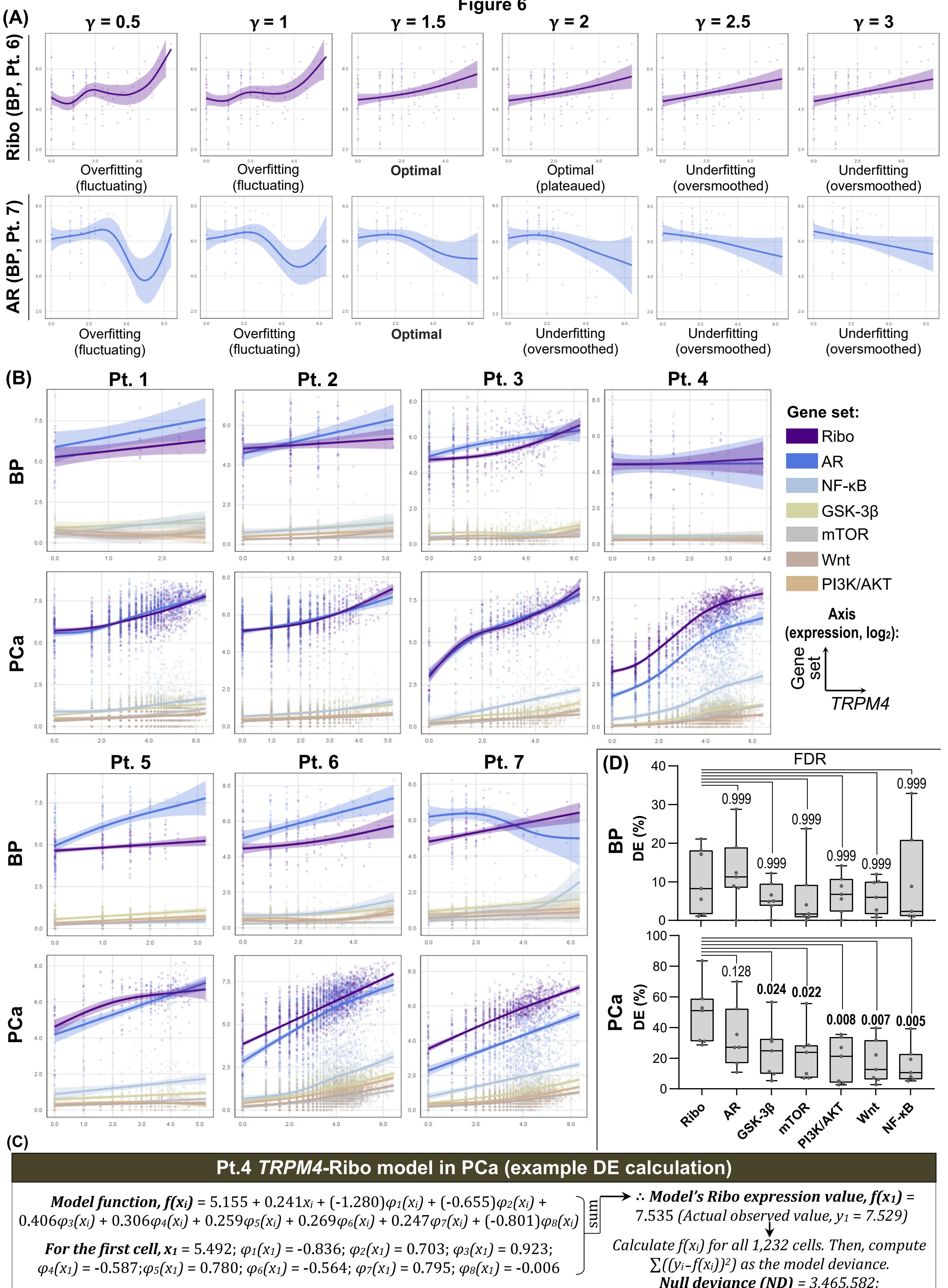

Figure 6
(A)
γ = 0.5
γ = 1
γ = 1.5
γ = 2
γ = 2.5
γ = 3
Ribo (BP, Pt. 6)
Overfitting (fluctuating)
Overfitting (fluctuating)
Optimal
Optimal (plateaued)
Underfitting (oversmoothed)
Underfitting (oversmoothed)
AR (BP, Pt. 7)
Overfitting (fluctuating)
Overfitting (fluctuating)
Optimal
Underfitting (oversmoothed)
Underfitting (oversmoothed)
Underfitting (oversmoothed)
(B)
Pt. 1
Pt. 2
Pt. 3
Pt. 4
BP
PCa
Gene set:
Ribo
AR
NF-κB
GSK-3β
mTOR
Wnt
PI3K/AKT
Axis (expression, log2):
Gene set
TRPM4
Pt. 5
Pt. 6
Pt. 7
BP
PCa
(D)
FDR
BP
DE (%)
0.999
0.999
0.999
0.999
0.999
0.999
PCa
DE (%)
0.128
0.024
0.022
0.008
0.007
0.005
Ribo
AR
GSK-3β
mTOR
PI3K/AKT
Wnt
NF-κB
(C)
Pt.4 TRPM4-Ribo model in PCa (example DE calculation)
**Model function, $f(x_i)$** = $5.155 + 0.241x_i + (-1.280)\varphi_1(x_i) + (-0.655)\varphi_2(x_i) + 0.406\varphi_3(x_i) + 0.306\varphi_4(x_i) + 0.259\varphi_5(x_i) + 0.269\varphi_6(x_i) + 0.247\varphi_7(x_i) + (-0.801)\varphi_8(x_i)$
**For the first cell, $x_1$** = $5.492$; $\varphi_1(x_1) = -0.836$; $\varphi_2(x_1) = 0.703$; $\varphi_3(x_1) = 0.923$; $\varphi_4(x_1) = -0.587$; $\varphi_5(x_1) = 0.780$; $\varphi_6(x_1) = -0.564$; $\varphi_7(x_1) = 0.795$; $\varphi_8(x_1) = -0.006$
sum
*$x_i$ = TRPM4 value for each cell; $\varphi_i(x_i)$ = Basis function value*
***PCa cells in Pt. 4:** n=1,232 (each with own TRPM4 value)*
*∴ **Model's Ribo expression value, $f(x_1)$** = 7.535 (Actual observed value, $y_1$ = 7.529)*
*Calculate $f(x_i)$ for all 1,232 cells. Then, compute $\sum((y_i - f(x_i))^2)$ as the model deviance.*
***Null deviance (ND)** = 3,465.582;*
***Model deviance (MD)** = 573.051*
*∴ **Deviance explained (DE)** = 83.46%*

**Figure 6**. GAM modeling in BP and PCa cells. (A) Six different gamma values ($\gamma$=0.5-3) were tested for *TRPM4*-Ribo (Pt. 6 adopted as the reference) and *TRPM4*-AR (Pt. 7 adopted as the reference) in BP cells, where balanced smoothing was observed utilizing $\gamma$=1.5 in the fitting process; (B) GAM modeling in BP (top graphs) and PCa (bottom graphs) cells in each of the seven patients for Ribo, AR, GSK-3β, mTOR, NF-κB, PI3K/AKT, and Wnt gene sets; (C) Example DE calculation from *TRPM4*-Ribo modeling in Pt.4 PCa cells. Up to three decimal points were used in this calculation for simplicity, and the original values of the basis functions (**Supplementary Table 9**) contain values up to 15 decimal points; minor rounding differences in predicted values (≤0.002) between figures are attributable to this truncation (D) DE comparison of *TRPM4*-Ribo versus *TRPM4* modeling with other gene sets in BP and PCa cells, and FDR <0.05 is in bold.

### 3.8 CEP-IP framework validation: CEP-classified TREP cells are well-predicted by GAM's DE

As GAM's DE indicated how much variability in Ribo expression can be captured by *TRPM4* expression, the next question to address was which individual cells were most well-predicted by the fitted model, and these cells may be biologically distinct from the remaining cells. GAM's DE was a metric aggregated from all cells, in which ND and MD were the sum of all cells' squared null and model residuals, respectively. To decompose this aggregate measure into cell-level, individual cell's NDC, MDC, and EP were computed. Each cell was then ranked from the highest to the lowest according to their EP values, and the top DE% of cells were selected and classified as TREP cells (**Supplementary Table 10**). For comparison with the CEP classification of cells into TREP or non-TREP, cells were instead randomly classified or through the leverage-based classification, where their GAM plots showed dissimilar pattern of TREP and non-TREP cells distribution (**Figure 7A**).

**Figure 7**

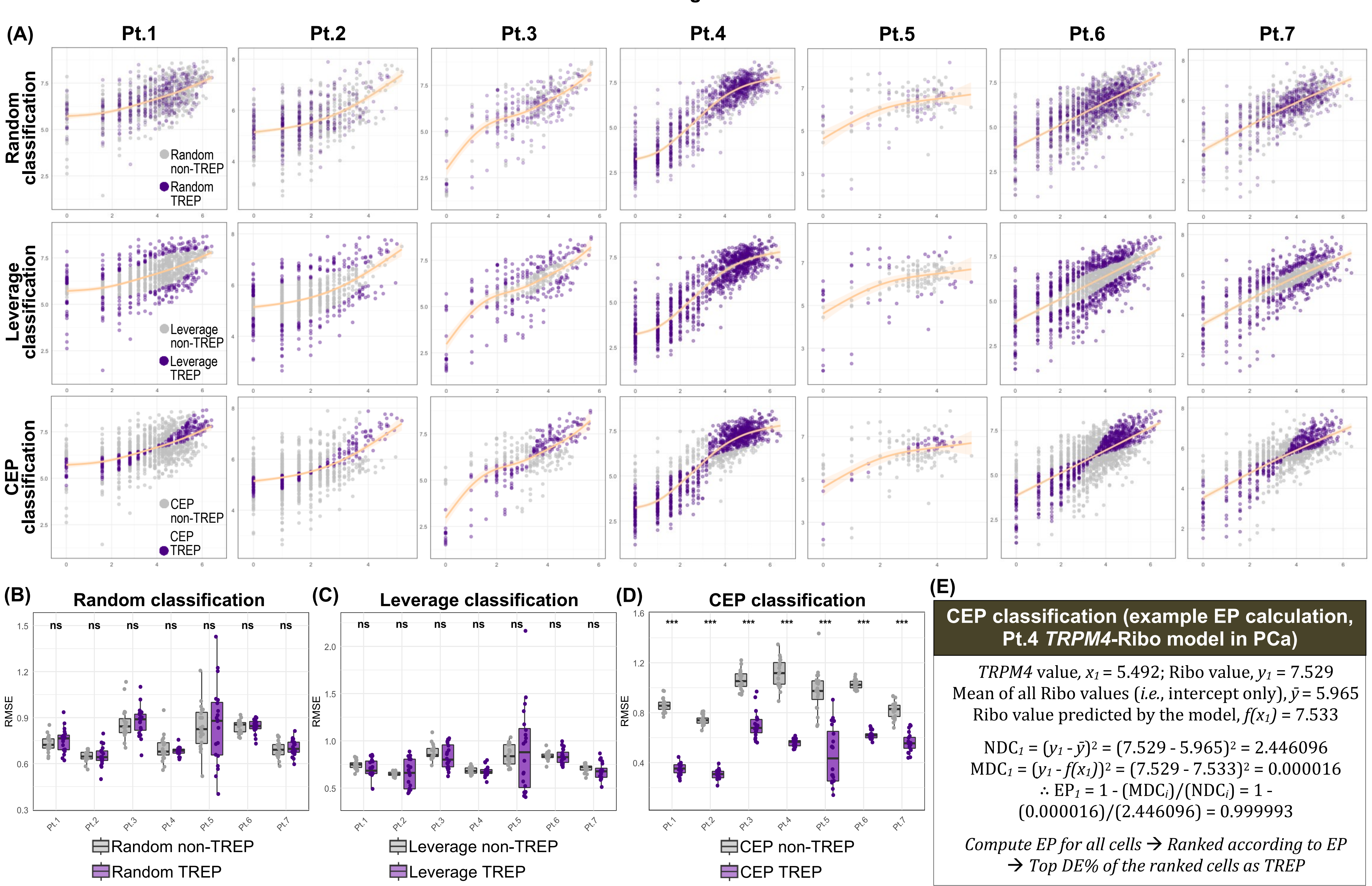

**Figure 7**. GAM plot distribution patterns and MCCV of TREP and non-TREP cells classified by random, leverage, or CEP classification. (A) Distribution of TREP and non-TREP cells based on the three classification systems; (B-D) MCCV of the performance of random (B), leverage (C), or CEP (D) classification, where lower RMSE in test set's TREP compared with non-TREP group indicated better performance. ***: FDR<0.001; ns: not significant FDR; (E) Example calculation of EP value of a CEP-classified TREP cell, given the actual observations ($x_1$ and $y_1$), mean ($\bar{y}$) and $f(x_1)$ model-predicted Ribo values, to compute $\mathrm{NDC}_1$, $\mathrm{MDC}_1$, and $\mathrm{EP}_1$.

MCCV (train:test split of 70:30 for 20 randomized iterations) showed that random or leverage-based classification did not yield significant difference in RMSE of their assigned TREP or non-TREP cells in the test set (**Figures 7B** and **7C**). Essentially, CEP-classified TREP cells showed significantly lower RMSE than non-TREP cells across all seven PCa patients in the test set (FDR<0.001; **Figure 7D** and **Supplementary Table 11**). For explainability, example calculation of a CEP-classified TREP cell's EP value is shown in **Figure 7E**.

To further compare CEP classification with other techniques, cells were categorized using Cook's distance, a standard metric that integrates leverage and standardized residuals to identify data points that skew a regression line. The same MCCV pipeline was adopted (70:30 train-test ratio, 20 iterations). The Cook's distance-based classification showed no significant difference in RMSE between assigned TREP and non-TREP cells in any patient (all FDR >0.05 in the test set; **Supplementary Table 11**), which is in line with the leverage and random control results. Collectively, neither random assignment, leverage-based classification, nor Cook's distance-based classification could replicate the out-of-sample RMSE separation attained by CEP classification in all seven PCa patients. This demonstrates that EP identifies a cell population (*i.e.*, those with a robust *TRPM4*-Ribo relationship) that remains obscure via standard influence diagnostics.

An additional concern is whether *TRPM4*-Ribo relationship may be influenced by technical rather than biological effects. To address this, four lines of evidence are provided. First, the post-QC metrics showed that filtering had been performed accurately: the nCount and nFeature distributions were unimodal, the ribosomal and mitochondrial content thresholds were set at the 90th percentile for each dataset, and the nFeature and nCount values showed a positive linear relationship, indicating minimal doublets post-QC (**Supplementary Figure 4**).

Secondly, post-hoc pairwise testing showed that IBP cells (cluster 16, median ribosomal content percentage=27.3%) had significantly higher ribosomal content than three of five PCa clusters (clusters 6, 9, or 19; all Dunn's FDR<0.001) and was comparable to the other two PCa clusters (clusters 11 or 14; FDR=0.495 and 0.045, respectively). However, no ribosomal genes passed the dual-filter in IBP cells, demonstrating that ribosomal content level alone does not predict TRPM4-Ribo co-expression. Similarly, BP cells (cluster 3, median=22.7%) showed heterogeneous ribosomal content compared with PCa clusters, where it was significantly lower than clusters 11 or 14 but significantly higher than clusters 6 or 19 (FDR<0.001) (**Supplementary Figure 4**).

Next, analysis of the top 2,000 highly variable genes (HVGs) showed that only one ribosomal gene (*RPL3L*, ranked 1,533rd out of 2,000) was present in PCa clusters and none in the NonCa control clusters (**Supplementary Table 12**). Essentially, *RPL3L* was not one of the seven dual-filtered ribosomal genes that constituted the Ribo composite. This indicates that ribosomal genes did not dominate the variable gene space, and downstream genes selection process was not biased toward them. Lastly, none of the seven ribosomal genes passed the Spearman-Kendall dual-filter in IBP or BP control cells despite identical QC processing with the PCa group, but they passed the dual-filter specifically in PCa cells as shown previously in **Figure 1D**. Hence, Ribo's elevated expression and monotonic co-expression with *TRPM4* was specific to PCa cells, and ribosomal content or filtering as a technical confounder is unlikely.

### 3.9 EP reflects the integration of both null and model deviances

One concern pertaining to the CEP framework is whether EP values could be inflated when NDC is small, potentially causing bias in TREP classification. The formula $\mathrm{EP}_i = 1 - (\mathrm{MDC}_i / \mathrm{NDC}_i)$ indicates that if $\mathrm{NDC}_i$ is small, $\mathrm{EP}_i$ value will be lower due to the subtraction. This aligns with $\mathrm{ND} = \Sigma(y_i - \bar{y})^2$ which indicates that cells with higher ND (not smaller ND) are likely more relevant biologically due to their higher deviation from the average values. To further illustrate that EP is dependent on an individual cell's ratio of MDC:NDC, and not simply either NDC or MDC values, the distributions of NDC, MDC, MDC:NDC ratio, and EP values were examined in TREP and non-TREP cells for all seven PCa patients (**Figure 8**). NDC and MDC individually showed substantial overlap between TREP and non-TREP cells in all PCa patients (NDC: OVL range 88.4-98.6%; MDC: OVL range 82.3-96.7%). This demonstrates that neither deviance component alone could distinguish TREP from non-TREP cells, and TREP classification is not driven solely by extreme NDC values.

Instead, it is the ratio between an individual cell's MDC and NDC (and together with GAM's DE) that determines whether the cell should be classified as TREP; TREP cells consistently exhibited higher NDC alongside lower MDC (*e.g.*, Pt.4 TREP NDC IQR=1.14-4.24 vs MDC IQR=0.02-0.35), yielding a low MDC:NDC ratio that achieved 0% overlap between TREP and non-TREP cells in all seven patients (TREP MDC:NDC IQR=0.01-0.16 vs Non-TREP MDC:NDC IQR=1.89-16.40). EP values of each patient's TREP cells showed 0% overlap with their non-TREP cells' EP values (**Figure 8**). This verifies that TREP identification emerges from the relationship between both deviance components rather than either in isolation.

Another concern is potential circularity, whereby the same model fit is used both to define TREP cells and to claim those cells are well-predicted. In the MCCV analysis (Section 3.8), GAM was fitted exclusively on training cells and applied to held-out test cells whose TREP or non-TREP status was assigned using the training set's DE threshold. This DE threshold differed from the value derived from all cells via GAM fitting. Nonetheless, CEP-classified TREP cells showed significantly lower RMSE than non-TREP cells in the test set in all seven patients (FDR<0.001), whereas random and leverage-based classification yielded no significant RMSE difference (**Figures 7B-D**). Collectively, these two lines of evidence (*i.e.*, deviance assignment showing that EP is determined by the MDC:NDC ratio rather than NDC magnitude, and out-of-sample MCCV validating that TREP cells are genuinely better predicted on unseen data) support CEP classification as a non-circular framework.

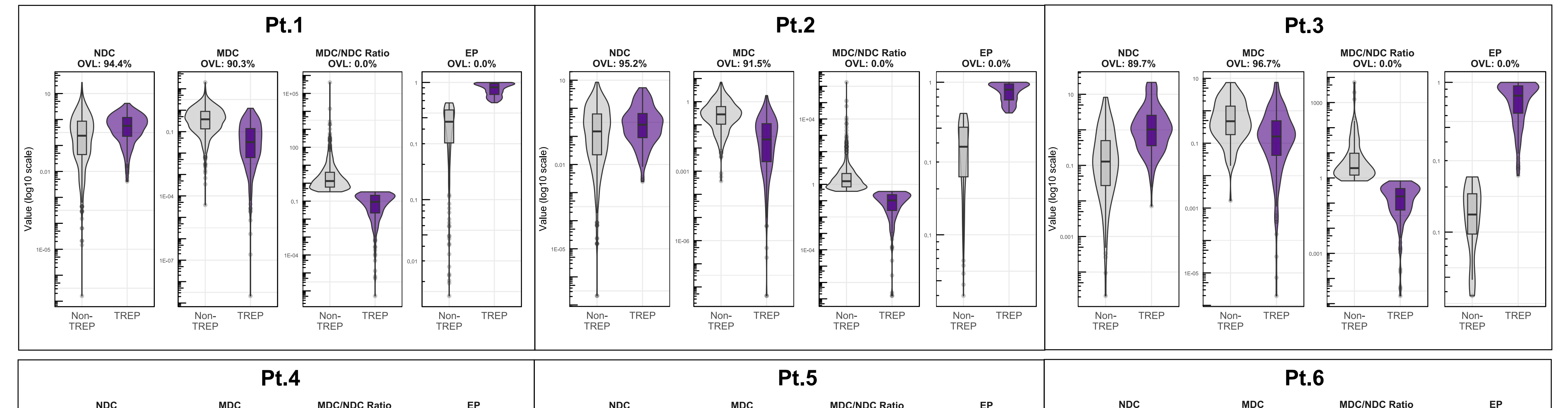

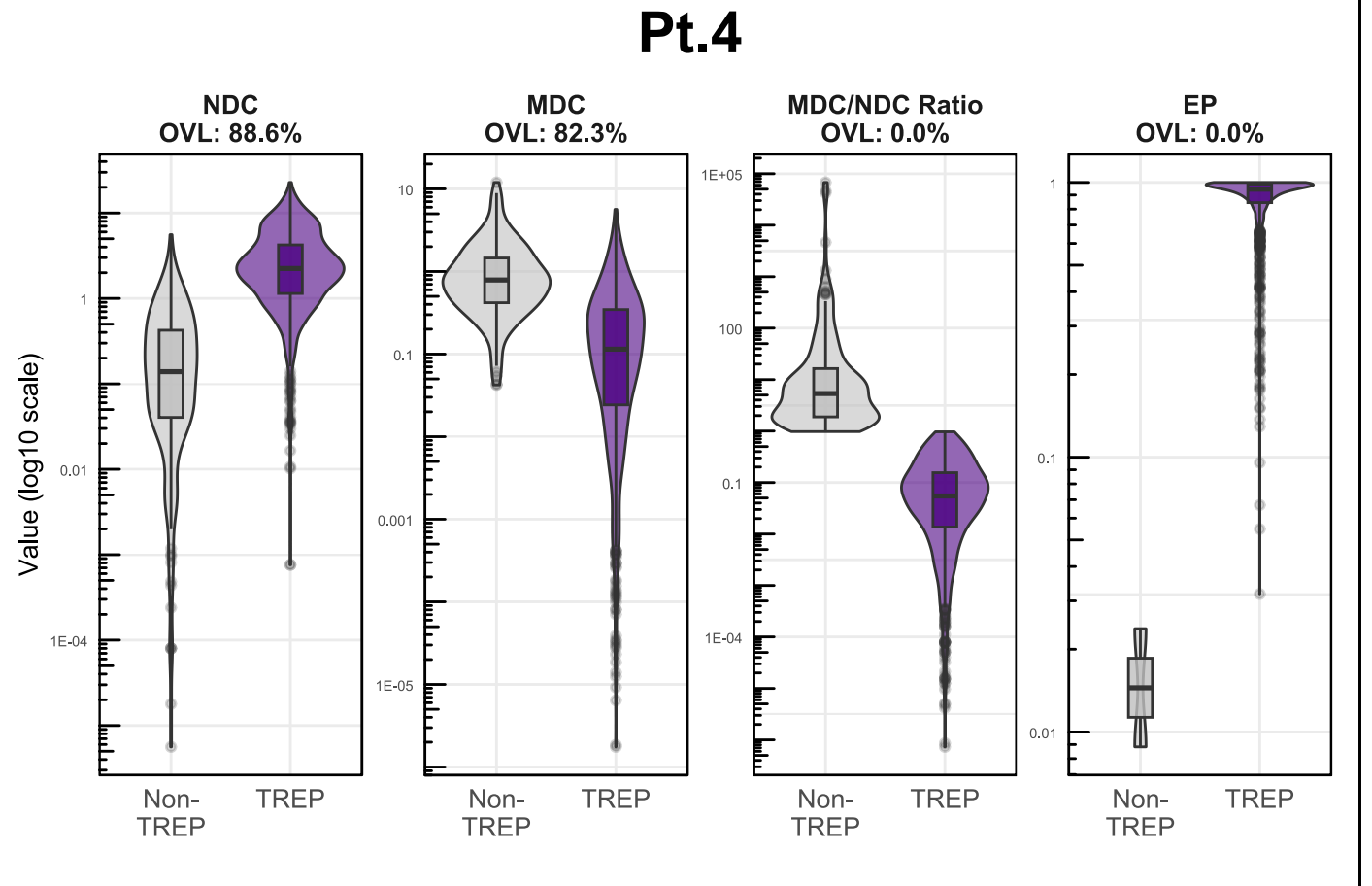

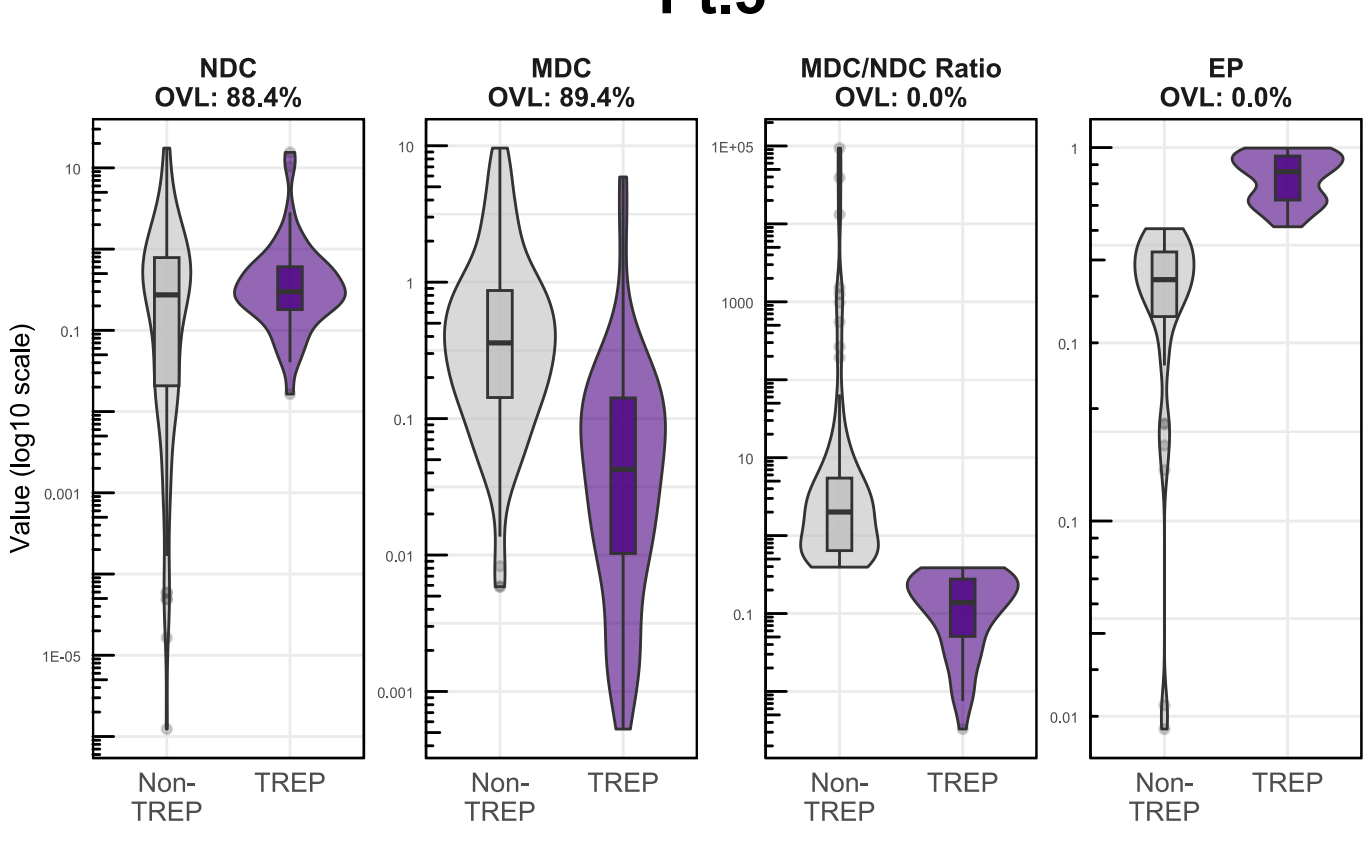

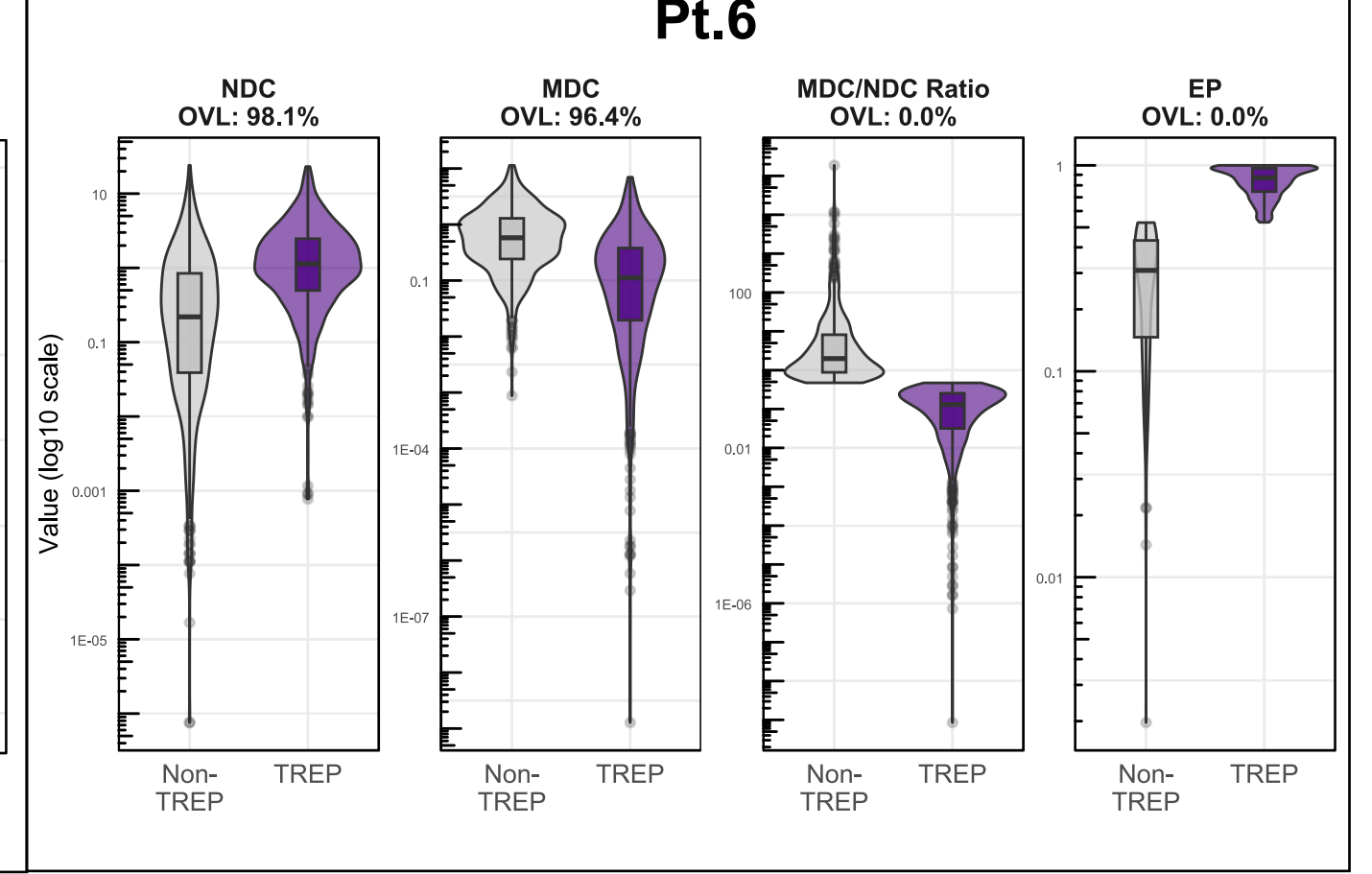

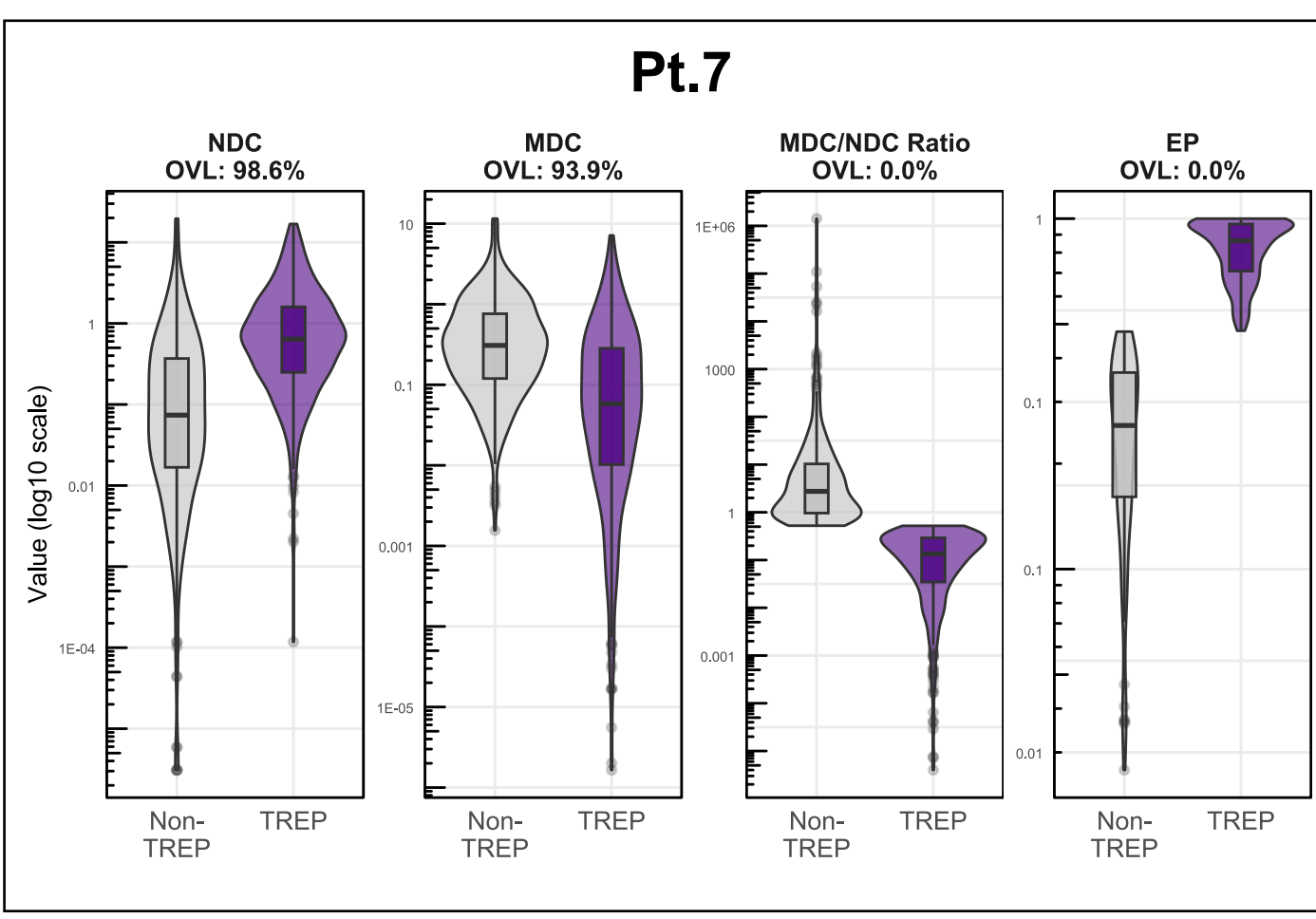

| Patient | Metric | TREP (IQR) | Non-TREP (IQR) | OVL (%) |
|---|---|---|---|---|
| Pt.1 | NDC | 0.23 - 1.20 | 0.04 - 0.86 | 94.44 |
| | MDC | 0.01 - 0.14 | 0.13 - 0.88 | 90.26 |
| | MDC:NDC | 0.02 - 0.21 | 0.60 - 4.04 | 0 |
| | EP | 0.79 - 0.98 | -3.04 - 0.40 | 0 |
| Pt.2 | NDC | 0.09 - 0.65 | 0.02 - 0.63 | 95.21 |
| | MDC | 0.00 - 0.11 | 0.11 - 0.62 | 91.48 |
| | MDC:NDC | 0.03 - 0.23 | 0.70 - 4.63 | 0 |
| | EP | 0.77 - 0.97 | -3.63 - 0.30 | 0 |
| Pt.3 | NDC | 0.36 - 2.56 | 0.03 - 0.50 | 89.70 |
| | MDC | 0.04 - 0.49 | 0.18 - 1.39 | 96.72 |
| | MDC:NDC | 0.05 - 0.38 | 1.30 - 9.61 | 0 |
| | EP | 0.62 - 0.95 | -8.61 - -0.30 | 0 |
| Pt.4 | NDC | 1.14 - 4.24 | 0.04 - 0.42 | 88.60 |
| | MDC | 0.02 - 0.35 | 0.42 - 1.46 | 82.28 |
| | MDC:NDC | 0.01 - 0.16 | 1.89 - 16.40 | 0 |
| | EP | 0.84 - 0.99 | -15.40 - -0.89 | 0 |
| Pt.5 | NDC | 0.18 - 0.61 | 0.02 - 0.79 | 88.43 |
| | MDC | 0.01 - 0.14 | 0.14 - 0.87 | 89.35 |
| | MDC:NDC | 0.05 - 0.28 | 0.64 - 5.47 | 0 |
| | EP | 0.72 - 0.95 | -4.47 - 0.36 | 0 |
| Pt.6 | NDC | 0.50 - 2.47 | 0.04 - 0.85 | 98.05 |
| | MDC | 0.02 - 0.38 | 0.24 - 1.29 | 96.44 |
| | MDC:NDC | 0.03 - 0.25 | 0.88 - 8.17 | 0 |
| | EP | 0.75 - 0.97 | -7.17 - 0.12 | 0 |
| Pt.7 | NDC | 0.25 - 1.60 | 0.02 - 0.37 | 98.64 |
| | MDC | 0.01 - 0.28 | 0.12 - 0.76 | 93.85 |
| | MDC:NDC | 0.03 - 0.29 | 0.96 - 10.34 | 0 |
| | EP | 0.71 - 0.97 | -9.34 - 0.04 | 0 |

**Figure 8**

**Figure 8**. Distribution of NDC, MDC, MDC:NDC ratio, and EP values in TREP and non-TREP cells in PCa patients (Pt1-Pt.7). Violin plots with embedded boxplots display the distribution of each metric in TREP (purple) and non-TREP (gray) cells. OVL denotes the percentage of overlap between TREP and non-TREP distributions for each metric. Values are displayed on $\log_{10}$ scale to enable visual comparison of distributions. All statistical comparisons and TREP classifications were performed on untransformed values. IQR values for each metric and patient are summarized in the inset tables.

### 3.10 CEP-classified TREP cells exhibit preference for consistent *TRPM4* and Ribo expression states in pre-IP and post-IP regions

Implementation of the complete CEP-IP framework required combining CEP classification with IP analysis to create distinct cell subpopulations. In *TRPM4*-Ribo modeling, although the DEs from all PCa patients were significantly higher than other gene sets (except AR), a wide range of DE was captured across the patients ranging from the lowest 28.6% (Pt.2) to the highest 83.5% (Pt.4), with median DE of 51.1% (**Figure 9**). In order to explain why certain patients showed low or high DE in the *TRPM4*-Ribo modeling, the following key concepts were first investigated:

i) CEP classification of TREP and non-TREP cells.

ii) Visualizing the distribution pattern of TREP (colored in purple) and non-TREP (colored in gray) cells in scatter plots with the fitted GAM curve.

iii) Determining the IP *i.e.*, the point on the x-axis (*TRPM4* value) where the distribution pattern of TREP and non-TREP cells shifted, and this was obvious visually only when concepts #1 and #2 above had been implemented.

Based on the aforementioned concepts, it was observed that TREP cells were more frequently found below the GAM curve before the IP (pre-IP *i.e.*, the region with lower *TRPM4* levels), and this pattern was inverted after the IP (post-IP *i.e.*, the region with higher *TRPM4* levels) where more TREP cells were found above the GAM curve, as exemplified by Pt.1 and Pt.4 in **Figure 9A** (FDR<0.001 for all proportion comparison shown in mosaic plots). In contrast, non-TREP cells were more frequently found above or below the GAM curve in pre-IP or post-IP, respectively. This distribution pattern was more obvious in Pt.4, with a sigmoidal GAM curve, where 100% of non-TREP cells were found above or below the pre-IP or post-IP region, respectively (FDR<0.001; **Figure 9A**).

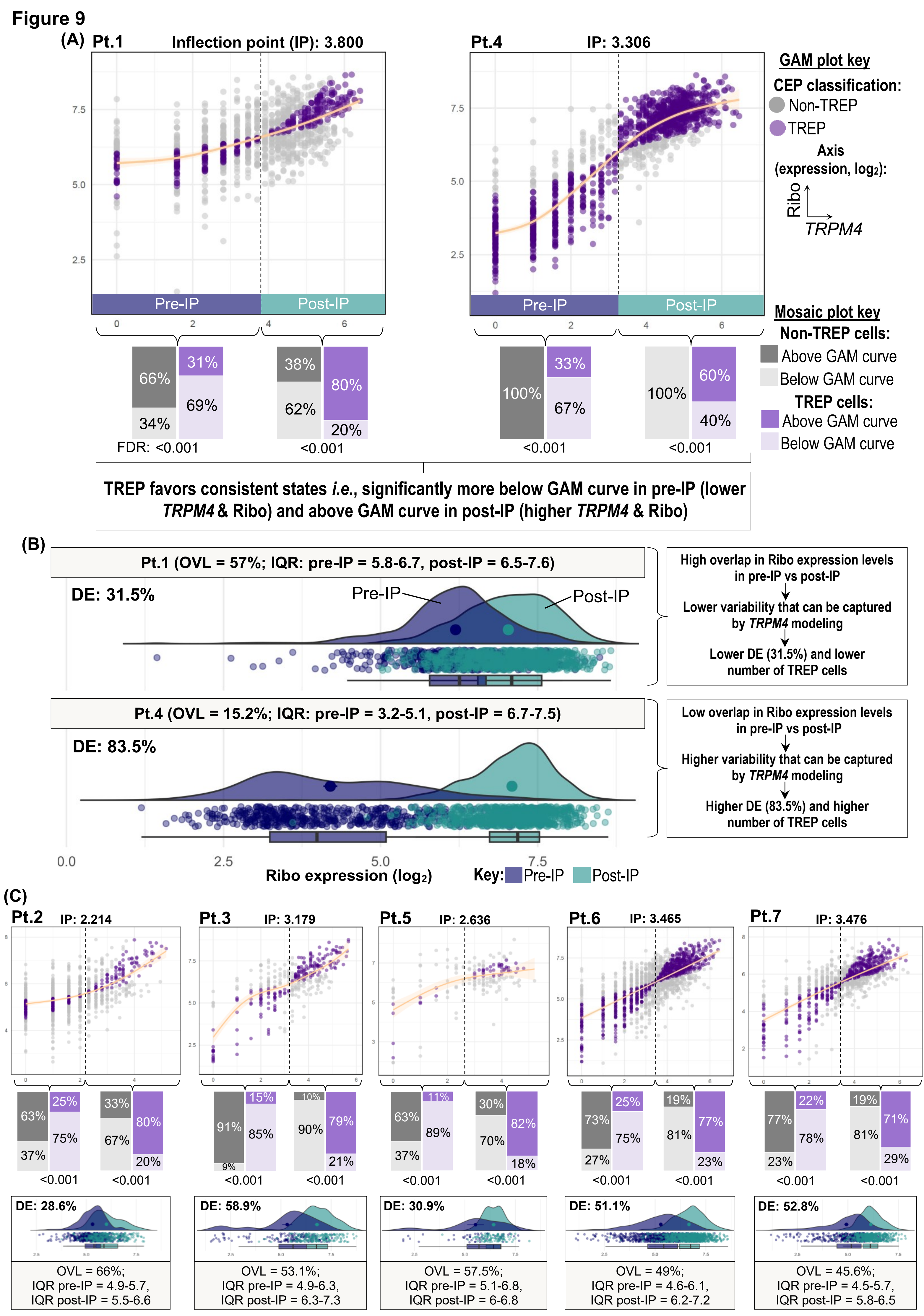
Figure 9
(A) Pt.1
Inflection point (IP): 3.800
Pt.4
IP: 3.306
GAM plot key
CEP classification:
Non-TREP
TREP
Axis
(expression, log2):
Ribo
TRPM4
Pre-IP
Post-IP
Mosaic plot key
Non-TREP cells:
Above GAM curve
Below GAM curve
TREP cells:
Above GAM curve
Below GAM curve
66%
34%
31%
69%
38%
62%
80%
20%
100%
33%
67%
100%
60%
40%
FDR: <0.001
<0.001
<0.001
<0.001
TREP favors consistent states i.e., significantly more below GAM curve in pre-IP (lower TRPM4 & Ribo) and above GAM curve in post-IP (higher TRPM4 & Ribo)
(B)
Pt.1 (OVL = 57%; IQR: pre-IP = 5.8-6.7, post-IP = 6.5-7.6)
DE: 31.5%
Pre-IP
Post-IP
High overlap in Ribo expression levels in pre-IP vs post-IP
Lower variability that can be captured by TRPM4 modeling
Lower DE (31.5%) and lower number of TREP cells
Pt.4 (OVL = 15.2%; IQR: pre-IP = 3.2-5.1, post-IP = 6.7-7.5)
DE: 83.5%
Low overlap in Ribo expression levels in pre-IP vs post-IP
Higher variability that can be captured by TRPM4 modeling
Higher DE (83.5%) and higher number of TREP cells
0.0
2.5
5.0
7.5
Ribo expression (log2)
Key: Pre-IP Post-IP
(C)
Pt.2 IP: 2.214
Pt.3 IP: 3.179
Pt.5 IP: 2.636
Pt.6 IP: 3.465
Pt.7 IP: 3.476
63% 37% 25% 75% 33% 67% 80% 20%
91% 9% 15% 85% 10% 90% 79% 21%
63% 37% 11% 89% 30% 70% 82% 18%
73% 27% 25% 75% 19% 81% 77% 23%
77% 23% 22% 78% 19% 81% 71% 29%
<0.001
DE: 28.6%
DE: 58.9%
DE: 30.9%
DE: 51.1%
DE: 52.8%
OVL = 66%; IQR pre-IP = 4.9-5.7, IQR post-IP = 5.5-6.6
OVL = 53.1%; IQR pre-IP = 4.9-6.3, IQR post-IP = 6.3-7.3
OVL = 57.5%; IQR pre-IP = 5.1-6.8, IQR post-IP = 6-6.8
OVL = 49%; IQR pre-IP = 4.6-6.1, IQR post-IP = 6.2-7.2
OVL = 45.6%; IQR pre-IP = 4.5-5.7, IQR post-IP = 5.8-6.5

**Figure 9** Characteristics of TREP and non-TREP cells in relation to GAM curves and insights on why DE% differs across patients. (A) Distribution of TREP and non-TREP cells, and the fitted GAM curve. Mosaic plots to compare the difference in the proportion of TREP vs non-TREP cells below or above the GAM curve in pre-IP (left mosaic plot placed below pre-IP region) or post-IP (right mosaic plot placed below the post-IP region). For instance, in Pt.1 (or Pt.4), significantly more TREP cells were found below or above the GAM curve in pre-IP or post-IP, respectively (FDR<0.001); (B) Raincloud plots of each cell's Ribo expression values ($log_2$) according to pre-IP or post-IP in Pt.1 (top graph) and Pt.4 (bottom graph). OVL assessed the proportion of cells (%) with overlapping range of Ribo values in pre-IP and post-IP; IQR of Ribo expression values in pre-IP or post-IP; DE% of *TRPM4*-Ribo GAM modeling; (C) Distribution patterns of TREP and non-TREP cells along the fitted GAM curves (top graphs), mosaic plots of the proportion of TREP or non-TREP cells in pre-IP or post-IP (middle graphs), and raincloud plots of Ribo expression for each cell (bottom graphs) in the rest of the PCa patients (Pt.2, Pt.3, Pt.5, Pt.6, and Pt.7).

In other words, TREP favored consistent states *i.e.*, significantly more below GAM curve in pre-IP (lower *TRPM4* and Ribo) and above GAM curve in post-IP (higher *TRPM4* and Ribo).

### 3.11 Pre-IP/post-IP overlap and Ribo expression variability predict *TRPM4*-Ribo modeling outcomes

Next, to gain insights on why patients demonstrated a wide range of DE, raincloud plots were used to visualize the distribution pattern and overlap of Ribo expression levels in pre-IP versus post-IP, in Pt.1 (low DE, 31.5%) and Pt.4 (high DE, 83.5%) PCa cells (**Figure 9B**). In Pt.1, high overlap in Ribo expression levels in pre-IP and post-IP was observed (OVL=57%). In contrast, Pt.4 with high DE showed low overlap of Ribo expression (OVL=15.2%), and the pre-IP region showed a wider range of Ribo expression levels (Pt.4 pre-IP IQR: 3.2-5.1 versus Pt.1 pre-IP IQR: 5.8-6.7; **Figure 9B**). These observations indicate that higher variability in Ribo expression levels, when modeled according to *TRPM4*, resulted in higher DE. The remaining patients also showed similar characteristics where higher OVL and lower variability in Ribo expression (in pre-IP and/or post-IP) resulted in lower DE, such as in Pt.2 (OVL=66%; IQR pre-IP: 4.9-5.7, IQR post-IP: 5.5-6.6; DE=28.6%) (**Figure 9C**). The exact number of TREP and non-TREP cells in pre-IP and post-IP regions, below or above GAM curve, for each patient is presented in **Supplementary Table 13**.

### 3.12 CEP-IP framework reveals distinct biological pathways: Ribosomal, translation, and cell adhesion GOs enriched in post-IP TREP cells

GO enrichment analysis of genes significantly upregulated in TREP or non-TREP cells, in pre-IP or post-IP, was conducted to test the potential GO differences among these four populations of PCa cells in each patient. Post-IP TREP cells in six of the seven PCa patients consistently showed significant enrichment of ribosomes, translation, and cell adhesion GOs (FDR<0.01; **Figure 10**). The upregulated genes in these three GOs primarily involved ribosomal genes, including the seven ribosomal genes within Ribo gene set such as *RPL10*, *RPL27*, *RPS2*, or *RPS8* (**Supplementary Table 14**). Post-IP TREP cells of Pt.5 showed enriched ribosome GOs only (FDR=0.015) without translation or cell adhesion GOs enrichment, potentially due to low number of TREP cells (post-IP TREP cells, n=39, while the number of post-IP TREP cells in all other patients were >50). In pre-IP TREP cells, major histocompatibility complex (MHC) class I and II GOs were significantly enriched (FDR<0.01) in three PCa patients (Pt. 3, Pt.4, and Pt.6), whereby MHC class I and II genes such as *HLA-A*, *HLA-B*, or *CD74* were significantly upregulated. Non-TREP cells in pre-IP or post-IP yielded enrichment of mitochondria and other cellular structures (endoplasmic reticulum, lysosomes, or organelles), respectively, across majority of the PCa patients (**Figure 10** and **Supplementary Table 13**).

To compare CEP-IP subpopulations with Seurat clusters, the four CEP-IP subpopulations (pre-IP TREP, pre-IP non-TREP, post-IP TREP, and post-IP non-TREP) were mapped onto existing Seurat cluster identities in all 5,855 PCa cells spanning the five PCa-enriched clusters (**Supplementary Figure 5**). CEP-IP subpopulations were non-randomly distributed across Seurat clusters (global chi-square test of independence: $X^2$ = 858.62, $p$<0.0001, Monte Carlo simulation, B=9,999 permutations). Essentially, normalized Shannon entropy within each cluster ranged from 0.81 to 0.99, indicating that all four CEP-IP subpopulations were represented by every cluster with high or near-maximal diversity. These findings show that CEP-IP stratification identifies cell states distributed in multiple clusters, capturing heterogeneity not apparent to conventional scRNA-seq clustering.

### 3.13 Pre-IP TREP cells formed distinct population of cells with post-IP TREP cells in Monocle3 trajectory

As shown in the previous section, GAM's DE potentially identified population of cells that progressed along a continuous trajectory in the *TRPM4*-Ribo transcriptional space. To test this further, Monocle3 trajectory analysis was performed to observe if pre-IP TREP cells formed distinct, yet continuous, population of cells with post-IP TREP cells along Monocle3 trajectory *i.e.*, well-separated clusters of TREP cells with intermediate TREP cells connecting both distinct populations. Qualitatively, visual inspection of the Monocle3 trajectory in UMAP1 and UMAP2 space for each patient showed distinct clusters of pre-IP TREP cells (colored in magenta) that were well-separated from post-IP TREP cells (colored in indigo), with intermediate pre-IP or post-IP TREP cells (**Figure 11**). These observations also corroborated with ridgeline plots of cells density in UMAP1 space (the Monocle3 UMAP dimension that captured the most variability) showing dissimilar yet coherent distribution of TREP cells between the pre-IP and post-IP regions. Quantitatively, magnitude of their separation in UMAP1 space was assessed whereby all patients showed significant separation of pre-IP TREP cells from post-IP TREP cells (FDR<0.05). Pt.3, Pt.4, and Pt.6 showed large magnitude of separation that was highly significant ($|\delta| \geq 0.43$; FDR<0.001).

Figure 10

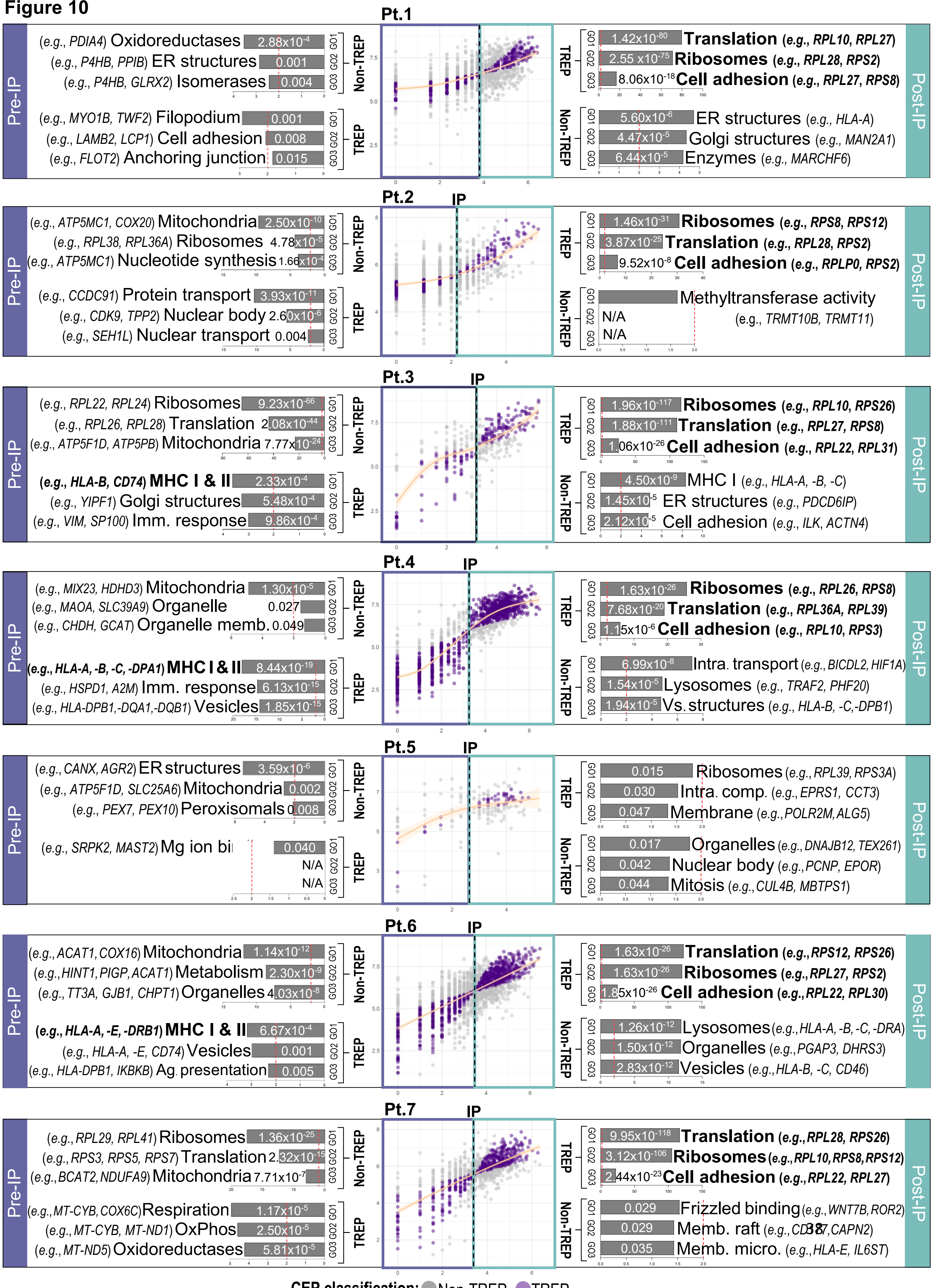

**Figure 10** GOs enriched in TREP (purple) or non-TREP (gray) cells in pre-IP or post-IP of each PCa patient. The IP for each patient represents the point where more TREP cells transitioned from being found more frequently below the GAM curve to above GAM curve *i.e.*, the point separating pre-IP (purple box) from post-IP (green box). For each pre-IP and post-IP region, the cells were binarized into TREP or non-TREP cells, with each population of cells containing their own enriched GOs. Three significant GO groups (GO1-GO3) are represented in each bar graph (x-axis: $\log_{10}$FDR) with dotted red line representing FDR=0.01 [-$\log_{10}(0.01)=2$] to illustrate a more stringent cut-off; note that the actual significance threshold applied for GO shortlisting was FDR<0.05. GO groups representing ribosomes, translation, cell adhesion, and MHC class I and II GOs are in bold.

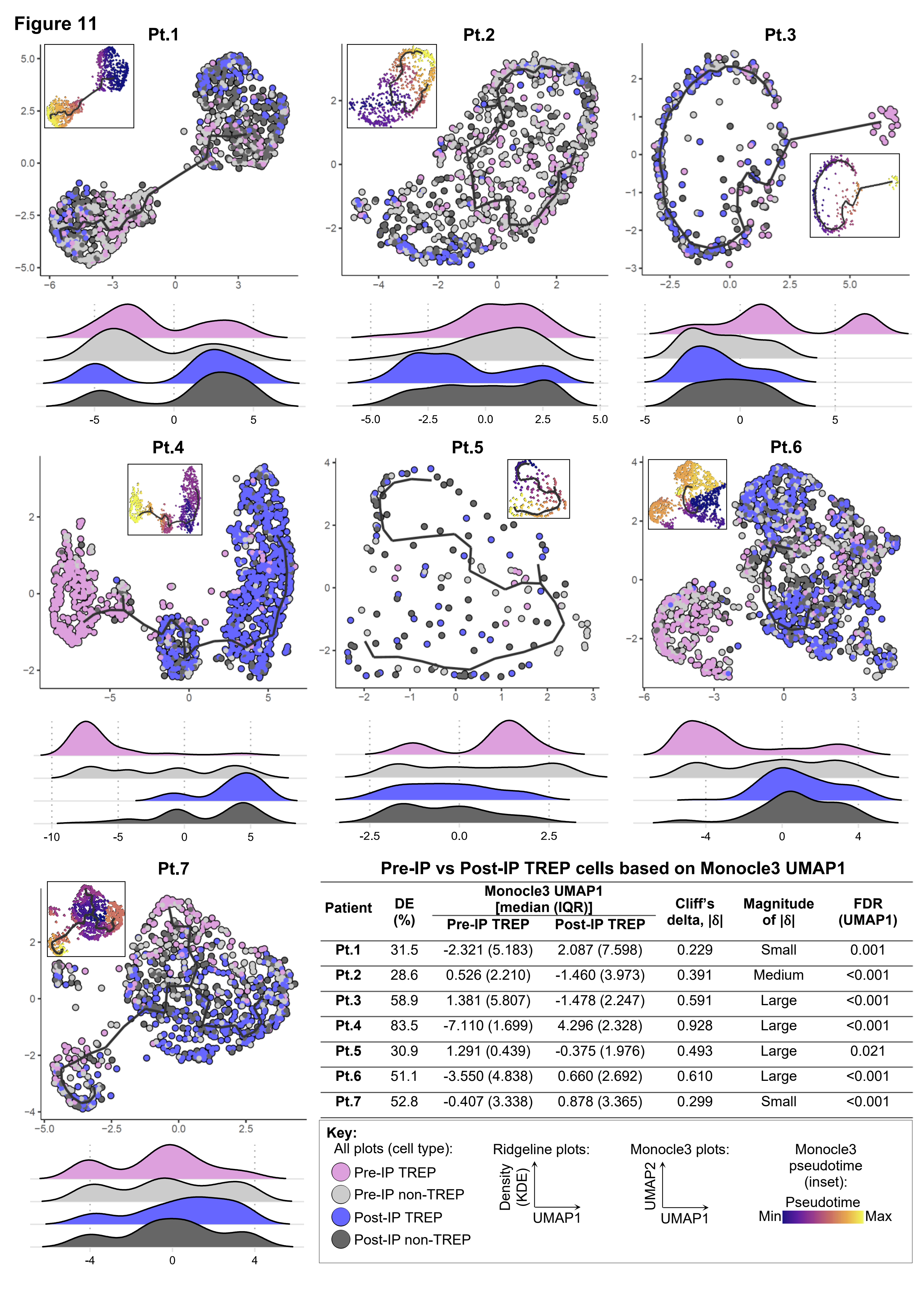

**Pre-IP vs Post-IP TREP cells based on Monocle3 UMAP1**

| Patient | DE (%) | Monocle3 UMAP1 [median (IQR)] | | Cliff's delta, \|δ\| | Magnitude of \|δ\| | FDR (UMAP1) |
|---|---|---|---|---|---|---|
| | | Pre-IP TREP | Post-IP TREP | | | |
| **Pt.1** | 31.5 | -2.321 (5.183) | 2.087 (7.598) | 0.229 | Small | 0.001 |
| **Pt.2** | 28.6 | 0.526 (2.210) | -1.460 (3.973) | 0.391 | Medium | <0.001 |
| **Pt.3** | 58.9 | 1.381 (5.807) | -1.478 (2.247) | 0.591 | Large | <0.001 |
| **Pt.4** | 83.5 | -7.110 (1.699) | 4.296 (2.328) | 0.928 | Large | <0.001 |
| **Pt.5** | 30.9 | 1.291 (0.439) | -0.375 (1.976) | 0.493 | Large | 0.021 |
| **Pt.6** | 51.1 | -3.550 (4.838) | 0.660 (2.692) | 0.610 | Large | <0.001 |
| **Pt.7** | 52.8 | -0.407 (3.338) | 0.878 (3.365) | 0.299 | Small | <0.001 |

**Figure 11**. Distribution patterns of TREP cells in pre-IP (magenta) and post-IP (indigo) according to Monocle3 cell trajectory of the PCa cells of each patient (Pt.1-Pt.7). Monocle3 pseudotime plot (root cell: highest *TRPM4* expression) is included as an inset of the main trajectory plot. Ridgeline plot is shown below the trajectory plot according to Monocle3's UMAP1 (x-axis) and density (kernel density estimation, KDE; y-axis). Comparison of Monocle3's UMAP1 values of TREP cells in pre-IP versus post-IP is shown and magnitude of |δ| is according to established cut-offs [46]. Note that the ridgeline plots display KDE of UMAP1 coordinate distributions, which may appear shifted relative to raw coordinates in the trajectory plot, but their distribution patterns (*i.e.*, density across UMAP1 axis) remain similar.

### 3.14 CEP-IP automated IP detection vs. visually determined IPs in *TRPM4*-Ribo dataset

Automated IP detection was applied to the *TRPM4*-Ribo dataset to verify the visually determined IPs used in all preceding analyses (sections 3.10-3.12). The CEP-IP automated detection pipeline and IPRS framework are detailed in **Figure 12**, and results for all seven patients are illustrated in **Supplementary Figure 6**. Automated IPs were concordant with visually determined values in the patients, with a median absolute difference of 0.083 *TRPM4* units (IQR: 0.045-0.144; range: 0.004-0.250), representing <2% of the observed *TRPM4* expression range. Differences were bidirectional across the patients, indicating absence of consistent over- or under-estimation.

Visually determined IPs were within the automated IPs' 95% CI in six of seven patients. The sole exception was Pt.4, where the manual IP (3.306) marginally exceeded the upper CI bound by 0.012 *TRPM4* units, reflecting the narrow CI of Pt.4 (width: 0.138). All seven patients were classified as strong tier (IPRS: 0.631-0.769), with no sample reaching very strong (IPRS ≥0.80) tier, reflecting the deliberate stringency in the design of the IPRS algorithm. Detailed IPRS results including the calculation to reach every component C1-C5 for each PCa patient are presented in **Supplementary Table 15**. The CEP-IP framework and IPRS were subsequently applied to two independent validation datasets that contained more challenging cases with fewer cells, non-obvious expression transitions, and intratumoral heterogeneity (GBM dataset).

Figure 12

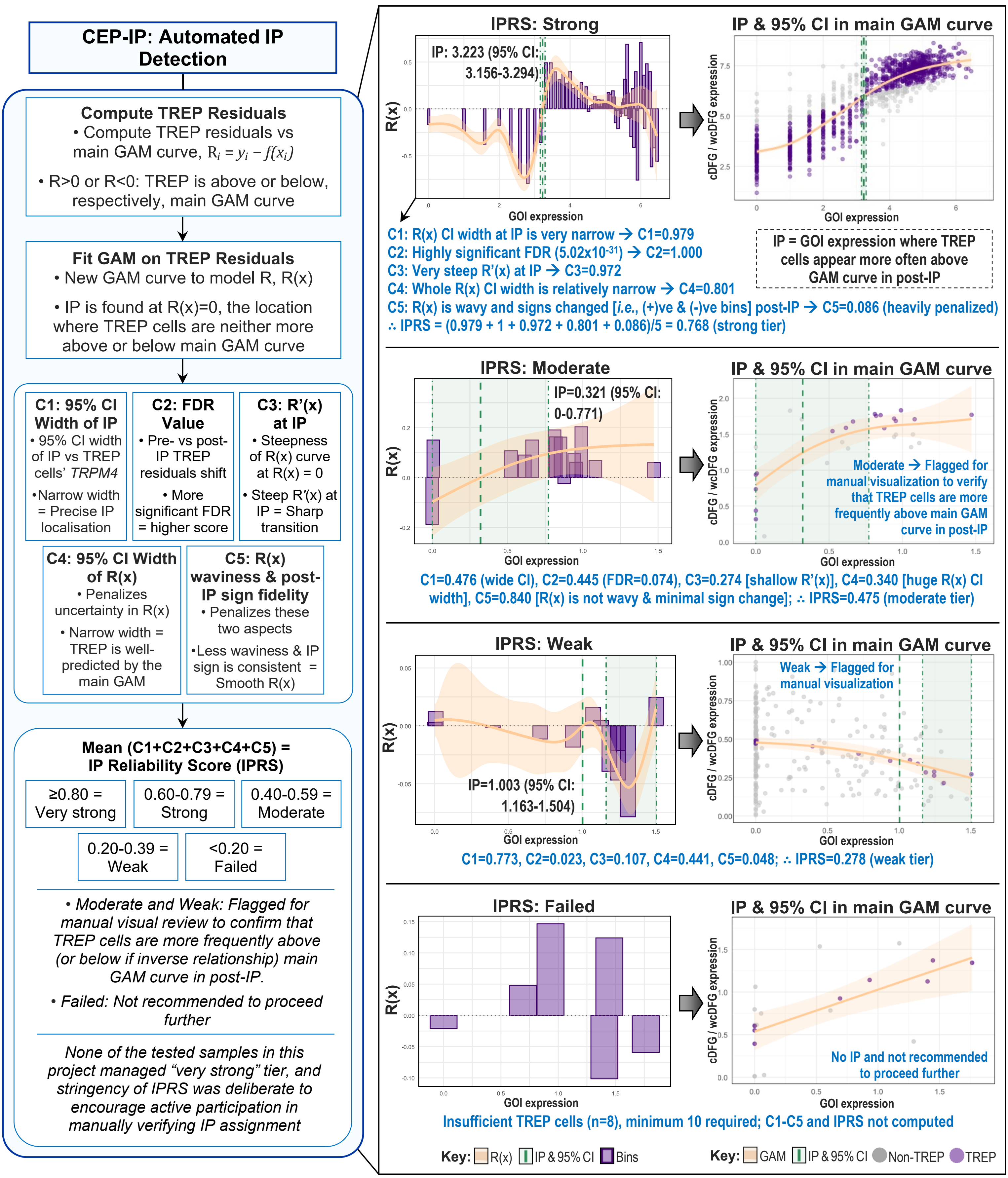

**Figure 12.** CEP-IP automated IP detection and IPRS workflow. *Left panel*: Schematic of the automated IP detection and IPRS pipeline. TREP residuals are computed against the main GAM curve, and a secondary GAM, R(x), is fitted to these residuals. The IP is identified at the first negative-to-positive zero-crossing of R(x), the GOI expression value at which TREP cells transition from predominantly below to above the main GAM curve, with a 95% CI derived via Brent's method for root-finding. The IPRS is the unweighted mean of five components C1-C5. IPRS moderate and weak tiers are flagged for manual visual review; Failed samples are excluded from downstream analysis; *Right panel*: Representative examples of four IPRS tiers shown as paired R(x) diagnostic plots and main GAM plots with the automated IP and 95% CI.

### 3.15 CEP-IP framework validation in the Allen MTG brain dataset: Neuronal projection and synaptic GOs in post-IP TREP cells

The Allen Human MTG SMART-seq dataset comprised of 24 clusters (clusters 0-23) representing different cortical cell types *e.g.*, excitatory neurons (clusters 0-4, 6, 9-10, 12-14, 21), inhibitory interneuron subtypes (clusters 5, 7-8, 11, 17-19, 23), astrocytes (cluster 15), oligodendrocytes (cluster 16), oligodendrocyte progenitor cells (OPCs; cluster 20), and microglia (cluster 22), as determined by ToppGene Coexpression database enrichment of the top 50 cluster markers. *CARM1P1* was most highly expressed in cluster 9 (Exc L3-5 RORB eC1 neurons), with significantly lower expression in all other clusters (FDR<0.01), including cluster 3 (Exc L3 RORB CARTPT neurons) (**Figure 13A**). Following the standard Seurat QC workflow, six of the eight donors were excluded due to insufficient cells (<50 cells per sample post-processing), retaining MTG.1 (H200.1030; n=242 cells) and MTG.2 (H200.1023; n=160 cells) for downstream CEP-IP analysis.

Five DFGs were monotonically correlated with *CARM1P1* (*i.e.*, *CLMN*, *EPHA3*, *EPHA6*, *LOC101928964*, *ROBO2*) (**Figure 13A**). These genes were averaged into a cDFG score vs. cHKG (the comparator group). GAM fitting yielded higher DE for cDFG than cHKG in both samples (MTG.1: 41.2% vs 6.1%; MTG.2: 51.1% vs 6.7%) (**Figure 13B**). Automated IP detection identified strong tier IPs in both samples (MTG.1: IP=0.202, IPRS=0.703; MTG.2: IP=0.435, IPRS=0.701; **Figure 13C**). CEP-IP framework implementation revealed consistent enrichment of TREP cells above the GAM curve in the post-IP region for both samples.

GO enrichment analysis of post-IP TREP cells yielded significant enrichment of neuron projection, neuron compartment, and synapse organisation GOs (FDR<0.05), while the rest of the cell subpopulations enriched for distinct GOs in both patients (**Figure 13D**). Detailed GAM performance metrics and list of enriched GOs of both validation datasets (including GBM dataset detailed in the next section 3.16) are presented in **Supplementary Table 16**. Monocle3 trajectory analysis showed clear separation of post-IP TREP vs. pre-IP TREP cells qualitatively in UMAP plots. Significant separation of post-IP from pre-IP TREP cells was demonstrated along the UMAP1 space in both MTG.1 (Cliff's δ=0.414; FDR=0.004) and MTG.2 (Cliff's δ=0.565; FDR=$6.05 \times 10^{-6}$) (**Figure 13E).**

**Figure 13**

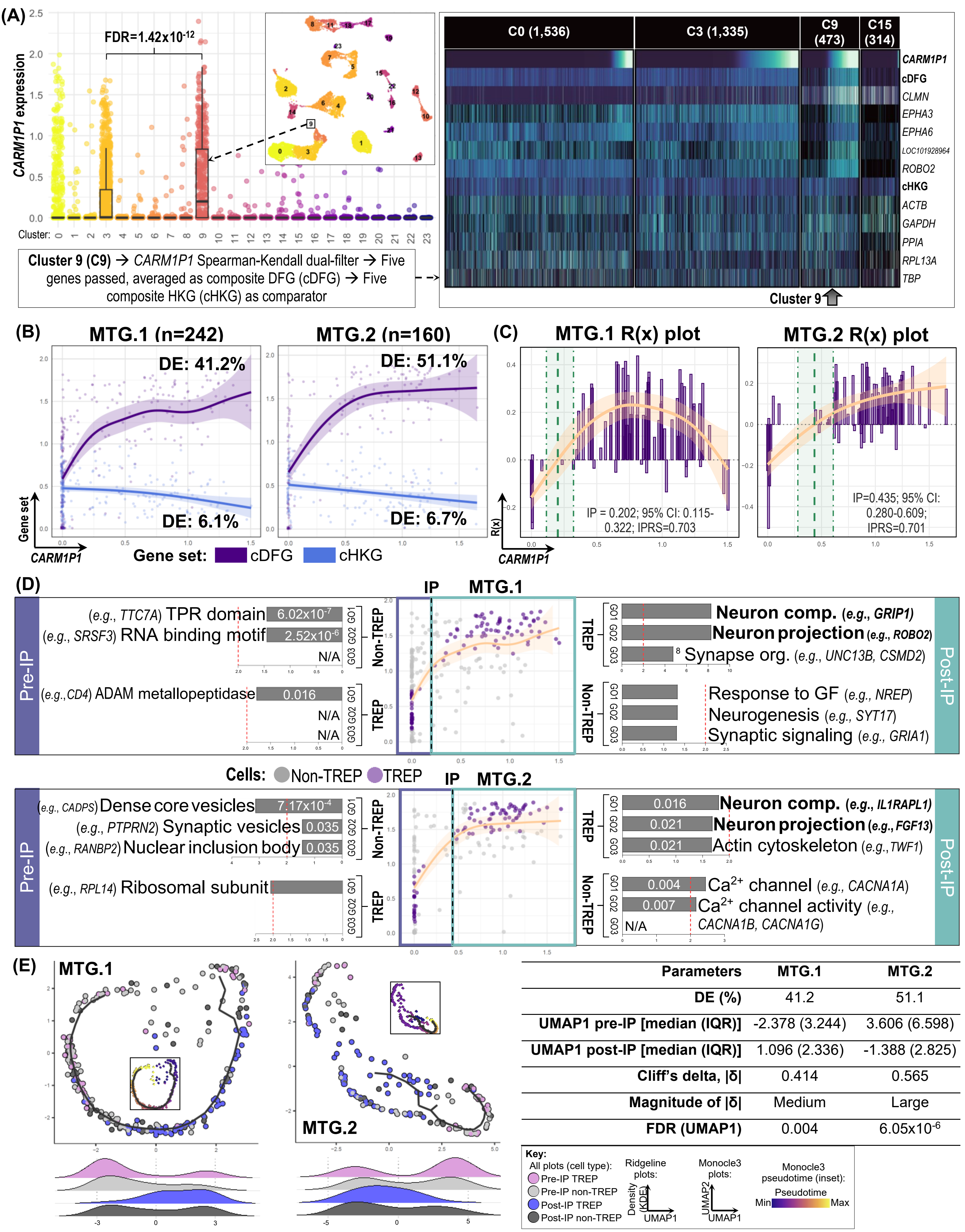

| Parameters | MTG.1 | MTG.2 |
|---|---|---|
| DE (%) | 41.2 | 51.1 |
| UMAP1 pre-IP [median (IQR)] | -2.378 (3.244) | 3.606 (6.598) |
| UMAP1 post-IP [median (IQR)] | 1.096 (2.336) | -1.388 (2.825) |
| Cliff's delta, \|δ\| | 0.414 | 0.565 |
| Magnitude of \|δ\| | Medium | Large |
| FDR (UMAP1) | 0.004 | 6.05x10^-6 |

**Figure 13.** CEP-IP framework validation in the Allen MTG dataset. (A) *CARM1P1* was most highly expressed in cluster 9 (Exc L3-5 RORB eC1 neurons). Five DFGs passed dual-filtering and averaged into cDFG. Heatmap shows cDFG and cHKG expression in representative clusters ordered by ascending *CARM1P1* expression; (B) GAM scatter plots of *CARM1P1* versus cDFG and cHKG in MTG.1 and MTG.2; (C) Automated IP detection R(x) diagnostic plots showing strong tier IPRS classifications in both samples; (D) CEP-IP GAM scatter plots with top enriched GOs for each cell subpopulation in cluster 9; (E) Monocle3 trajectory plots with ridgeline distributions and summary statistics.

### 3.16 *FOXM1* is the sole GOI with MES-state DFGs in adult GBM cells

In the second CEP-IP validation dataset of GBM cells, *FOXM1* was initially selected as the GOI due to its emerging roles in GBM mesenchymal transition, a hallmark aggressive feature [52]. In the Neftel GBM dataset, *FOXM1* expression was most highly expressed in Louvain clusters 1, 6, and 9. However, each of these clusters contained a mixture of all four canonical GBM cell states (MES, AC, NPC, OPC), consistent with the well-documented transcriptional heterogeneity of GBM [45]. Among *FOXM1*$^{+}$ cells, the four cell states were relatively evenly distributed in all clusters combined (MES 33%, AC 30%, NPC 15%, OPC 22%), indicating that *FOXM1* expression was not confined to a single cell state. In view of this intratumoral heterogeneity, 659 GOIs upregulated in GBM versus normal brain derived from Prasad *et al.* [53] were screened for DFGs within each cell state independently (**Supplementary Table 17**). *FOXM1* was the sole GOI yielding DFGs in the MES cell state (the most aggressive GBM cell state) with 7 MES DFGs (*BIRC5, MKI67, CENPF, TOP2A, PBK, TROAP,* and *NUSAP1*). All other GOIs with DFGs showed enrichment in non-MES GBM cell states. Expression of these seven genes were integrated into a wcDFG composite with frequency-based weights proportional to the number of GBM cell states each gene was identified in (**Figure 14**).

As 19 of 20 patients had fewer than 50% *FOXM1*$^{+}$ cells (**Figure 15A**), jointly-zero cells may inflate $r_s$ and $\tau$ via zero-concordance. This was evident when GAM was fitted to model cDFG and *FOXM1* relationship in all cells within each patient, where fluctuating GAM curves occurred in multiple cases (*e.g.*, MGH66, MGH110, MGH121) (**Supplementary Figure 7**) due to zero-inflation that distorts the curves. This necessitated monotonicity validation within *FOXM1*$^{+}$ cells only (*i.e.*, within-positive monotonicity) by computing $r_s$ and $\tau$ between *FOXM1* and wcDFG restricted to *FOXM1*$^{+}$ cells in each patient separately. Three patients (MGH152, MGH66, MGH110) passed the dual-filter, indicating genuine within-positive monotonicity. Two patients (MGH100, MGH121) achieved partial pass (either Spearman or Kendall filter), while eight patients showed weak within-positive signal (did not pass either filter). Ten patients were excluded due to insufficient *FOXM1*$^{+}$ cells (n<50 yields insufficient TREP cells per quadrant *i.e.*, pre- and post-IP of TREP vs. non-TREP). Patients passing dual-filter or partial-pass criteria were selected for downstream CEP-IP framework analysis.

**Figure 14**

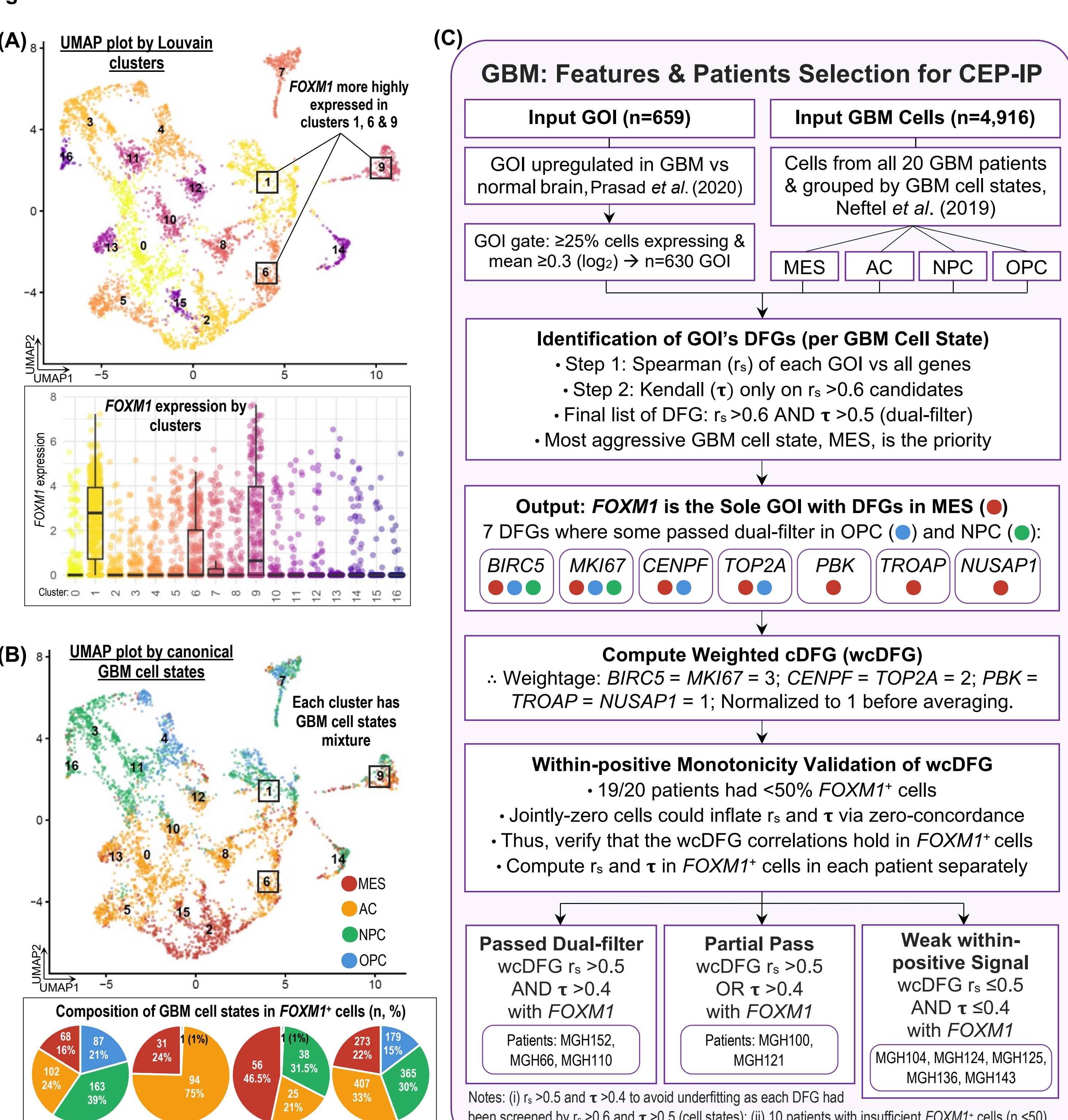

**Figure 14.** Feature and patient selection for the CEP-IP framework in GBM dataset. (A) UMAP of 4,916 adult malignant GBM cells colored by Louvain clusters (top) and *FOXM1* expression by cluster (bottom), showing enrichment in clusters 1, 6, and 9; (B) UMAP colored by canonical GBM cell state (top) and pie charts of cell state composition in *FOXM1*$^+$ cells per cluster (bottom), showing MES predominance in clusters 1 and 6; (C) Feature selection flowchart: 659 GBM-upregulated GOIs screened by expression gate (n=630 passed) before cell state-stratified for dual-filter ($r_s$ >0.6 and $\tau$ >0.5), identifying *FOXM1* as the sole GOI with MES-state DFGs (*BIRC5, MKI67, CENPF, TOP2A, PBK, TROAP, NUSAP1*), combined into a wcDFG. Within-positive monotonicity validated wcDFG-*FOXM1* correlations in *FOXM1*$^+$ cells.

### 3.17 CEP-IP reveals cell division and microtubule programs in post-IP TREP cells of *FOXM1*$^{+}$ GBM

The heatmap of *FOXM1*, wcDFG, individual DFG members, and cHKG in all 20 patients ordered by ascending *FOXM1* expression visually showed co-expression of *FOXM1* with wcDFG and its constituent genes (**Figure 15A**). GAM fitting yielded DE values where the three cases that passed dual-filter showed the highest DE (36.3-55.9%). The fitted GAM curves for all 20 cases showed that smooth GAM curves instead of fluctuating ones were obtained that justified the focus on *FOXM1*$^{+}$ cells for GAM fitting (**Supplementary Figure 8**), and dual-filter-passed cases showed strong IPRS tier (**Supplementary Figure 9**). GO enrichment analysis of post-IP TREP cells in all three dual-filter-passed patients showed consistent enrichment of cell division, cell cycle, and microtubule GOs with highly significant FDR ($5.63 \times 10^{-5}$ to $2.70 \times 10^{-23}$) (**Figure 15B**), a pattern absent in partial pass and weak within-positive signal patients. Pre-IP non-TREP cells of dual-filter-passed patients also showed enrichment of cell cycle and microtubules GOs but at lower FDR values (0.006 to $5.21 \times 10^{-16}$) in each patient.

### 3.18 3D UMAP trajectories reveal a continuous structure in dual-filter-passed GBM patients

Monocle3 trajectory analysis of 2D UMAP embeddings showed significant separation of post-IP from pre-IP TREP cells in dual-filter-passed patients MGH66 (FDR UMAP1: $3.04 \times 10^{-5}$) and MGH110 (FDR UMAP1: $9.0 \times 10^{-4}$; UMAP2: 0.047) except MGH152. Despite having the highest DE (55.9%), MGH152 showed insignificant separation in 2D (FDR UMAP1: 0.487; UMAP2: 0.117), but with two visually distinct populations of post-IP TREP cells clustered at opposing edges of UMAP1 and UMAP2 (**Figure 15C**). This was suggestive of the presence of a continuous 3D path compressed into two separated clusters in 2D. Extension of Monocle3 trajectory analysis to 3D UMAP embeddings verified this interpretation, revealing that post-IP TREP cells formed a continuous path in MGH152 (**Figure 16**). Interestingly, despite PERMANOVA demonstrating significant separation of pre-IP and post-IP TREP cells (FDR=0.002), both populations formed a loop-like continuum, where the edges of each population converged at transition-like points. This was supported by significant PERMDISP (FDR=0.031) indicating internal variability, implying internal dispersion due to a continuous loop trajectory between pre- and post-IP TREP cells. MGH66 and MGH110 similarly showed significant PERMANOVA separation (FDR=0.003 each), supported by significant individual UMAP axes, verifying robust 3D trajectory separation in all three dual-filter-passed patients. In contrast, MGH100 and MGH104 showed no significant separation across any metric (all FDR >0.05), consistent with their partial-pass and weak within-positive signal classifications, respectively. MGH121 showed significant PERMANOVA (FDR=0.004) and UMAP1/2 separation, but this should be interpreted with caution due to <10 post-IP TREP cells (n=8).

Figure 15

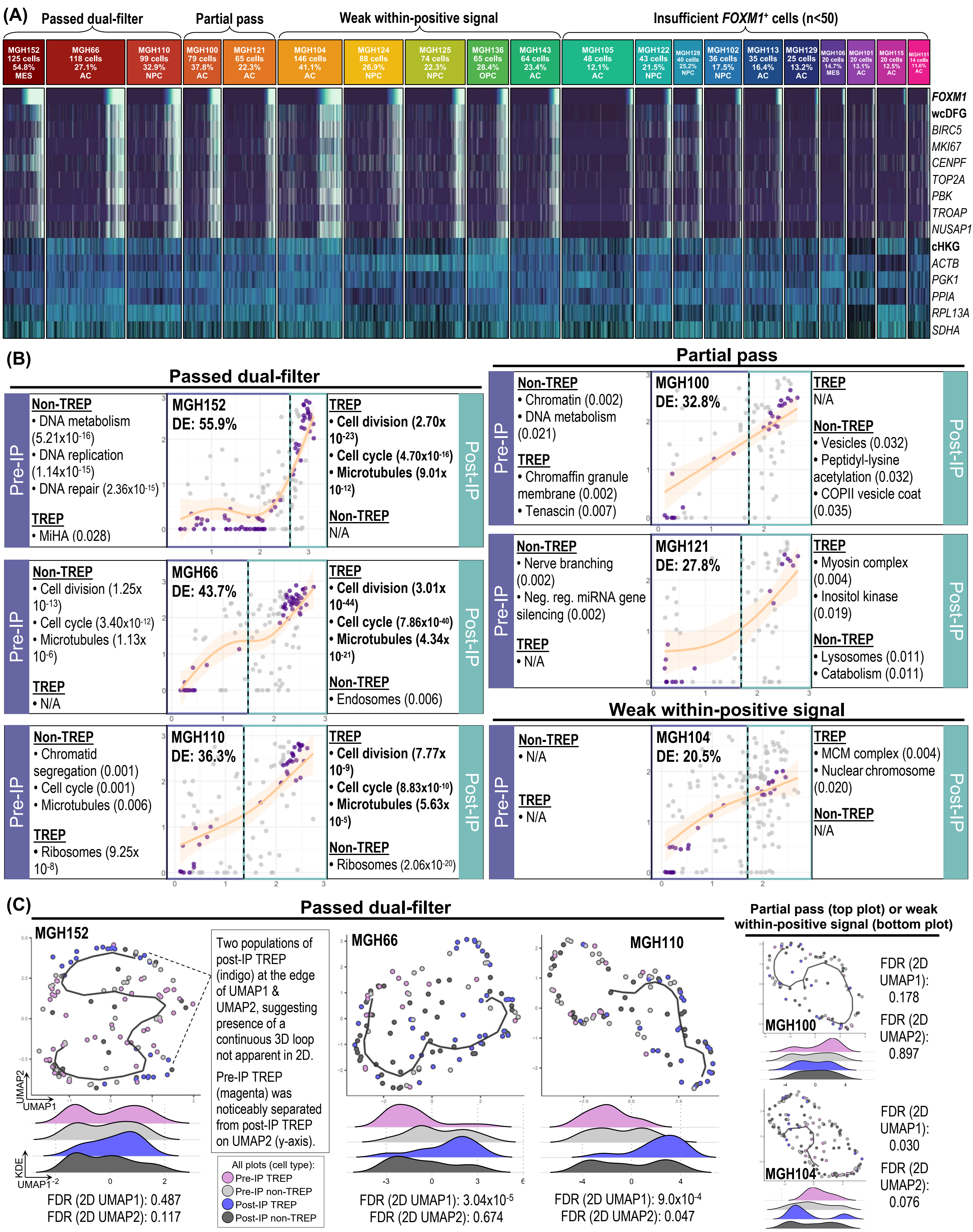

**Figure 15.** CEP-IP framework application in *FOXM1*$^{+}$ adult malignant GBM cells. (A) Heatmap of *FOXM1*, wcDFG, individual DFG and cHKG members in all 20 patients ordered by ascending *FOXM1* expression, grouped by within-positive monotonicity classification. Expression represented as z-score; (B) GAM scatter plots of *FOXM1* versus wcDFG expression for dual-filter-passed (MGH152, MGH66, MGH110), partial-pass (MGH100, MGH121), and weak within-positive signal (MGH104) patients. TREP cells (purple) and non-TREP cells (gray) are shown with the IP dashed line separating pre-IP and post-IP regions. Top enriched GO terms (FDR<0.05) for each subpopulation are annotated; (C) Monocle3 2D UMAP trajectory plots for dual-filter-passed and partial-pass or weak within-positive signal patients, with ridgeline plots of UMAP1 distributions and FDR values for post-IP versus pre-IP TREP separation. MGH121 is omitted due to space constraints, and its 3D UMAP results are reported in the next section. COPII: Coat protein complex II; MCM: Minichromosome maintenance complex; MiHA: Minor histocompatibility antigens.

**Figure 16**

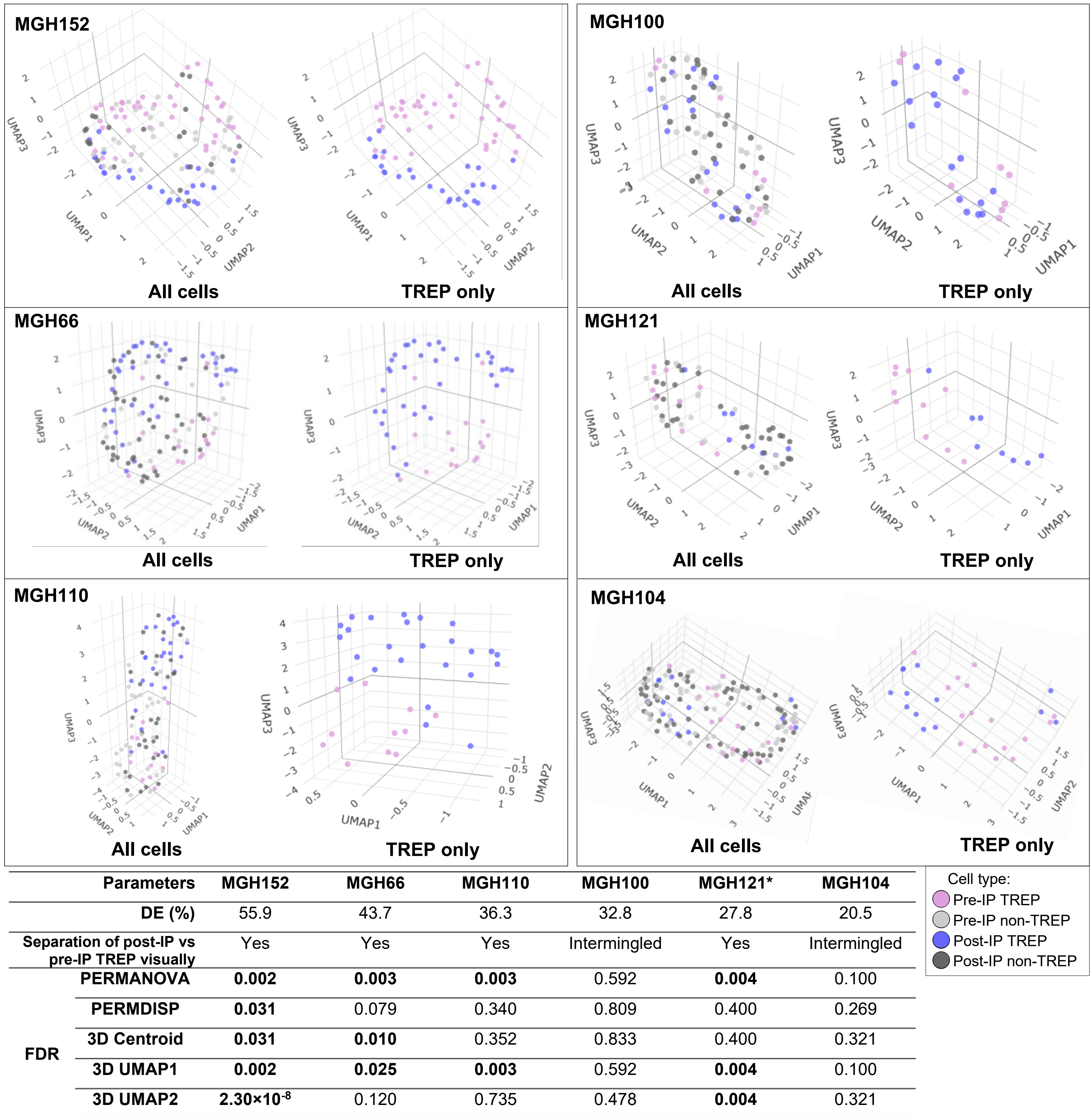

| | Parameters | MGH152 | MGH66 | MGH110 | MGH100 | MGH121* | MGH104 |
|---|---|---|---|---|---|---|---|
| | **DE (%)** | 55.9 | 43.7 | 36.3 | 32.8 | 27.8 | 20.5 |
| | **Separation of post-IP vs pre-IP TREP visually** | Yes | Yes | Yes | Intermingled | Yes | Intermingled |
| **FDR** | **PERMANOVA** | **0.002** | **0.003** | **0.003** | 0.592 | **0.004** | 0.100 |
| | **PERMDISP** | **0.031** | 0.079 | 0.340 | 0.809 | 0.400 | 0.269 |
| | **3D Centroid** | **0.031** | **0.010** | 0.352 | 0.833 | 0.400 | 0.321 |
| | **3D UMAP1** | **0.002** | **0.025** | **0.003** | 0.592 | **0.004** | 0.100 |
| | **3D UMAP2** | **$2.30\times10^{-8}$** | 0.120 | 0.735 | 0.478 | **0.004** | 0.321 |
| | **3D UMAP3** | **0.037** | **$5.79\times10^{-5}$** | **$3.36\times10^{-5}$** | 0.833 | **0.071** | 0.322 |

**MGH121 contains <10 post-IP TREP cells (n=8) and caution is advised when interpreting the significant FDR values.*

**Figure 16.** 3D UMAP separation of post-IP and pre-IP TREP cells in *FOXM1*$^{+}$ GBM samples. 3D UMAP plots show all cells (left) and TREP cells only (right) for each sample, colored by cell type. Samples are grouped according to dual-filter passed (MGH152, MGH66, MGH110; left column) and partial pass or weak within-positive signal (MGH100, MGH121, MGH104; right column). The accompanying table summarizes DE (%) by the fitted GAM, visual assessment of post-IP vs pre-IP TREP separation, and FDR values for PERMANOVA, PERMDISP, 3D centroid distance, and each UMAP axis independently. Bold values indicate FDR <0.05.

For further inspection of pre-IP and post-IP TREP cell separation in 3D, or for the presence of loop-like continuum, interactive 3D UMAP plots of all six GBM samples are provided as **Supplementary Figure 10** (HTML format; PDF shows static screenshot only). Finally, a high-level summary of the CEP-IP framework, its sequential steps, criteria and recommendations for application are presented in **Figure 17**.

Figure 17

# CEP-IP Framework

**❶ Target: GOI**
Overexpressed genes (the GOI) in specific cell populations

**❷ Module: DFGs**
Spearman + Kendall filters to identify the module (the DFGs) that not simply co-express but "move" in the same direction as GOI

**❸ Relationship: GAM**
Penalized splines to identify the strength of their relationship (DE)

**❹ Target-Module Coupling: TREP Cells**
Assign EP to single cells + ranking by DE to identify TREP cells *i.e.*, strong GOI-DFG relationship

**❺ Spatial: TREP Cells' IP**
Identify IP and separates Target-Module space into pre-IP & post-IP

***Pre-IP***: Target-Module is suppressed or transitioning

***Post-IP***: Target-Module is in full force; potentially transitioned

**❻ Dynamics: Pre-IP + Post-IP TREP Cells**
Monocle3 UMAP for trajectory & potential transition points of Target-Module program globally

**Criteria and Recommendations**

1) GOI is overexpressed or screened from a longlist of GOIs for DFG hits.

2) GOI contains multiple Spearman-Kendall DFGs.

3) Avoid reliance on sole DFG due to:
(i) Composite of multiple DFGs = Module tied with GOI;
(ii) Compensatory expression of module's gene when another is low.

4) DFG module as the response variable:
(i) Unweighted averaging of DFGs for homogeneous cell populations;
(ii) Weighted averaging of DFGs for heterogeneous cell populations.

5) Samples with majority cells (>50%) without GOI expression, conduct modeling on GOI-positive cells.

**Figure 17.** High-level summary of the CEP-IP framework, criteria and recommendations for application. The framework proceeds sequentially through six steps, guided by the operational criteria and recommendations for application.

## 4. Discussion

The CEP-IP framework developed in this study represents an explainable ML approach in single-cell analysis. In this section, the biological findings of the study were discussed before addressing the ML aspects.

Ribosomes consist of the large (60S) and small (40S) subunits, which contain ribosomal proteins of the large subunit (RPL) and small subunit (RPS) families, respectively. Aberrant expression and derailed functions of RPL and RPS proteins have been implicated in multiple types of cancer including PCa [54-56]. For instance, RPL10 protein is decreased in early stage PCa but increased in late stage PCa cases, suggesting that reduced RPL10 expression may play roles in the development of early PCa and increased in more aggressive phenotype [57]. RPS2 is overexpressed in PCa tissues compared with BP cases [54]. In immunodeficient mice with PCa cells, tumor growth and metastasis were inhibited by targeting RPS2 and eventually eradicated, where their survival was improved by inhibiting RPS2 in a dose-dependent manner [58].

There is a paucity of information linking TRPM4 with ribosomes in PCa. However, TRPM4 has been associated with ribosomes in calcified aortic valves, where TRPM4 is frequently elevated and promotes inflammation in the disease [59, 60]. *TRPM4* mRNA is post-transcriptionally regulated by the m6A reader protein YTHDF1, in which binding of YTHDF1 with the m6A site of *TRPM4* mRNA enhances stability of the transcript and increases TRPM4 translation. This is thought to be due to increased ribosomes occupancy for *TRPM4* transcript that promotes its translation, attributable to the established roles of YTHDF1 that induces ribosomal occupancy and enhances translation [59, 61].

AR signaling pathway is a main therapeutic target in advanced PCa such as the use of AR inhibitors abiraterone and enzalutamide [62]. Accumulating evidence has demonstrated the link between AR and ribosomal proteins. In advanced PCa cases, ribosomal proteins amplification was observed, and treatment of aggressive PCa cells with histone acetyltransferase inhibitors derailed AR signaling characterized by the downregulation of ribosomal proteins [63]. Likewise, activation of AR is also dependent on the integrity of the translational machinery. For instance, inhibition of translation processes via ribosome collisions led to reduction in AR translation, enhancing AR-targeted therapy in PCa [64].

Essentially, mass spectrometry of PCa cells showed that proteins involved in translation regulation were enriched. In particular, AR transactivation was suppressed in hormone-sensitive PCa cells through translational inhibition [65]. In terms of association with TRPM4, treatment of PCa cells with TRPM4 inhibitor suppressed the expression of AR protein in a dose-dependent manner, suggesting the involvement of TRPM4 protein in AR signaling in PCa [48].

As observed in this study, the *TRPM4*-Ribo's DE was significantly higher than other signaling pathways implicated with TRPM4 previously including GSK-3β, mTOR, NF-κB, PI3K/AKT, and Wnt but insignificant when compared with *TRPM4*-AR's DE. The patterns of *TRPM4*-Ribo and *TRPM4*-AR GAM plots were also similar in PCa cells such as concave up (Pt.1 and Pt.2), logit (Pt.3), sigmoidal (Pt.4), and linear (Pt.6 and Pt.7). Taken together, there may exist a TRPM4/AR/Ribo axis in PCa where the expression of AR and ribosomal genes promotes one another, leading to the stabilization of *TRPM4* transcripts in PCa, which in turn may promote AR signaling and ribosomal translation. However, this remains hypothetical and requires experimental verification.

In addition, cell adhesion GOs were consistently upregulated in TREP/post-IP cells. This is comparable with the reported function of TRPM4 in mediating cellular adhesion such as promoting focal adhesion (FA) disassembly for cellular migration [66], and interaction of TRPM4 with end-binding proteins also promotes TRPM4-mediated FA disassembly and cellular invasion [67]. In PCa cells, TRPM4 induces PCa migration and invasion by suppressing the cell adhesion protein E-cadherin and promoting epithelial-mesenchymal transition [20]. Knockdown of the ribosomal gene *RPL19* altered the expression of cell adhesion genes in prostate cancer [68], and cell adhesion was enhanced by stabilizing integrin β1 mRNA for translation [69], demonstrating the links between cell adhesion and ribosomes. On the other hand, MHC class I and II GOs were upregulated in TREP/pre-IP cells of Pt.3, Pt.4 and Pt.6, and these three PCa patients exhibited the largest difference between TREP/pre-IP versus TREP/post-IP in Monocle3 UMAP1 space. This indicates that PCa cells with MHC expression showed more distinct trajectory or transcriptome that may be amenable to MHC/antigen immunotherapy [70, 71], while TRPM4 inhibitors [48, 72, 73] or ribosomal subunit inhibitors [74] may be recommended to target TREP/post-IP cells.

In terms of ML aspects, monotonic relationship of *TRPM4* expression with other genes in the transcriptome was utilized as the feature selection method for modeling with GAM. Instead of more sophisticated or non-monotonic approaches [75-77], robust monotonic relationships shortlisted via dual-filtering approach were adopted for explainability, and positive monotonic relationships that indicate unidirectionality is suited for additive-based modeling such as GAM. TRPM4 is also highly expressed in PCa cells [17, 21], suggesting the presence of other genes that share similar monotonic expression profile (co-expression) for functional synergy.

In the GAM workflow, the smooth terms dimension k was treated as a hyperparameter to optimize in PRSS, where the exploration phase tested a range of k values followed by verification phase to validate that the best k value was found. Each PRSS iteration (*i.e.*, each specific k value) was given the opportunity to optimize the smoothing parameter $\lambda$ in its nested REML loop. This differs from the current approach where default or user-specified k value is used according to the `gam.check()` function that indicates whether a tested k value needs to be increased [30, 78]. PRSS was instead adopted in this study as the objective function to obtain the best k value due to its objective nature for model comparison, without the requirement for manual re-specification and refitting with another k value. By treating k as a search space, it allows for the exploration of a broad range of k values, and this approach complements existing diagnostic tools.

The penalty multiplier $\gamma$ was optimized by testing a range of $\gamma$ values in 0.5 increments and visualizing the GAM curves for signs of suboptimal fitting. Automated optimization such as treating $\gamma$ as a hyperparameter space, or with metrics such as generalized cross-validation, was not conducted in favor of visual inspection. As the end-product is the fitted GAM curve, visual inspection provides direct assessment of whether the fitted curve fluctuates or with steep slopes, particularly in areas with sparse cell numbers that may represent artifacts in scRNA-seq data [79, 80]. The selected $\gamma$ value of 1.5 in this study aligns with the $\gamma$ value of 1.4 recommended in `mgcv` package documentation to reduce overfitting [43].

GOs enriched in PCa cells utilizing Spearman-Kendall dual-filter, without segregating into subtypes based on TREP/IP status, showed that TRPM4 may play roles in the regulation of protein localization and transporter complexes with significance level of approximately FDR=0.01 (**Figure 1E**). When the PCa cells were subtyped into four populations, translation, ribosomal and cellular adhesion processes were consistently upregulated in TREP/post-IP cells with stronger level of significance ($FDR<1x10^{-5}$) (**Figure 7**). This aligns with GAM-identified TREP/IP cells which formed spatially dissimilar clusters in Monocle3 trajectory space. Moreover, GAM's DE prioritized cells in consistent states where both *TRPM4* and Ribo were either lower or higher, giving less weight to cells with inconsistent *TRPM4*-Ribo relationship that may represent transitional cells, outliers, or with complex variation influenced by other factors beyond the relationship being modeled. It was also observed that GAM did not simply take into account cells located closely along the GAM curve to map for TREP cells, but instead considered cells that contributed to the overall shape of the fitted curve globally. This may improve the likelihood of identifying biologically relevant cells with preserved functional relationships, and to map for cells that may follow a continuous developmental trajectory into distinct states. Collectively, these observations imply that GAM-derived cell states are potentially biologically relevant, not captured by conventional correlation thresholds but revealed by GAM and TREP/IP grouping in *TRPM4*-Ribo space.

The IP represents the key transition point (*i.e.*, specific *TRPM4* value) where the distribution pattern of TREP cells shifted, stratifying the cells into the pre-IP and post-IP regions. This enables the following insights: (i) GAM's DE prioritizing cells with consistent expression states; (ii) Explanation of why patients demonstrated lower or higher DE in their modeling; (iii) Stratification of cells into four groups for GO enrichment analysis to inform potential targeted therapies for each group of cells. The IP status differs from conventional gene expression analysis approaches, where fixed thresholds are often used such as fold-change cut-offs or fixed ranks [81-83]. In GAM analysis, the IP cut-off is rooted in the overall shape of the fitted curve that leverages the flexibility of splines to account for non-linearity in the modeled relationship. Instead of adopting uniform thresholds across the patients, the IP approach produces patient-specific transition points, reflecting potential underlying biological relationship, cellular populations heterogeneity, and progression trajectory.

The automated IP detection pipeline and IPRS introduced in this study address concerns regarding the reproducibility and objectivity of IPs determined visually. By fitting a secondary GAM to TREP cell signed residuals and defining the IP as the first negative-to-positive zero-crossing of the smoothed residual R(x) curve, the CEP-IP framework minimizes subjectivity from IP identification. Validation against visually determined IPs in all seven PCa patients showed close concordance. The composite IPRS, derived from five equally weighted components capturing multiple aspects of the performance of the fitted R(x) curve, provides a quantitative reliability tier for each detected IP. Notably, no sample in any dataset utilized in this study achieved the very strong IPRS tier attributable to the algorithm's design to be more stringent than lenient. Rather than being fully reliant on automated IP detection, the IPRS framework was intentionally designed to encourage researchers to manually verify the IP visually, particularly for moderate and weak tier samples. This encourages a dual-layer check for this key decision point, minimizing errors for equivocal samples.

The CEP-IP framework was subsequently validated in two biologically distinct independent brain datasets: the Allen Human MTG SMART-seq neuronal dataset and the Neftel SMART-Seq2 adult malignant GBM dataset. These differ from the PCa cohort in tissue type, disease context, sequencing platform (PCa's newer 10x Genomics Chromium droplet-based scRNA-seq vs. older SMART-seq platforms), and cell count per sample (lower in the validation sets); consistent framework performance in these older platforms with lower cell yields suggests generalizability to modern platforms with higher cell throughput. Brain datasets were chosen as validation cohorts due to the well-established reports of transcriptome diversity and heterogeneity of neural cell types, including excitatory and inhibitory neurons, glial subtypes, and GBM malignant cell states characterized by highly diverse and dynamic transcriptome [84, 85]. Furthermore, GBM is one of the most deadly cancers, 5-year survival rate of 5-10% despite multimodal treatment with a median survival of 15-16 months [86], and it is considered incurable [87, 88]. These challenging datasets are ideal to stress test the generalizability of the CEP-IP framework.

In the Allen MTG dataset, *CARM1P1* was selected as the GOI as it defines a transcriptomically distinct human-specific deep layer 3 excitatory neuron subtype, characterized by long-range axonal projections and large dendritic arbors [89, 90]. *CARM1P1* provides a well-characterized functional context to validate whether CEP-IP's post-IP TREP cells demonstrate similar functions. Indeed, post-IP TREP cells were consistently enriched for neuron projection, neuron compartments, and synapse organisation GOs in MTG cells, aligning with the functions of *CARM1P1* in excitatory neurons. In the GBM dataset, *FOXM1* was identified as the sole GOI yielding DFGs specific to the MES GBM cell state, with post-IP TREP cells in all three dual-filter-passed patients consistently enriched for cell division, cell cycle, and microtubule GOs, pathways well-established by independent studies on FOXM1-driven proliferation in cancers [91-93].

In the MGH152 GBM sample, 3D UMAP trajectory analysis revealed that apparent separation failures observed in 2D embeddings may represent genuine continuous trajectories compressed by dimensionality reduction. The color-coded stratification of pre-IP and post-IP TREP cells was what first revealed two separated post-IP TREP populations in 2D UMAP, prompting the 3D investigation that confirmed their continuity. This underscores the importance of evaluating trajectories in higher dimensions when 2D projections yield ambiguous patterns. PERMANOVA further indicated significant separation between pre-IP and post-IP TREP cells in this dataset, but with a significant PERMDISP. These findings are consistent with the two populations occupying distinct yet internally dispersed transcriptional space, a feature that was noticeable through visual inspection of the 3D plot. These suggest that both population of cells were distributed along a continuous developmental trajectory rather than forming discrete clusters. Collectively, these validation analyses demonstrate that the CEP-IP framework is not limited to TRPM4 biology or a single cohort. The framework generalizes across distinct gene pairs, cell types, and sequencing platforms, provided that the genes survive the Spearman-Kendall dual-filter that confirms a genuine monotonic relationship, before GAM modeling to identify cellular populations that most strongly hold the GOI-DFGs relationship.

The Spearman-Kendall dual-filter was utilized as the feature selection strategy in all three datasets in this study (PCa, Allen MTG, and GBM). Essentially, the dual-filter's requirement that candidate genes satisfy both $r_s$ (sensitivity screen) and $\tau$ (specificity for monotonic concordance, and robust to outliers) imposes stringency that reduces false monotonic relationships that can arise from distributional artefacts common in scRNA-seq data [94, 95]. This increased stringency may account for the compatibility between the monotonic dual-filter and GAM, as GAM models smooth relationships along continuous values of a predictor [28]. Furthermore, GOI and its DFGs, whose gene members passed the dual-filter, have multiple biological interpretations and implications. The dual-filter screens the transcriptome blinded to prior knowledge on known pathways or mechanisms (*e.g.*, transcription factor activities) and it does not require DFGs to be downstream targets of the GOI. Instead, the dual-filter identifies genes whose expression moves consistently in the same direction with the GOI, based on purely the mathematics of rank concordance. Thus DFGs represent program co-members, where the genes co-activate or co-suppress alongside the GOI in cells harboring such molecular signature. The GOI and its DFGs may be participants in a specific cellular state rather than members of a strict hierarchical regulatory pathway.

GAM has recently been adopted to analyze scRNA-seq data, primarily for cellular trajectories analysis [37, 96, 97]. The tradeSeq algorithm models gene expression with pseudotime utilizing GAM to identify DEGs between lineages along pseudotime trajectories, allowing the inference of gene expression changes during development [37]. This is conceptually different from this study's approach, whereby the gene expression relationship between two variables (*e.g.*, *TRPM4* and Ribo) were first modeled to identify the underlying shape, before determining which specific cell contributed to this relationship via their individual contribution magnitude to the model's DE. This subsequently enables subtyping of cells into different biological states by incorporating the IP spatial dimension. In addition, tradeSeq requires pre-computed trajectories as input for GAM analysis to identify lineage-specific genes. This is distinct from this study that uses GOI's monotonic genes, the DFGs, as input for GAM to identify cell populations that drive (or do not drive) the modeled relationship, and implementing them as input for cross-platform trajectory validation.

The XAI landscape for single-cell analysis is an area of active progress. Recent deep learning models such as xSiGra employed hybrid graph transformer architectures with gradient-weighted class activation mapping to assign gene and cell importance scores for spatial cell type identification [98]. Additionally, perturbation-focused frameworks such as scPerb utilize style transfer-based variational autoencoders to separate perturbation-related variance from cell-type-specific patterns, addressing the black-box limitations of conventional neural network approaches [99]. In contrast, CEP-IP achieves explainability through the inherent mathematical transparency of GAMs, where spline basis functions and their penalization are interpretable. Moreover, cell-level contributions are assigned from aggregate model deviance via EP values. As an XAI framework, CEP-IP is tasked for pairwise gene-module relationship modeling to identify individual cells harboring their strong relations or progressing along a trajectory. The framework also aids translational efforts: clinical validation of a CEP-IP-identified GOI and its DFGs would require quantifying a handful of genes via qPCR, rendering the framework's outputs actionable without the need for transcriptome profiling.

The limitations of the current study, and the future opportunities that they represent, are acknowledged as follows:

i) The presence of a specific shape of the relationship being modeled was assumed, and such a relationship may not exist in the first place. However, without attempts to unravel the potential nature of their relationship, the investigation would stop at the simple monotonic level, which obscures potential non-linearity and underlying biology.

ii) Cell populations with insufficient number of cells (*e.g.*, <50 cells) resulted in unreliable enrichment of GOs, as exemplified by Pt.5's TREP/post-IP cell population that did not exhibit enrichment of GOs consistently shared by the rest of the patients. For such scenarios, it is suggested to adopt patient-specific DFGs, instead of the pan-patient filtering approach employed in this study.

iii) GAM's DE is an aggregate metric of residuals from individual cells. The CEP classification was designed as attempts to decompose the aggregate model performance into individual cell contributions, providing cell-level explainability for GAM models. However, this is not a complete decomposition due to individual EP values that do not sum exactly to GAM's DE. Hence, CEP is primarily a method to identify cells most well-predicted by the model, instead of identifying cells most contributive to the overall DE.

iv) Three patients in this study (Pt.3, Pt.4, and Pt.6), showed consensus GOs in their pre-IP and post-IP TREP cells, with clear separation of pre-IP and post-IP TREP cells in Monocle3 trajectory space. Hence, GAM modeling can delineate underlying biology of well-predicted cells (*i.e.*, TREP cells) in cases with strong pairwise relationships (*e.g.*, DE >50%). However, Pt.7 showed dissimilar GOs in their pre-IP TREP cells (no enrichment of MHC GOs) despite DE of 52.8% and with consensus GOs for post-IP TREP cells. This highlights the ongoing challenge posed by cancer cells heterogeneity, as well as the limitations of pairwise gene relationship modeling that does not take into account additional genes or regulatory factors that may further explain biological distinctions. Nevertheless, this study was designed to focus on pairwise gene relationship modeling, without diluting the focus with additional variables because another aim was to explore GAM's mechanisms and touted interpretability, aligning with the XAI approach where ML models are recommended to be explainable for safe deployment [100-103].

v) Experimental wet-lab validations beyond in silico modeling and pathway analysis are required to confirm that GAM-identified cell populations represent actual biological states. For instance, whether *TRPM4*-Ribo module reflects aggressive PCa biology or a generic high-translation state cannot be resolved through transcriptomic analysis alone, and perturbation experiments pairing *TRPM4* inhibition with ribosomal activity assays are required.

vi) Complete automation of the workflow in this study remains challenging, as manual methods were required such as patients selection and GO enrichment interpretation. Nonetheless, these manual interventions necessitate study-specific optimization, which encourages deeper understanding of the algorithm's mechanisms and explainability in the process.

In conclusion, the CEP-IP framework (*i.e.*, mapping for TREP or non-TREP cells in pre-IP or post-IP regions) produces quadrants of cell subpopulations. While demonstrated in this study with GOI-DFGs relationships in PCa, MTG, and GBM, the CEP-IP framework may be applicable for other settings.

**Data and Code Availability**

The processed Seurat objects `GSE185344_Seurat_processed.RData` (9.52 GB), `AllenMTG_Seurat_processed.RData` (8.22 GB) and `NeftelGBM_SS2_AdultMalignant_Seurat_processed.RData` (4.02 GB) are available on HuggingFace: https://huggingface.co/datasets/kahkengwong/CEP-IP_Framework

The code used in this project is available on GitHub:
https://github.com/kahkengwong/CEP-IP_Framework

The following blocks of code are available from the GitHub repository:

```
Part_1_scRNAseq_preprocessing_and_UMAP_clusters.r
Part_2_UMAP_Heatmap_Spearman-Kendall's-matrix.r
Part_3.01_Mean_Expression_Justifications.r
Part_3.02_Family_Distribution_Analysis.r
Part_3.03_GAM_REML_PRSS_Setup.r
Part_3.04_GAM_REML_PRSS_Analysis.r
Part_3.05_REML_Extraction_and_Convergence.r
Part_3.06_REML_PRSS_Plots_and_EDF_Analysis.r
Part_3.07_Validation_of_k_and_Lambda_Selection.r
Part_3.08_Visualize_TPRS_and_GAM_Components.r
Part_3.09_Extract_GAM's_ND_MD_DE.r
Part_3.10_Extract_TRPM4-Ribo_EP.r
Part_3.11_CEP-IP_MCCV_of_CEP_Classification.r
Part_3.12_CEP-IP_GAM_Plots.r
Part_3.13_CEP-IP_Mosaic_and_Raincloud_Plots.r
Part_3.14_CEP-IP_DEGs_Analysis.r
Part_3.15_CEP-IP_in_Monocle3_Trajectory.r
Part_3.16_CEP-IP_Validation_Allen_MTG_dataset.r
Part_3.17_CEP-IP_Validation_Neftel_GBM_dataset.r
```

The R packages and versions used in this study were saved in the `renv.lock` file included in the GitHub repository. This lockfile contains version information for all packages used with their dependencies. Clone the repository and run `renv::restore()` to install the identical package versions used in this study. Note that `renv` creates an isolated library and will not modify a system's existing R packages and setup. For manual or selected package installation, a simplified `r_packages_info.json` file is provided with the package names, versions, and sources (CRAN, Bioconductor, or GitHub).

**CRediT authorship contribution statement**

Kah Keng Wong: Conceptualization, Funding acquisition, Resources, Data curation, Investigation, Formal analysis, Software, Validation, Visualization, Supervision, Writing - original draft, Writing - review & editing.

**Declaration of competing interest**

The author declares that the research was conducted in the absence of any commercial or financial relationships that could be construed as a potential conflict of interest.

**Acknowledgements**

This work was supported by the Research University Grant (1001/PPSP/8012349), Universiti Sains Malaysia awarded to Kah Keng Wong. The author extends heartfelt thanks to the School of Medical Sciences and Research Creativity & Management Office (RCMO), Universiti Sains Malaysia, for their invaluable support.

**Supplementary Tables and Figures**

**Supplementary Table 1**. The number and proportion of cells before and after each QC step in the scRNA-seq processing workflow, and the top 50 marker genes representing each cluster of the PCa and NonCa cases.

**Supplementary Table 2**. Cell count by cluster, and comparison of *TRPM4* expression between PCa clusters in NonCa and PCa cases.

**Supplementary Table 3**. List of *TRPM4*-monotonic genes in IBP (internal BP cells in PCa cases) and PCa cells, and the gene sets enriched by DFGs in PCa cells.

**Supplementary Table 4**. List of *RPL10*-, *RPL27*-, *RPL28*-, *RPS2*-, *RPS8*-, *RPS12*-, and *RPS26*-monotonic genes in PCa cells (shortlisted via Spearman-Kendall dual-filter), and the *TRPM4*-containing gene sets enriched according to the seven ribosomal genes in PCa cells.

**Supplementary Table 5**. Assessment of the reliability to averaging ribosomal or AR genes into their averaged gene set (Ribo or AR) for downstream modeling.

**Supplementary Table 6**. Data distribution family comparison according to AIC, BIC, and DE, and the best distribution families according to AIC and BIC.

**Supplementary Table 7**. PRSS iterations in GAM of BP and PCa cells, REML iterations of the best GAM models of BP and PCa cells, best model's REML convergence information, and independent GAM fitting with predetermined $\lambda$ values and the resulting REML scores in *TRPM4*-Ribo modeling of Pt.6 PCa cells.

**Supplementary Table 8**. GAM model optimization results and statistical performance metrics across different $\gamma$ values tested in 0.5 increment ($\gamma$ tested: 0.5-3.0).

**Supplementary Table 9**. GAM statistical performance metrics, observed and GAM-predicted gene set values, and smooth basis functions' coefficient values for all modeled TRPM4-gene sets (AR, GSK-3β, mTOR, NF-κB, PI3K/AKT, and Wnt; $\gamma$=1.5) in PCa and BP cells.

**Supplementary Table 10**. GAM's ND, MD, and DE data for each *TRPM4*-gene set modeling, as well as each *TRPM4*-Ribo cells' EP value, ranking according to EP, and TREP or non-TREP cell status (n=5,855 cells).

**Supplementary Table 11**. MCCV analysis of random, leverage, Cook's distance, and CEP classification of PCa cells into TREP and non-TREP.

**Supplementary Table 12**. List of HVGs in PCa and NonCa cases.

**Supplementary Table 13**. Number of TREP (purple in GAM plots) and non-TREP (gray in GAM plots) cells in pre-IP and post-IP, below or above GAM curves, of *TRPM4*-Ribo modeling in PCa cells.

**Supplementary Table 14**. GOs enriched in four populations of cells (TREP or non-TREP in pre-IP or post-IP) of each PCa patient, and the upregulated DEGs in each population of cells contributing to their enrichment.

**Supplementary Table 15.** Detailed IPRS results of all three tested datasets (PCa, MTG, GBM).

**Supplementary Table 16.** GAM performance metrics and list of enriched GOs of the validation datasets (Allen MTG and Neftel GBM).

**Supplementary Table 17.** GOIs (n=659 initially and n=630 passed the expression gate) screened for DFGs in GBM cell states and monotonicity tests of wcDFG in each GBM patient's *FOXM1*+ cells.

# Supplementary Figure 1

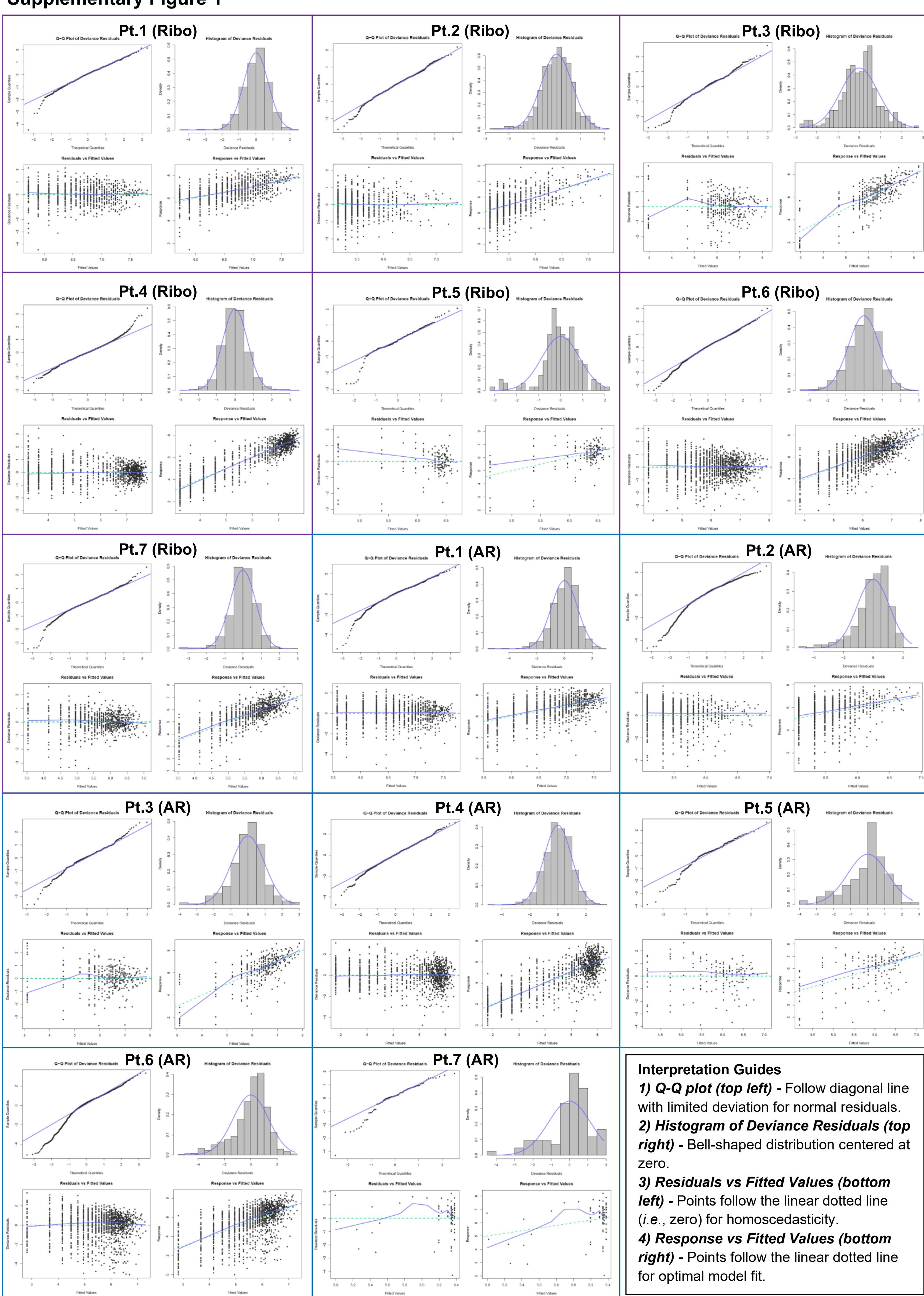

**Supplementary Figure 1**. The `gam.check()` plots of *TRPM4*-Ribo and *TRPM4*-AR modeling to assess if the data distribution was appropriate for Gaussian GAM analysis. Four types of plots were generated *i.e.*, Q-Q plot (top left), histogram of deviance residuals (top right), residuals versus fitted values (bottom left), and response versus fitted values (bottom right).

**Supplementary Figure 2**

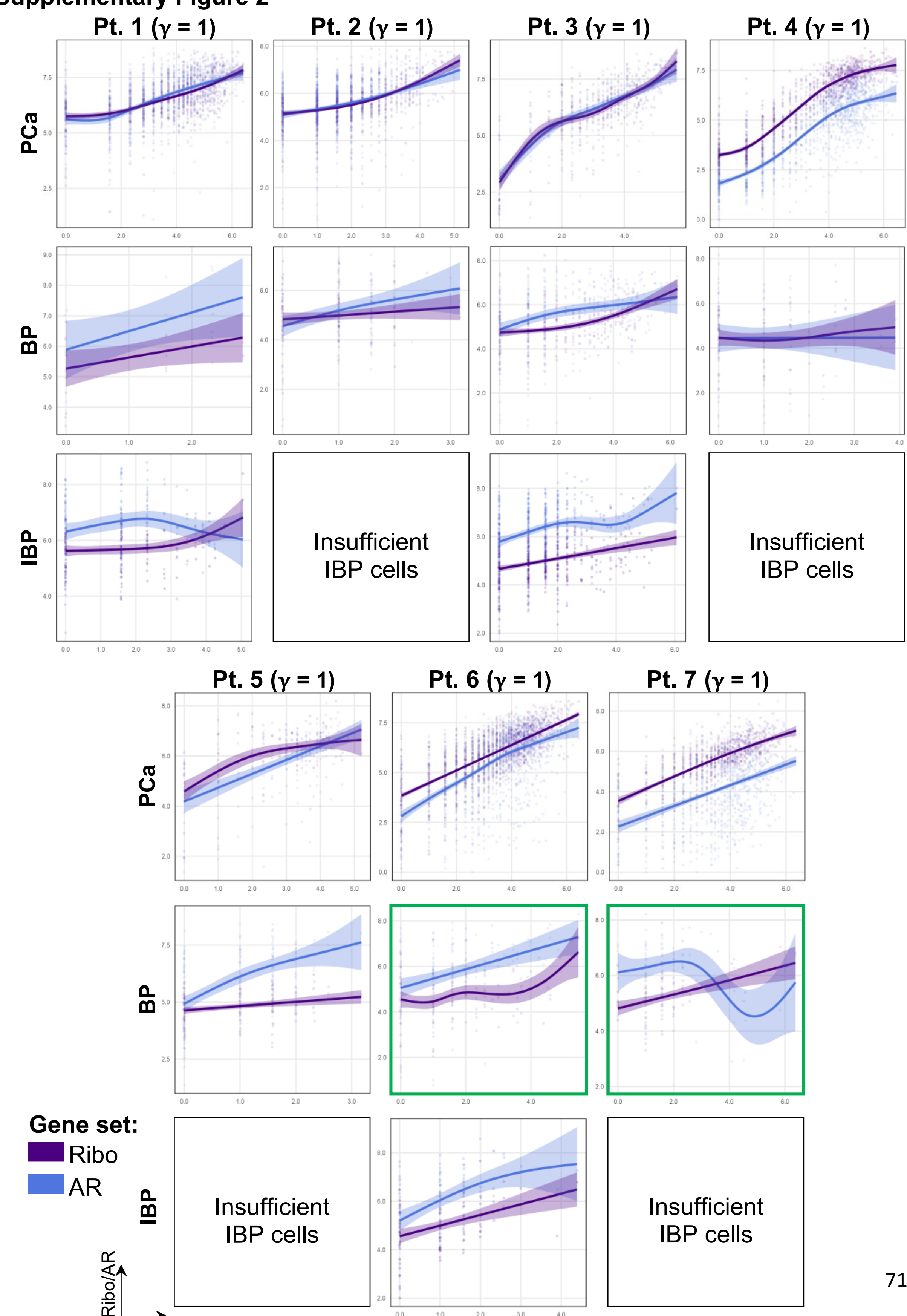

**Supplementary Figure 2**. GAM plots of *TRPM4*-Ribo and *TRPM4*-AR modeling in PCa, BP, and IBP cells, using the default γ value of 1 by `mgcv`.

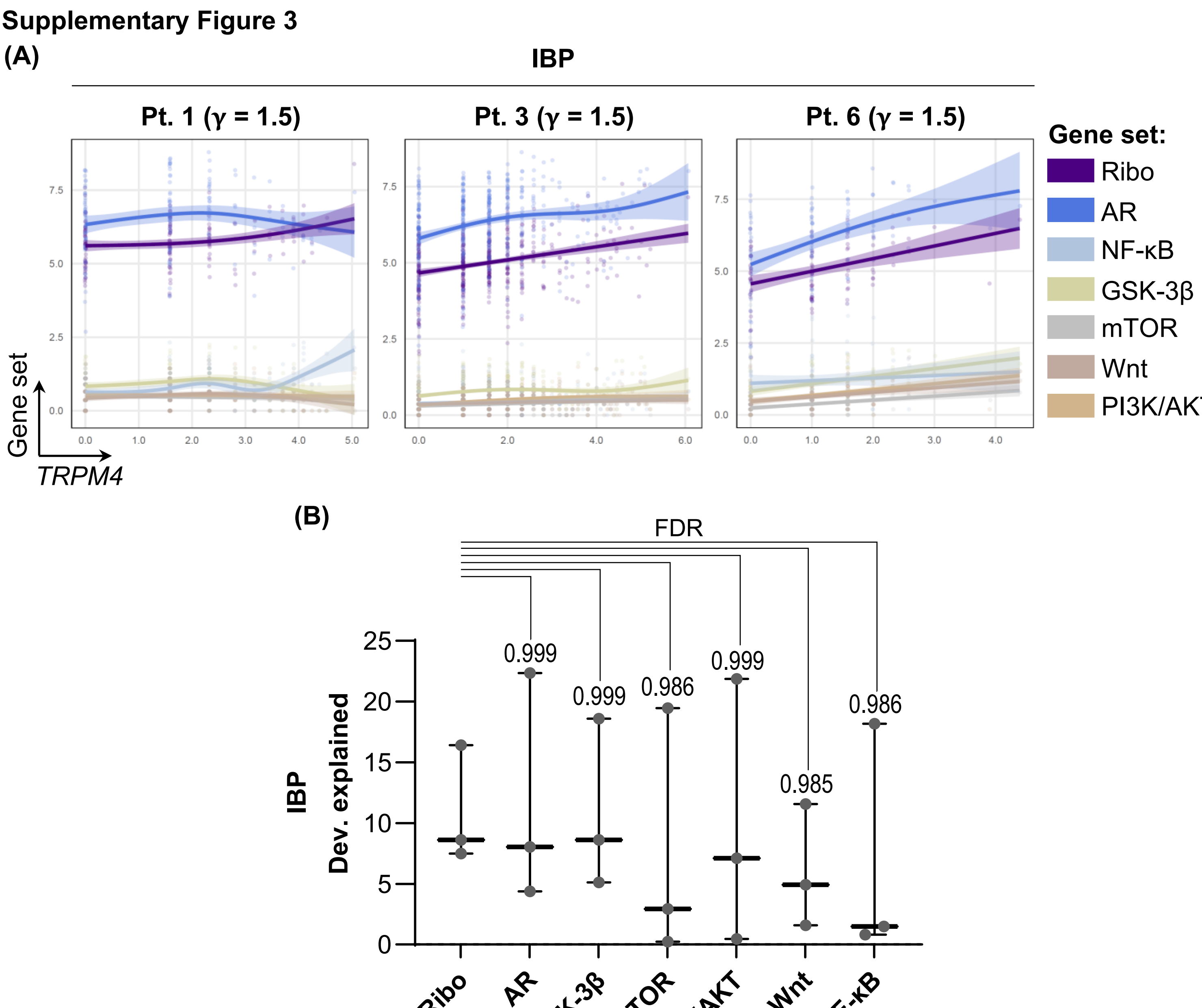

**Supplementary Figure 3**. *TRPM4*-gene sets modeling in IBP cells(A) GAM plots of *TRPM4* with multiple gene sets modeling in IBP cells (γ=1.5); (B) Comparison of DE between different gene sets modeling with *TRPM4* in IBP cells, and DE values from each gene set modeling were compared with DE of *TRPM4*-Ribo modeling.

**Supplementary Figure 4**

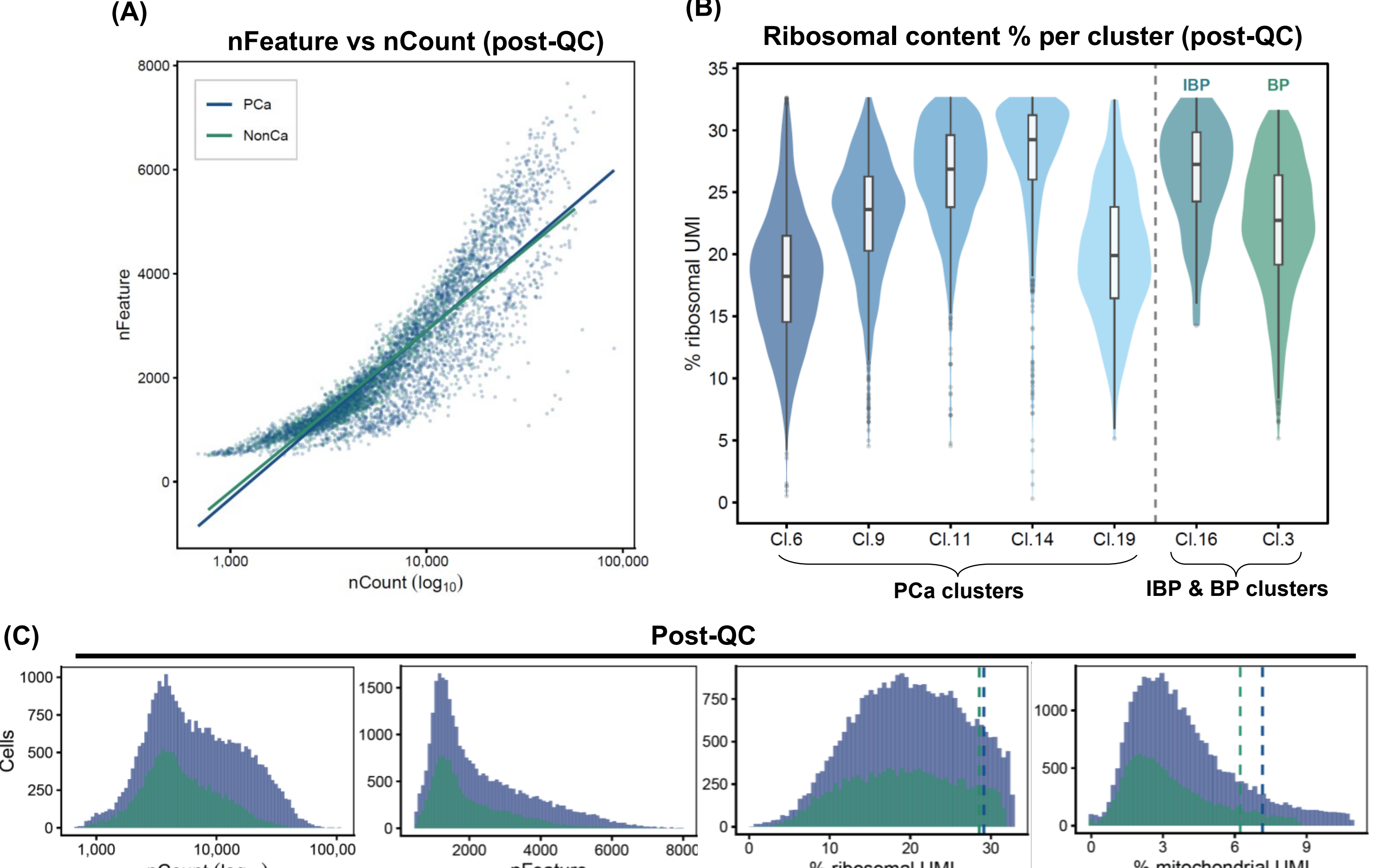

**Supplementary Figure 4.** Post-QC assessment of scRNA-seq datasets demonstrating data quality and absence of ribosomal gene bias. (A) Scatter plot of nFeature versus nCount ($\log_{10}$) in PCa (navy) and NonCa (teal-green) cells post-QC, showing a linear relationship; (B) Violin plots of residual ribosomal content (% of total UMI) in the five PCa clusters (6, 9, 11, 14, 19), IBP (16) and BP (3) clusters post-QC, demonstrating comparable ribosomal content distributions; (C) Post-QC distributions of nCount, nFeature, ribosomal content (%), and mitochondrial content (%) in PCa (navy, n=30,932 cells) and NonCa (teal-green, n=12,205 cells) datasets. Dashed vertical lines in the ribosomal and mitochondrial panels indicate the applied 90th percentile filtering thresholds.

**Supplementary Figure 5**

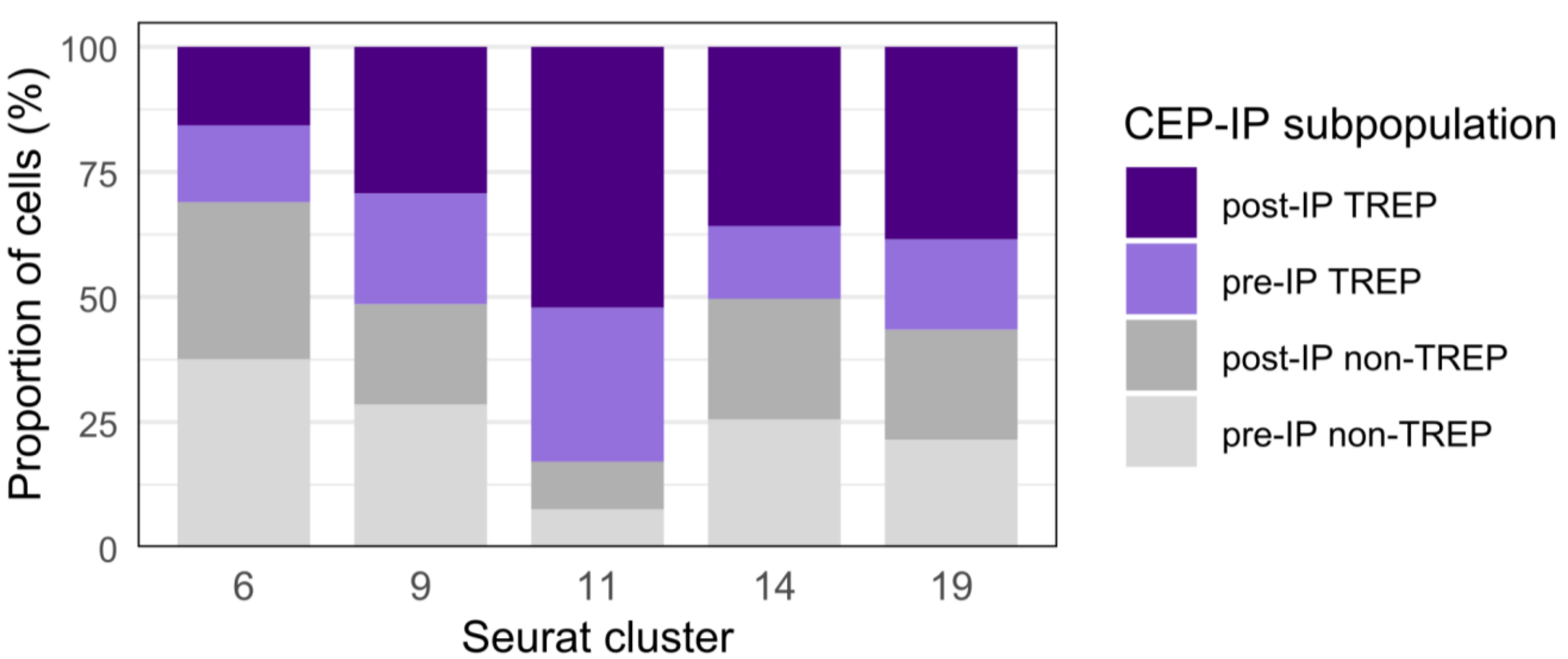

**(B)**

**CEP-IP subpopulation composistion in Seurat clusters of PCa cells**

| **Seurat cluster** | **Number of cells** | **Post-IP TREP (%)** | **Pre-IP TREP (%)** | **Post-IP non-TREP (%)** | **Pre-IP non-TREP (%)** | **Normalized entropy** |
|---|---|---|---|---|---|---|
| 6 | 1,869 | 15.7 | 15.4 | 31.4 | 37.6 | 0.945 |
| 9 | 1,618 | 29.3 | 22.1 | 20.1 | 28.5 | 0.991 |
| 11 | 1,222 | 52.1 | 30.8 | 9.6 | 7.5 | 0.809 |
| 14 | 824 | 35.8 | 14.6 | 24.2 | 25.5 | 0.967 |
| 19 | 322 | 38.5 | 18.0 | 22.0 | 21.4 | 0.966 |
| All clusters | 5,855 | 31.1 | 20.5 | 22.2 | 26.2 | - |

**Supplementary Figure 5.** CEP-IP subpopulation stratification is orthogonal to Seurat clustering in PCa cells. (A) Stacked bar chart showing the proportional composition of CEP-IP subpopulations within each of the five PCa-enriched Seurat clusters (6, 9, 11, 14, 19); (B) Summary table of CEP-IP subpopulation composition and normalized Shannon entropy per cluster. Entropy values approaching 1.0 indicate near-maximal mixing of CEP-IP subpopulations within a given cluster. Global chi-square test of independence: $X^2$ =858.62, $p$<0.0001 (Monte Carlo permutation, B=9,999).

## Supplementary Figure 6

**Pt.1**

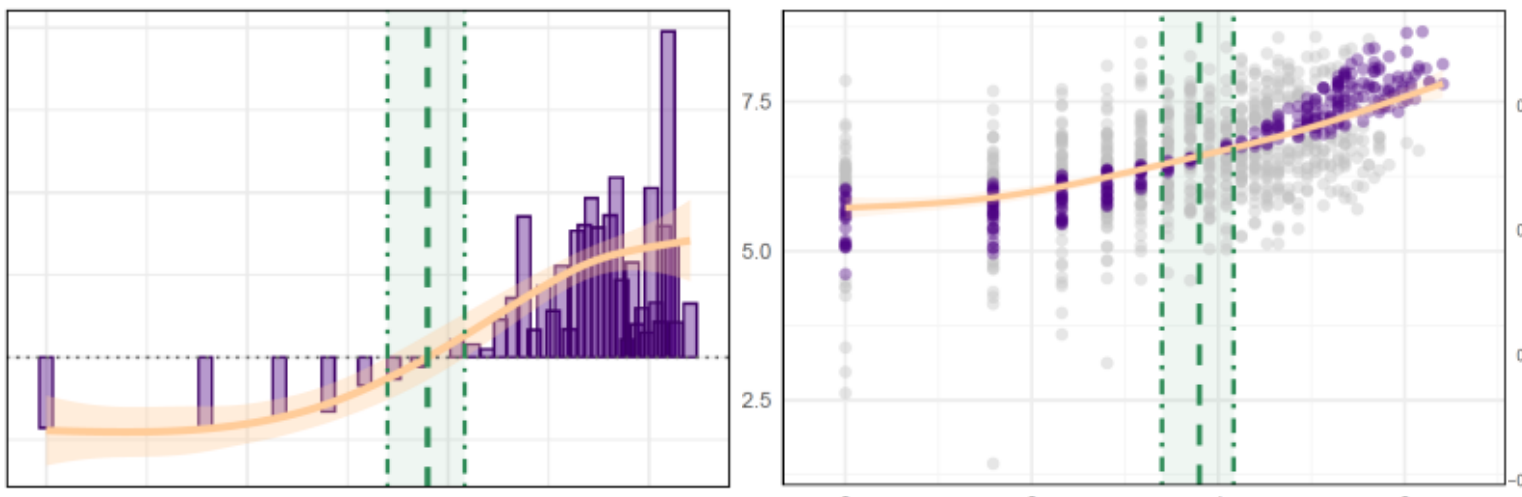

IP=3.796; 95% CI: 3.397-4.166
C1=0.880, C2=1.000, C3=0.170, C4=0.795, C5=0.633;
IPRS=0.696 (Strong)

**Pt.2**

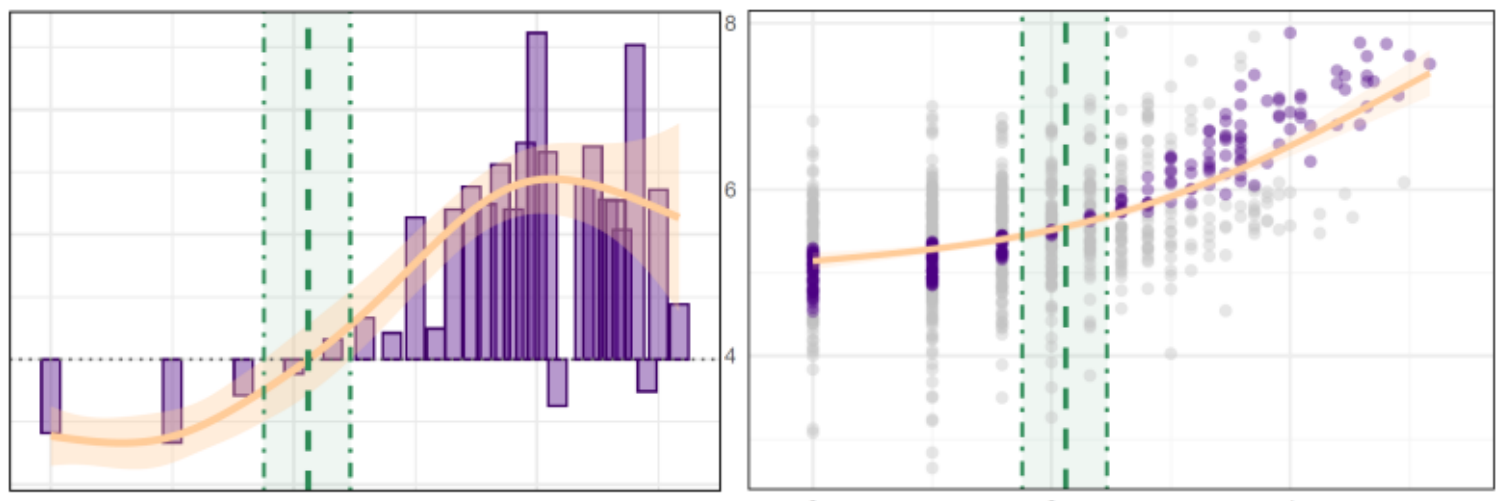

IP = 2.119; 95% CI: 1.753-2.466
C1=0.862, C2=1.000, C3=0.181, C4=0.769, C5=0.536;
IPRS=0.670 (Strong)

**Pt.3**

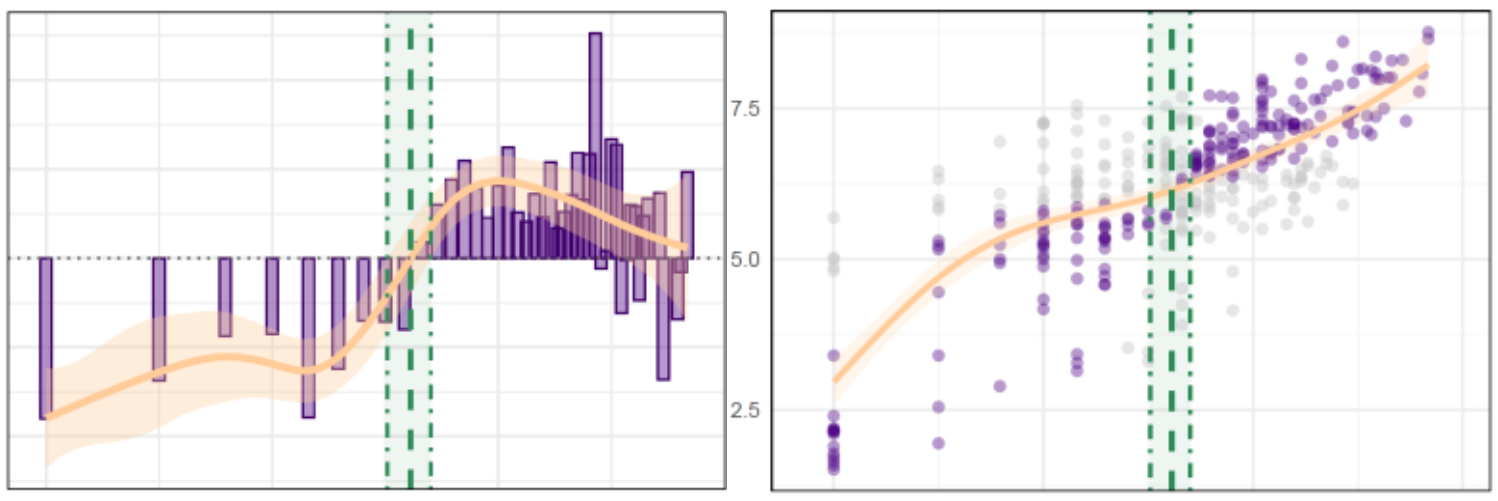

IP = 3.224; 95% CI: 3.018-3.402
C1=0.932, C2=1.000, C3=0.770, C4=0.723, C5=0.419;
IPRS=0.769 (Strong)

**Pt.4**

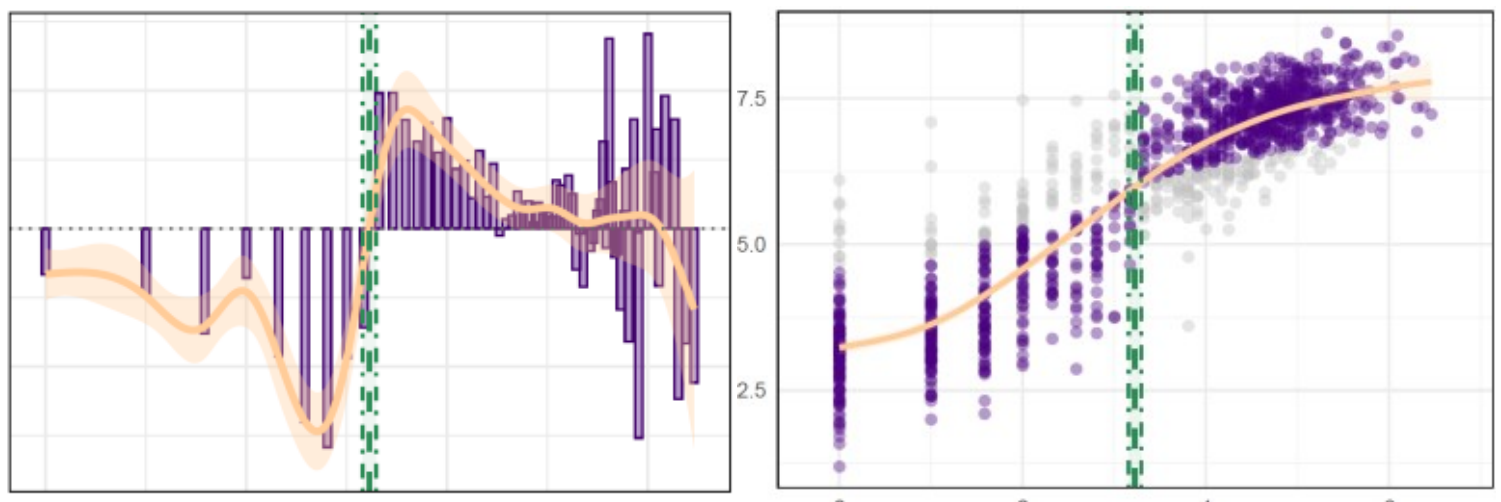

IP = 3.223; 95% CI: 3.156-3.294
C1=0.979, C2=1.000, C3=0.972, C4=0.801, C5=0.086;
IPRS=0.768 (Strong)

**Pt.5**

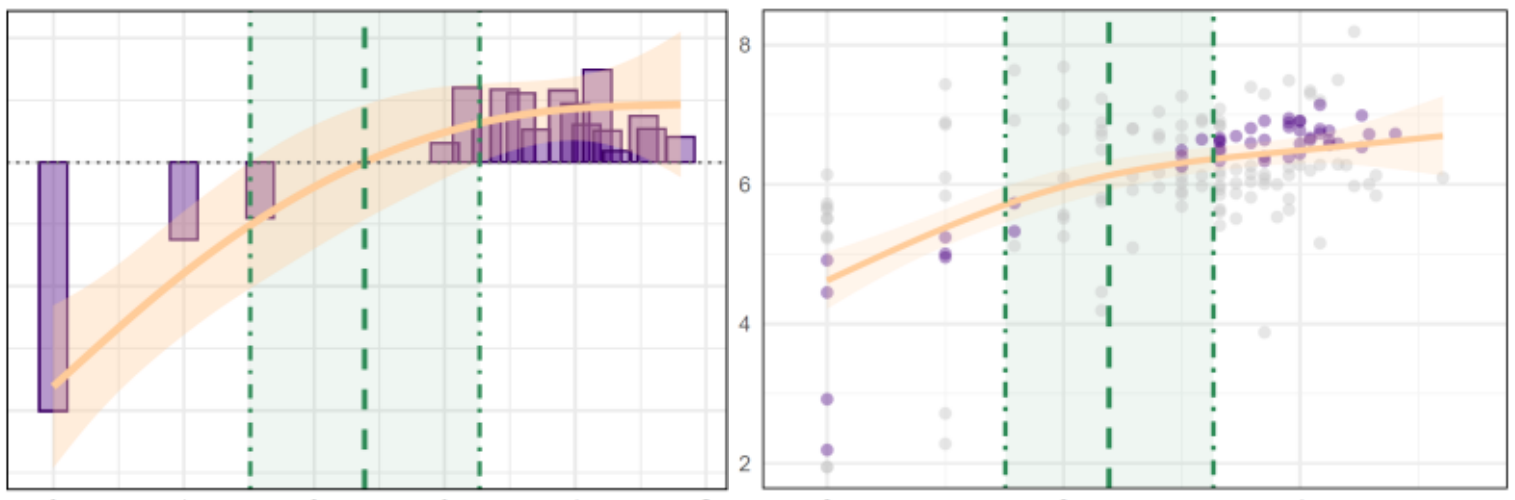

IP = 2.386; 95% CI: 1.509-3.267
C1=0.634, C2=0.934, C3=0.230, C4=0.683, C5=0.673;
IPRS=0.631 (Strong)

**Pt.6**

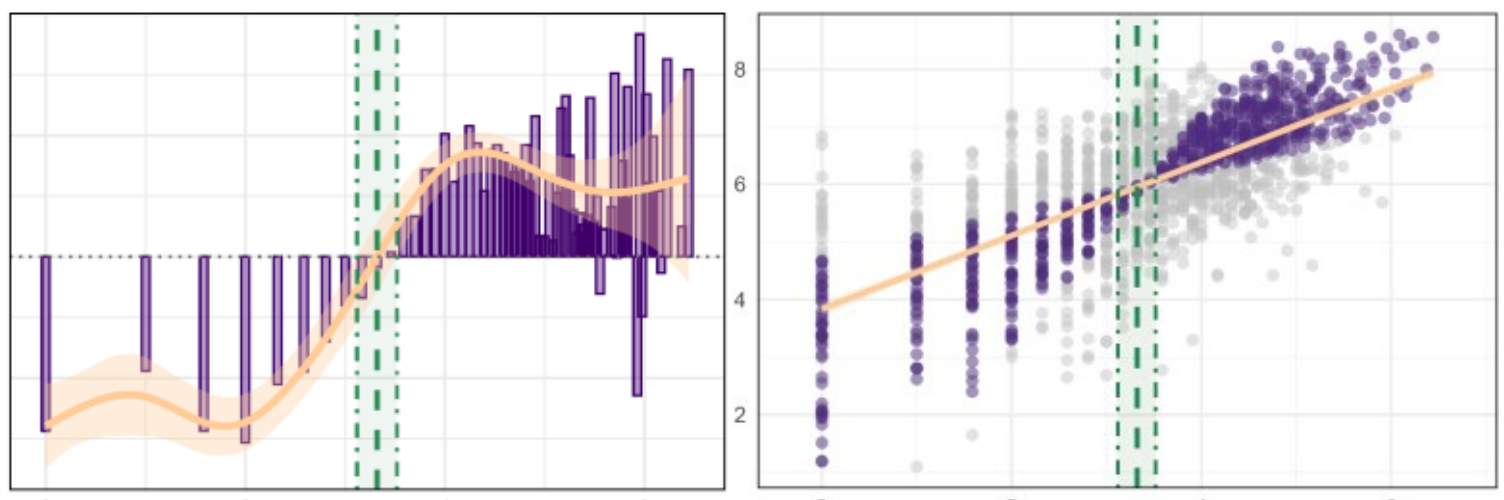

IP = 3.321; 95% CI: 3.119-3.520
C1=0.938, C2=1.000, C3=0.493, C4=0.782, C5=0.382;
IPRS=0.719 (Strong)

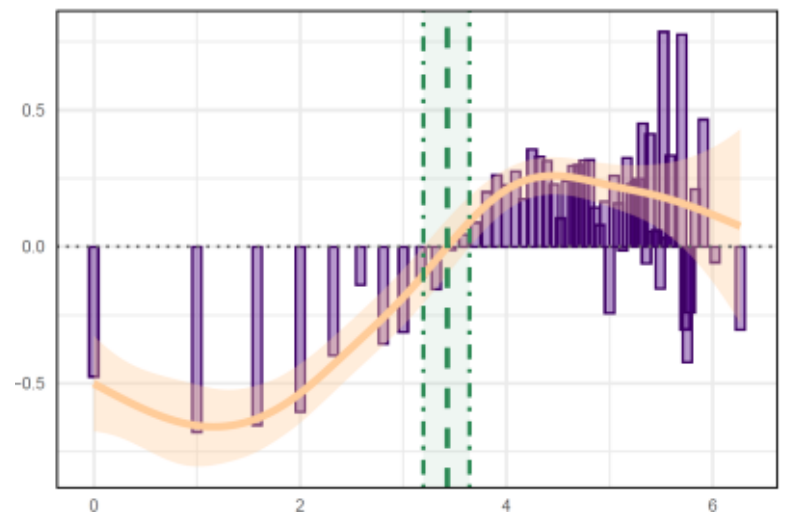

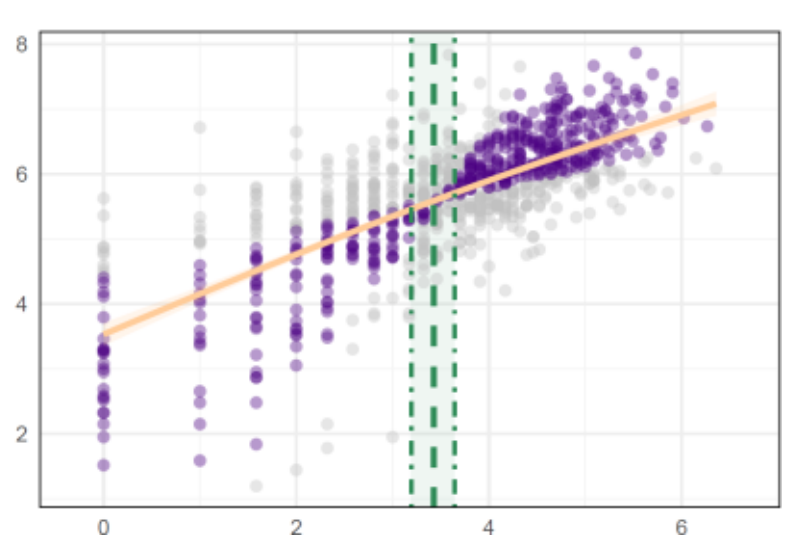

IP = 3.427; 95% CI: 3.195-3.644
C1=0.928, C2=1.000, C3=0.415,
C4=0.769, C5=0.341;
IPRS=0.691 (Strong)

**Comparison of manual vs. automated IP values**

| Patient | Manual IP | Automated IP | Difference | Automated IP 95% CI | Manual IP within automated IP 95% CI |
|---|---|---|---|---|---|
| Pt.1 | 3.800 | 3.796 | 0.004 | 3.397-4.166 | Yes |
| Pt.2 | 2.214 | 2.119 | 0.095 | 1.753-2.466 | Yes |
| Pt.3 | 3.179 | 3.224 | 0.045 | 3.018-3.402 | Yes |
| Pt.4 | 3.306 | 3.223 | 0.083 | 3.156-3.294 | No (by 0.012) |
| Pt.5 | 2.636 | 2.386 | 0.250 | 1.509-3.267 | Yes |
| Pt.6 | 3.465 | 3.321 | 0.144 | 3.119-3.520 | Yes |
| Pt.7 | 3.476 | 3.427 | 0.049 | 3.195-3.644 | Yes |

**Supplementary Figure 6**. CEP-IP automated IP detection and IPRS results for TRPM4-Ribo dataset (Pt.1-Pt.7). For each patient, paired R(x) diagnostic plots (left) and main GAM scatter plots (right) are shown. R(x) represents the expected signed residual of TREP cells at a given TRPM4 expression level, with binned mean residuals (purple bars), the smoothed R(x) curve and 95% confidence band (orange), and the IP and 95% CI indicated by green dashed and dot-dash vertical lines, respectively, with green shading denoting the CI interval. The main GAM scatter plots show TREP (purple) and non-TREP (gray) cells with the main GAM curve (orange), with IP and 95% CI overlaid with the same color scheme. The summary table (bottom right) compares visually determined and automated IPs, showing absolute differences and whether each manual IP is within the automated IP's 95% CI. All patients were classified as strong tier.

## Supplementary Figure 7

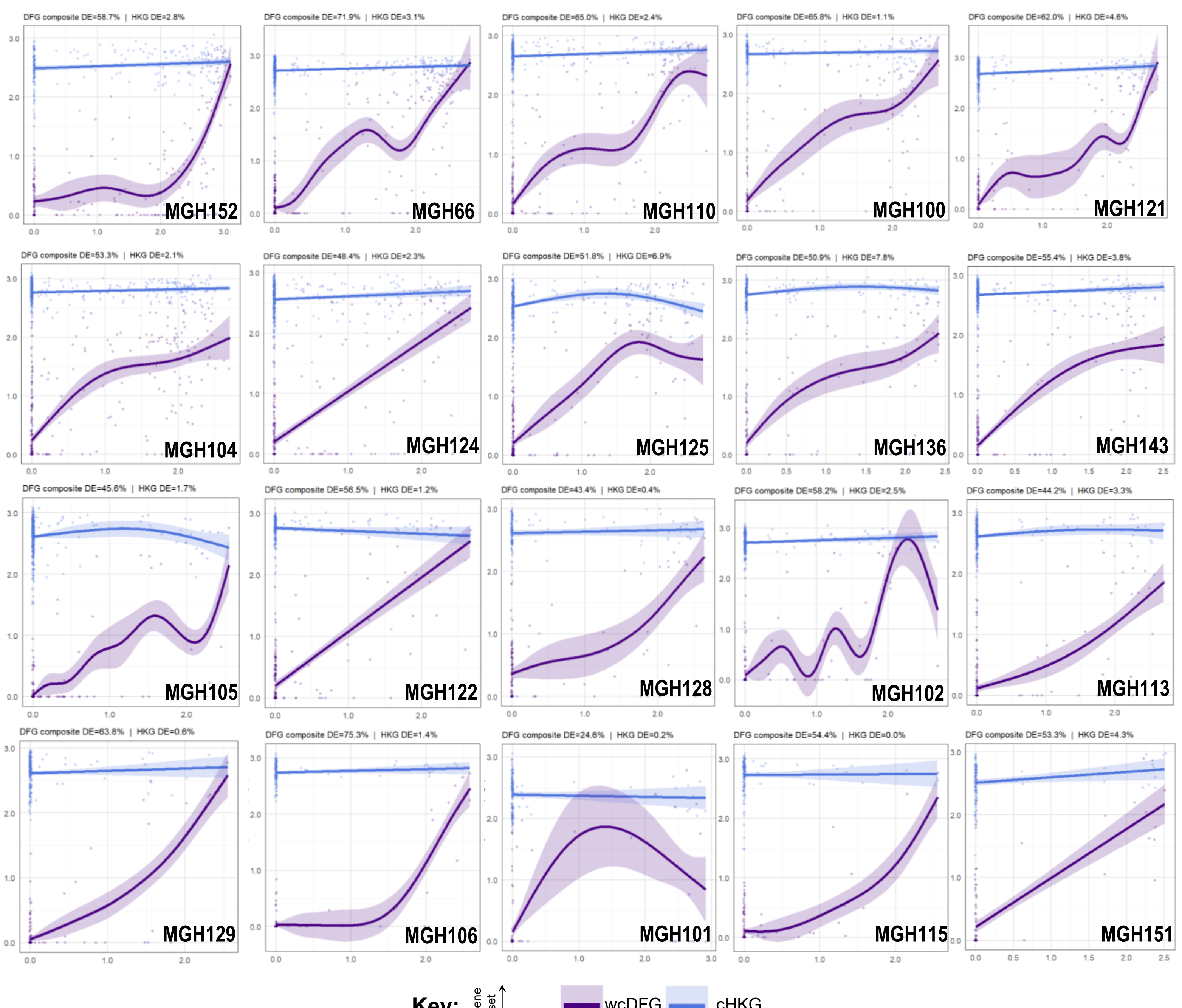

**Supplementary Figure 7.** GAM curves of *FOXM1* vs. wcDFG or cHKG expression fitted to all cells (including *FOXM1*$^-$ cells) in each patient, ordered by within-positive monotonicity classification. Fluctuating curves in multiple patients (*e.g.*, MGH66, MGH110, MGH121) reflect zero-inflation distortion due to jointly-zero cells, showing the necessity of restricting GAM fitting to *FOXM1*$^+$ cells only for downstream CEP-IP analysis.

## Supplementary Figure 8

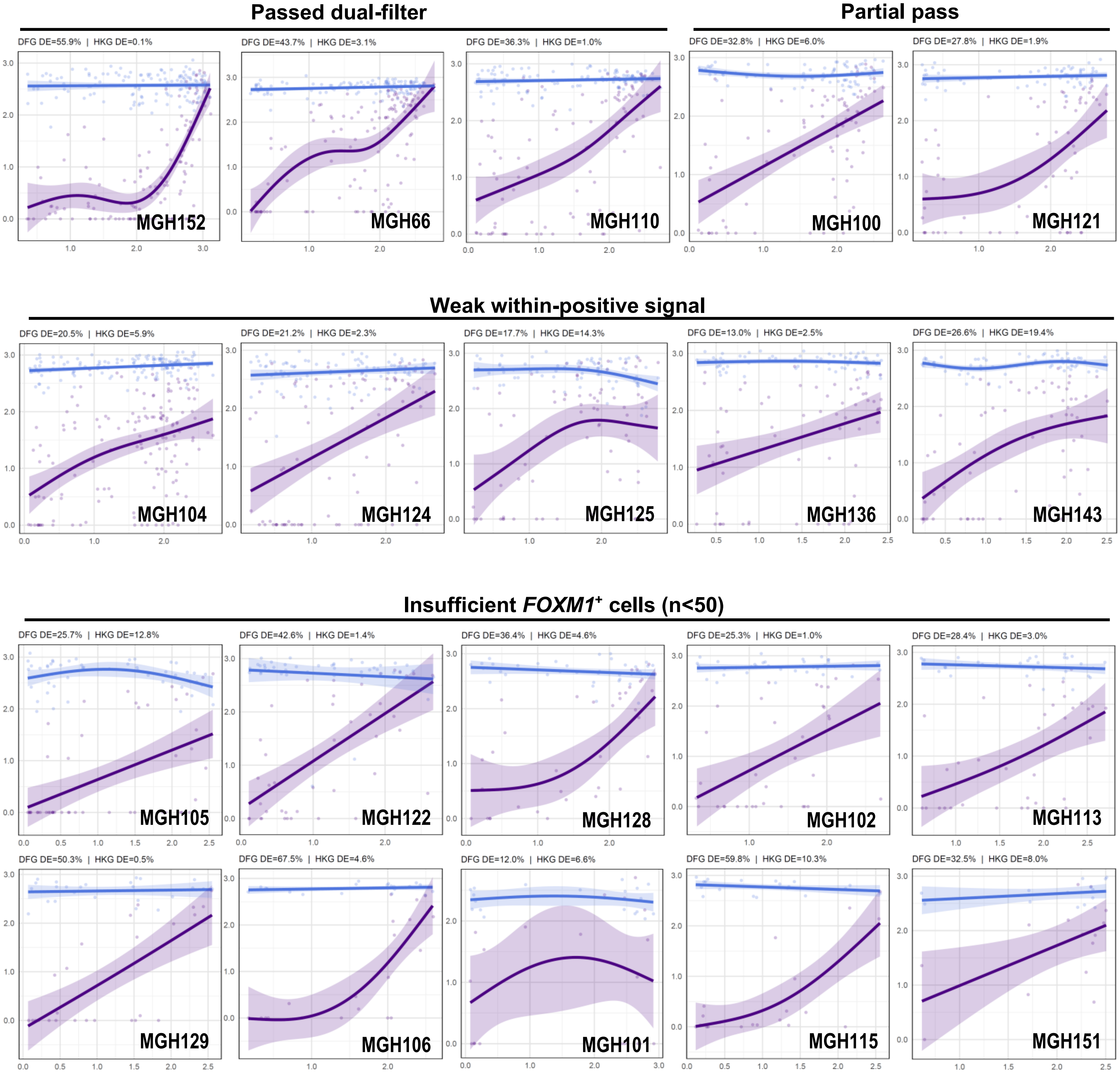

**Supplementary Figure 8**. Fitted GAM of wcDFG or HKG vs. *FOXM1* in *FOXM1*$^{+}$ cells of the Neftel GBM dataset. Take note that the wcDFG GAM curves here are smooth without fluctuations (unlike those seen in Supplementary Figure 7) due to absence of zero-inflation as all cells here were *FOXM1*$^{+}$. Passed dual-filter: Correlations of wcDFG with *FOXM1* passed the Spearman-Kendall dual-filter; Partial pass: Correlations of wcDFG with *FOXM1* passed either Spearman or Kendall filter; Weak within-positive signal: Correlations of wcDFG with *FOXM1* did not pass either Spearman or Kendall filter.

**Supplementary Figure 9**

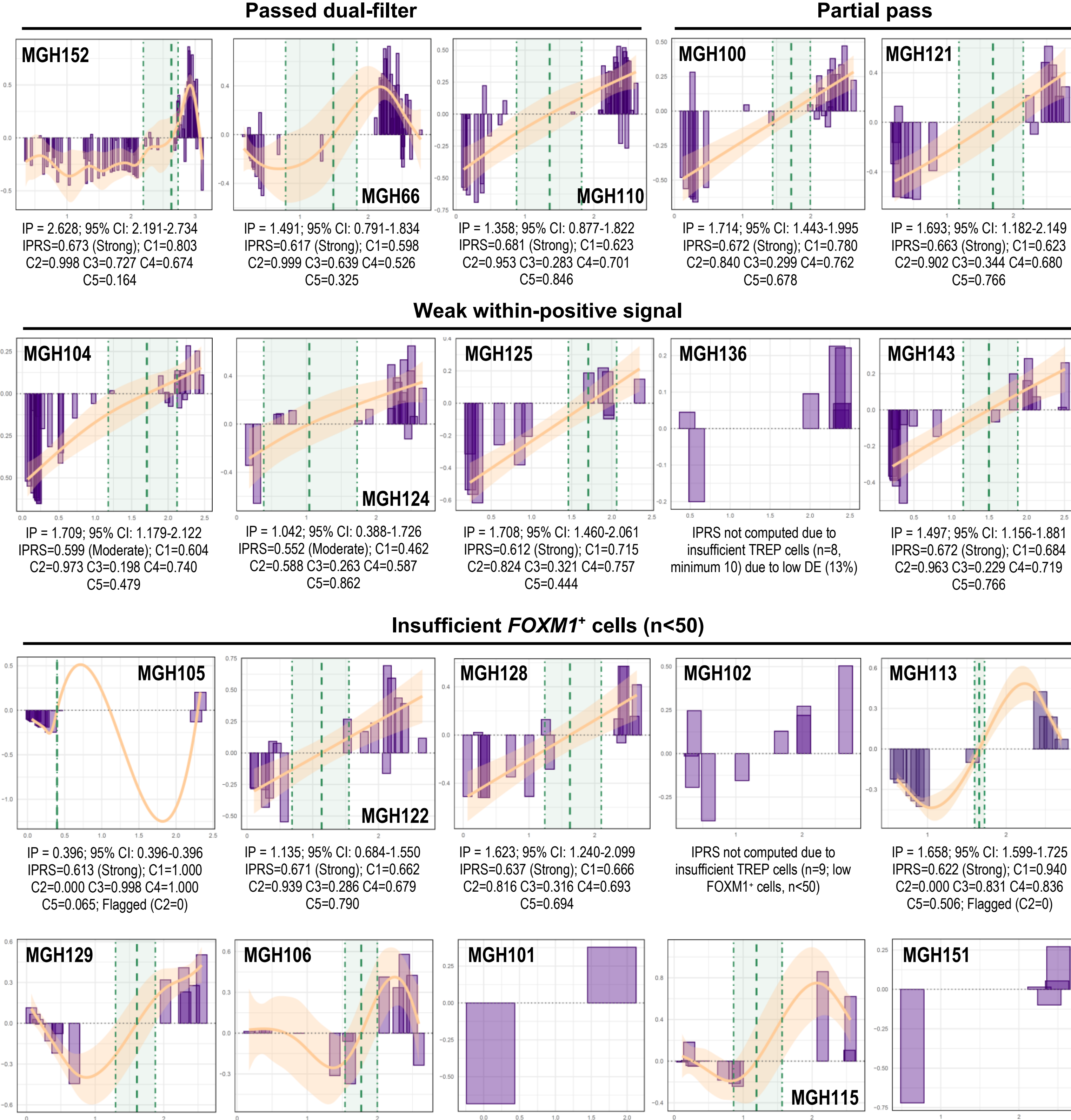

**Supplementary Figure 9**. Automated IP detection and IPRS scoring of *FOXM1*+ cells of the Neftel GBM dataset. Passed dual-filter: Correlations of wcDFG with *FOXM1* passed the Spearman-Kendall dual-filter; Partial pass: Correlations of wcDFG with *FOXM1* passed either Spearman or Kendall filter; Weak within-positive signal: Correlations of wcDFG with *FOXM1* did not pass either Spearman or Kendall filter.

**Supplementary Figure 10**

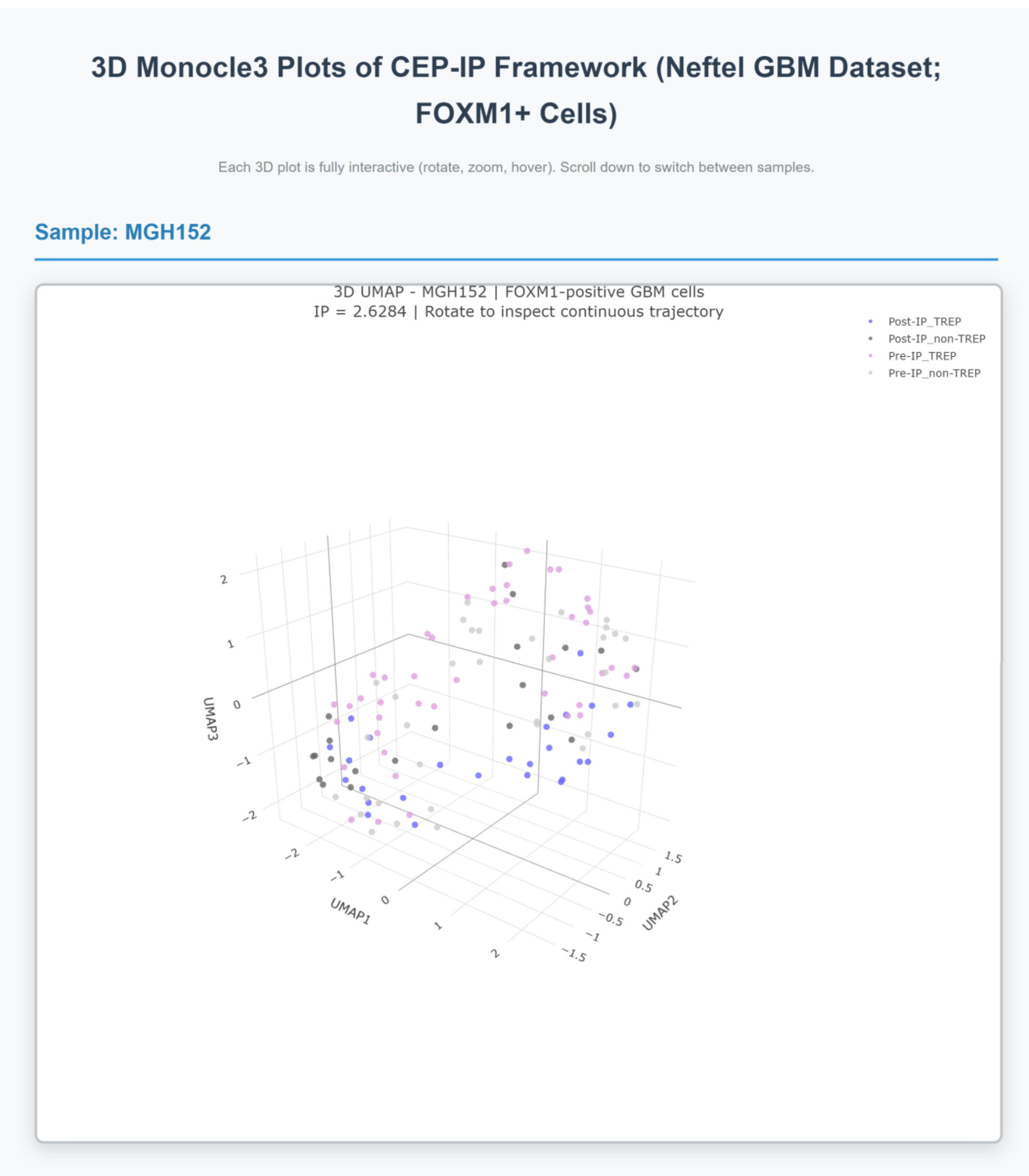

**Supplementary Figure 10**. *(Screenshot of the interactive HTML file available in the Supplementary Materials, where the plot is rotatable and scalable)*. 3D Monocle3 trajectories of TREP and non-TREP cells in dual-filter-passed (MGH152, MGH66, MGH110), partial pass (MGH100, MGH121), and weak within-positive (MGH104) cases.